\documentclass[sn-mathphys,Numbered]{sn-jnl}

\usepackage{graphicx}%
\usepackage{multirow}%
\usepackage{amsmath,amssymb,amsfonts}%
\usepackage{adjustbox} 
\usepackage{amsthm}%
\usepackage{mathrsfs}%
\usepackage[title]{appendix}%
\usepackage[dvipsnames]{xcolor}%
\usepackage{textcomp}%
\usepackage{manyfoot}%
\usepackage{booktabs}%
\usepackage{algorithm}%
\usepackage{algorithmicx}%
\usepackage{algpseudocode}%
\usepackage{listings}%
\usepackage{float}%
\usepackage{makecell}
\usepackage{orcidlink} 
\usepackage[nohyperlinks]{acronym}

\usepackage{multicol} 

\usepackage{subcaption}
\usepackage{mwe}

\begin{document}

\title[Article Title]{\bf Statistical Analysis of Scientific Metrics in High Energy, Cosmology, and Astroparticle Physics in Latin America}



\author*[1]{\fnm{Manuel} \sur{Morales-Alvarado}\,\orcidlink{0000-0002-3974-3448}}\email{mmorales@sissa.it}
\equalcont{These authors contributed equally to this work.}

\author[2,3]{\fnm{Fernando} \sur{Quevedo}\,\orcidlink{}}\email{fq201@cam.ac.uk}
\equalcont{These authors contributed equally to this work.}

\author[2]{\fnm{Mario} \sur{Ramos-Hamud}\,\orcidlink{}}\email{mr895@cam.ac.uk}
\equalcont{These authors contributed equally to this work.}

\author*[4]{\fnm{Diego} \sur{Restrepo}\,\orcidlink{}}\email{restrepo@udea.edu.co}
\equalcont{These authors contributed equally to this work.}

\affil[1]{\orgname{INFN, Sezione di Trieste}, \orgaddress{\street{SISSA, Via Bonomea 265}, \city{Trieste}, \postcode{34136}, \country{Italy}}}

\affil[2]{\orgdiv{Department of Applied Mathematics and Theoretical Physics}, \orgname{University of Cambridge}, \orgaddress{\street{Wilberforce Rd}, \city{Cambridge}, \postcode{CB3 0WA}, \state{Cambridgeshire}, \country{United Kingdom}}}

\affil[3]{\orgname{New York University Abu Dhabi}, \orgaddress{\street Saadiyat Island}, \city{Abu Dhabi}, \postcode{PO Box 128199}, \country{UAE}}

\affil[4]{\orgname{University of Antioquia}, \city{Medellín}, \country{Colombia}}


\abstract{We perform a comprehensive statistical analysis of key scientific metrics to evaluate the productivity and impact of research conducted in Latin American countries within the fields of \ac{HECAP}. Using data from the widely used open-access digital library INSPIRE-HEP, we provide a detailed assessment of the scientific contributions from the continent over the past 70 years. We provide data for the evolution of the overall productivity in the region relative to the rest of the world, comparing the productivity of each country, number of active researchers, number of publications, citations, h-index in total and relative to the population and number of researchers, as well as the productivity and impact compared to the percentage of \ac{GDP} invested in research, and the \ac{HDI} of each country. We also analyse collaborations among the different countries, as well as collaborations with the rest of the world. Additionally, we studied the gender gap evolution over the same period. This pioneering analysis, which relies solely on open data, can serve as an essential resource for researchers and policymakers alike. It aims to empower scientists with insights into the significance of their contributions to both regional and global research. Moreover, it provides both researchers and policymakers with critical quantitative data, strengthening their understanding of the progress in scientific productivity over the years to better support scientific endeavours in Latin America.}




\maketitle

\newpage
\tableofcontents 

\newpage
\section{Introduction}
\label{sec1}

Latin American countries share many common geographical and historical features to be considered as a unit in many regards, from culture, economics and politics to many strata of science.
Scientific activity in the region dates back to impressive achievements in mathematics and astronomy during pre-Columbian times, and over the centuries some areas of science such as medicine, physiology and related subjects have had a long tradition \cite{history}. 

However, research in physical sciences is relatively young in the region, essentially dating back to the early 20th century, when the effort of some local prominent figures, such as Manuel Sandoval-Vallarta, Marcos Moshinsky,  Juan Jose Giambiagi, and Jose Leite-Lopes led to the establishment of physics research, mostly nuclear and particle physics, in Mexico, Argentina and Brazil. Less activity took place in the smaller countries, with a few exceptions such as the role of Cesar Lattes at the Chalcataya observatory, in Bolivia, in confirming the discovery of pions in the late 1940s. But most developments can be traced to the era after the Second World War (see \cite{Masperi:2000yn,violini} for brief  descriptions of the history of Latin American science and \ac{HEP} in particular).  This is where our study begins.

The objective of this article is to analyse the evolution of physics research in Latin America by conducting a detailed study of its development over time. Ref. ~\cite{Sanchez:2017xqi} provided very valuable insights into theoretical high energy physics within a selected group of Latin American countries between 1990 and 2012. Our study expands both the geographical and temporal scope, considering a larger set of nations, scientific metrics (collective metrics of the continent, authors, analysis by gender, collaborations within and beyond Latin America, etc) and a broader range of research areas. This analysis is made possible by the availability of extensive bibliographic databases, which enable us to systematically extract information on how physics has evolved across these countries, particularly within the broad fields of \ac{HECAP}.

Knowing the historical origin of physics in the region, these areas are representative of the whole of physics that includes many other subjects such as condensed matter and atomic and molecular physics, as well as environmental sciences and other applied subjects. The main reason to keep to this area is the existence of a very complete database, INSPIRE-HEP~\cite{Ivanov:2010zzf}\footnote{\url{https://inspirehep.net}}, which has been available for the past 50 years, collecting all publications on these fields and going back to early 20th century publications. But, in the future, we plan to expand this study to all other areas of science using other related sources.

We will use this database to address the following aspects:
\begin{enumerate}
\item Evolution of the total scientific production and impact in these areas as a percentage of the world production over the past 70 years. 
\item Extract the number of active scientists at present as defined by those that have written scientific publications in the past 5 years (prior to the date of the database extraction in July 2022) and generate a database for each country with the contacts of all authors. 
\item Evolution of the number of publications per country since the year of their first publication. 
\item Compare the evolution of scientific productivity among the big countries (e.g. Brazil and Mexico), mid-size countries (e.g. Colombia and Venezuela) and the smaller countries (e.g. Guatemala and Bolivia). 
\item The total number of citations and h-index. To account for the different sizes of the Latin American countries in our study, we will also present these metrics under different normalisations such as their total population and the number of total and active authors. 
\item Extract correlations between scientific production and impact with investment in research and development as a percentage of the \ac{GDP} of each country, along with its \ac{HDI}. 
\item Extract information about gender representation among researchers to examine the relative proportion of active male and female scientists.  

\item Evaluate the level of collaborations among all the Ibero-American countries (including Portugal and Spain) and their collaboration with scientists based in other countries.
\end{enumerate}

At present, our study is restricted only to publications with 10 or fewer authors to identify the relevant information without being diluted by the countless publications with thousands of authors which are common in the experimental components of these areas. Including these long collaborations will be the subject of a future study; however, a curious reader interested in big collaborations may find useful \cite{florio2019, 10.1007/s11192-019-03030-1}.

This study was prompted by an initiative to organise the high energy and astro-particle physics community of Latin America, following a mandate given by the Ministers of Science of the Ibero-American countries. The mandate consisted of a pilot project to create a strategic forum towards planning research infrastructures in the region\footnote{\url{https://lasf4ri.org}} following in the footsteps of similar initiatives in Europe and North America. A first step was achieved by collecting almost 50 different projects. The result was assessed by a preparatory group composed of leading physicists from the region and reported to a \ac{HLSG} comprised of world-leading scientists. The report is published in 
\cite{Aihara:2021unm}. The main recommendation from the \ac{HLSG} was to create an official society, which now exists with more than 400 members\footnote{\url{https://www.ictp-saifr.org/laa-hecap}}. This project then came out as a concrete way to collect all the information regarding the composition and activities of all active scientists in this field.

The uses for this study are multi-fold. For scientists themselves, it is helpful to provide them with a better understanding of how they stand in terms of research in the region, to know the current active researchers, and to open up potential future collaborations. This analysis is also useful for policymakers to acknowledge the current situation regarding research in the region and how it keeps evolving. It is worth mentioning that basic research, from the commercial perspective, can be regarded as a public good, however, even though it has been proven to be highly beneficial for the development of societies to achieve national objectives \cite{oxfordhb}, it lacks an immediate monetizing mechanism and therefore seems less attractive to investors. This is precisely why new policies have to be implemented by the state in order to fund and achieve optimal results in basic science \cite{10.1093/oxrep/grx001}, particularly in theoretical \ac{HEP}. The impact of funding on scientific and economic development was recently studied in \cite{miao2023cooperation}. Thus, this work can be used as a quantitative metric to decide on future funding for scientific projects.

Naturally, this study can be extended in several ways and it is expected to be only the first step towards a bigger and more comprehensive analysis. Extending this work to other areas of science should be a natural next step, following on the lines of the ministerial meeting. Furthermore, this study hopes to inspire similar studies for other geographical regions. 

The structure of this work is as follows. In Sect. \ref{sec:methodology} we introduce the methodology of the analysis and aspects of the database extracted from INSPIRE-HEP. In Sect. \ref{sec:sup_inf} we introduce additional information that is useful for normalising the metrics of each country to account for their size, and studying the correlation between their scientific metrics and the amount of funds the countries allocate to \ac{RD} and their \ac{HDI}. After that, we discuss our main results in different aspects. Sect. \ref{sec:coll_latam} presents collective Latin American metrics,  addressing question 1. above; Sect. \ref{sec:auths_ranks} discusses authors and their ranks for each country, addressing question 2; Sect. \ref{sec:pubs} shows the status of the publications in each Latin American country, addressing questions 3 and 4, and Sect. \ref{sec:cits_hindex} discusses the citation and h-index of the articles that each country has published which corresponds to question 5. In Sect. \ref{sec:corr_latam} we discuss how the scientific output of each country correlates with their investment on \ac{RD}, addressing question 6. Then, Sect. \ref{sec:gender} displays a gender analysis of the authors in each country, addressing question 7. Sect. \ref{sec:collab}  presents collaboration metrics among Latin American countries and with Ibero-America in general, addressing question 8. Finally, in Sect. \ref{sec:summary} we summarise the main findings of this study, our conclusions, and possible directions in which the analysis can be extended. Apps. \ref{app:pub}, \ref{app:cit}, and \ref{app:ranks} provide supplementary information on the publications, citations, and ranks of each country, respectively. Apps. \ref{app:arxiv} and \ref{app:acro} provide respectively some information regarding primary arXiv categories and acronyms used throughout the text.

\section{Methodology and database}
\label{sec:methodology}
In this section, we describe the data used in this study and the reasoning for our choices
of restrictions and subsets of this data. 

For this analysis, we use the INSPIRE-HEP~\cite{Ivanov:2010zzf} database, which indexes all the relevant literature in the field of high energy, cosmology, and astroparticle physics. Full programmatic access to the database is provided through the INSPIRE-HEP REST API~\cite{Moskovic:2021zjs}, with well-documented entities for authors and institutions based on the ORCID\footnote{\url{https://orcid.org/}} and ROR\footnote{\url{https://ror.org/}} identifiers, respectively.
The platform provides an author disambiguation system that generates automatic author profiles, including full names and current affiliated institutions.
Authors can access these profiles using their ORCID credentials to correct and augment information regarding their careers, 
advisors across various educational stages, and personal data. All this information is accessible through the same REST API. 
Our analysis indicates that around $60\%$ of the authors have used this feature to enhance their profiles on the website.   
For a similar study using Web of Science, see~\cite{Sanchez:2017xqi}. 

We have built a database of the authors of articles with Latin American affiliations in INSPIRE-HEP. For that, 
we scroll through the full list of around 11 thousand institutions in INSPIRE-HEP, \url{https://inspirehep.net/institutions}. 
For each of them, we check for a Latin American country and capture its exact form name. With this, 
we can get the full literature associated with each institution,  and from that, can analyse the affiliation data of each article, as well as
to extract the full information about the authors and their profiles in INSPIRE-HEP.

To avoid the bias from hyper-authorships in experimental \ac{HEP}~\cite{10.1002/asi.1097,10.1162/qss_a_00092}, 
particularly in large-scale collaborations such as those at the LHC, we apply the built-in filter in INSPIRE-HEP 
to focus on articles with 10 authors or fewer\footnote{For a more general analysis of the impact of Latin American countries in international 
collaborations using the INSPIRE-HEP data, see~\cite{10.1007/s11192-019-03030-1} and \cite{arXiv:2411.10278}.}. This approach allows us to better assess the individual contributions 
of researchers from a given country, as the attribution of impact can become less clear in highly multi-authored publications. 
The complete dataset is available at~\cite{10.5281/zenodo.14852123} for the run in July 2022.

From the dataset, we obtain access to a wide range of bibliometric information, including authors, articles, citations, institution and country affiliations, among others. The database is constructed using the INSPIRE-HEP API, which, while comprehensive, is subject to minor inconsistencies, such as occasional missing metadata (articles lacking a publication year, for example) or small retroactive updates to the database (for instance, a change in the number of citable articles in 1950 from 132 to 133 during our study). Additionally, particularly for older records, INSPIRE-HEP occasionally indexes articles that, while relevant, may fall slightly outside the core scope of \ac{HECAP}. However, such instances are rare and overall these minor issues do not affect the robustness of the analysis.

In total, our dataset includes bibliometric information for the following Latin American countries\footnote{Although Puerto Rico appears in the dataset, its entries presented inconsistencies in INSPIRE-HEP at the moment of the extraction. A significant portion of its contributions is indexed under United States affiliations, and some records are misclassified as belonging to institutions of other countries. Due to these issues, we exclude Puerto Rico from the present analysis.}:  
\begin{multicols}{2}
\begin{enumerate}
    \item Argentina
    \item Bolivia
    \item Brazil
    \item Chile
    \item Colombia
    \item Costa Rica
    \item Cuba
    \item Dominican Republic
    \item Ecuador
    \item El Salvador
    \item Guatemala
    \item Honduras
    \item Mexico
    \item Panama
    \item Paraguay
    \item Peru
    \item Uruguay
    \item Venezuela
\end{enumerate}
\end{multicols}
In this study, we obtain individual scientific metrics for the countries above, as well as joint metrics for the entire continent.


\section{Supplementary information}
\label{sec:sup_inf}

To refine the metrics of scientific output and their impact, it is beneficial to introduce additional data that normalises these metrics. This includes the population of each country and the proportion of their \ac{GDP} invested in research and development.

The population of the Latin American countries in our analysis is sourced from \ac{UN} data\footnote{Checked in July 2023.} on
\begin{center}
\url{http://data.un.org}.
\end{center}
Our method to obtain the metrics per capita consists of simply dividing \ac{PMI} using the values in Table \ref{table:population}. 
\begin{table}[h]
\begin{tabular}{@{}ll@{}}
\toprule
Country & Population \\
\midrule
Argentina    & 45,606,000   \\
Bolivia    & 11,833,000  \\
Brazil   & 213,993,000   \\
Chile    & 19,212,000 \\
Colombia    & 51,266,000\\
Costa Rica  & 5,139,000\\
Cuba    & 11,318,000   \\
Dominican Republic & 10,954,000  \\
Ecuador  & 17,888,000   \\
El Salvador & 6,518,000   \\
Guatemala   & 18,250,000   \\
Honduras  & 10,063,000  \\
Mexico & 130,262,000    \\
Panama   & 4,382,000   \\
Paraguay  & 7,220,000  \\
Peru & 33,359,000    \\
Uruguay  & 3,485,000   \\
Venezuela  & 28,705,000  \\
\botrule
\end{tabular}
\caption{Population of each country according to the most up-to-date data provided by the \ac{UN} in 2021.}\label{table:population}%
\end{table}

The \ac{GDP} percentages allocated to research and development by countries in our analysis come from World Bank data\footnote{Information validated in July 2023.} on \url{https://datos.bancomundial.org/indicador/GB.XPD.RSDV.GD.ZS}. We use the values in the second column in Table \ref{table:PIB}, which consist of the most recent information available for each country. The decision to use these values is simply a prescription to describe potential correlations between investment in science and scientific output. Alternative prescriptions could involve taking the average \ac{GDP} for \ac{RD} over the years that have an entry (third column of Table \ref{table:PIB}), or taking the \ac{GDP} values shifted by a couple of years to account for the fact that a potential increase/decrease of the investment in science might not have effects in the same year. Another difficulty is that not all countries have provided information on their \ac{GDP} values. The entries vary greatly: some countries do not have a single entry, others that have very sparse entries, and a few that are mostly complete. To have a streamlined comparison, we will simply use the most up-to-date value on the World Bank data. At the time we have examined the website, all countries but the Dominican Republic have had at least one entry. 

The number in parenthesis signifies that the values used are from 10 years ago or older.
\begin{table}[htp]
\begin{tabular}{@{}lll@{}}
\toprule
Country & \makecell{Percentage of GDP \\ (most recent)} & \makecell{Average\\ (1996-2021)} \\
\midrule
Argentina & 0.46 & 0.49\\
Bolivia & 0.16 & 0.28 \\
Brazil & 1.21 & 1.12 \\ 
Chile & 0.34 & 0.35 \\
Colombia & 0.29 & 0.22 \\
Costa Rica & 0.37 & 0.4  \\
Cuba & 0.52 & 0.47 \\
Dominican Republic & (No info available) & (No info available)\\
Ecuador & 0.44 (2014)& 0.2 \\
El Salvador & 0.17 & 0.11 \\
Guatemala & 0.03 & 0.04 \\
Honduras & 0.04 (2004) & 0.04 \\ 
Mexico & 0.3 & 0.36 \\
Panama & 0.15 & 0.2 \\
Paraguay & 0.14 & 0.09 \\
Peru & 0.17 & 0.11 \\
Uruguay & 0.48 & 0.34 \\
Venezuela & 0.34& 0.29 \\
\botrule
\end{tabular}
\caption{Percentage of \ac{GDP} invested in Research and Development.}
\label{table:PIB}%
\end{table}

Table \ref{table:PIB} shows an extended range of investment levels, from as low as 0.11\% to as high as 1.12\% in the averages, indicating significant variation in \ac{RD} prioritisation across the region. Notable is Brazil's average investment at 1.12\%, the highest listed, whilst for several countries it is less than 0.2\%. The data for some countries, like the Dominican Republic, is not available, and for others like Ecuador and Honduras, the figures are based on data from previous years, 2014 and 2004 respectively. 

In comparison to industrialised nations, Latin American countries typically invest a smaller percentage of their \ac{GDP} in \ac{RD}. Table \ref{table:PIB_orgs} illustrates a selection of \ac{RD} investment figures from various global entities.
\begin{table}[h]
\begin{tabular}{@{}ll@{}}
\toprule
Entity & Percentage of \ac{GDP} \\
\midrule
China    & 2.24   \\
European Union average & 2.22  \\
OECD average   & 2.67  \\
\ac{UN} average  & 2.33 \\
United States   & 3.17 \\
\botrule
\end{tabular}
\caption{Percentage of \ac{GDP} invested in Research and Development. Note that the \ac{UN} value is clearly an overestimation (see the argument in the main text below).}\label{table:PIB_orgs}%
\end{table}

Note, however, that the values presented in Table \ref{table:PIB_orgs} for global entities' investment in \ac{RD} may not accurately reflect the true figures. These averages are computed from the data available to organisations such as the World Bank. Often, countries with lower \ac{RD} investments do not have their data included in the World Bank database. Consequently, when a value is missing, it is omitted from the average, which can result in an overestimation of the actual investment levels, as it happens in the case of the \ac{UN} average.

\begin{table}[htp]
\begin{tabular}{@{}lllll@{}}
\toprule
Year & Citeable articles  & Citations & h-index & Citations/article (avg) \\
\midrule
1946 & 49 & 6118 & 26 & 124.9 \\
1947 & 121 & 9619 & 38 & 79.5 \\
1948 & 121& 16166& 51 & 133.6 \\
1949 & 137 & 20202 & 54 & 147.5 \\
1950 & 133 & 15331 & 54 & 115.3 \\
1951 & 138 & 22163 & 50 & 160.6 \\
1952 & 177 & 15343 & 48 & 86.7 \\
1953 & 339 & 16782 & 59 & 49.5 \\
1954 & 226 & 22796 & 62 & 100.9 \\
1955 & 286 & 18841 & 58 & 65.9 \\
1956 & 423 & 26183 & 68 & 61.9 \\
1957 & 473 & 42566 & 84 & 90 \\
1958 & 483 & 38311 & 90 & 79.3 \\
1959 & 504 & 30087 & 75 & 59.7 \\
1960 & 429 & 35145 & 90 & 81.9 \\
1961 & 440 & 59921 & 85 & 136.2 \\
1962 & 562 & 57409 & 101 & 102.2 \\
1963 & 674 & 55066 & 99 & 81.7 \\
1964 & 819 & 91087 & 107 & 111.2 \\
1965 & 1263 & 53016 & 107 & 42 \\
1966 & 1312 & 74784 & 113 & 57 \\
1967 & 3269 & 124667 & 134 & 38.1 \\
1968 & 4478 & 119420 & 145 & 26.7 \\
1969 & 7257 & 153964 & 151 & 21.2 \\
1970 & 6874 & 144197 & 147 & 21 \\
1971 & 7510 & 159664 & 156 & 21.3 \\
1972 & 7747 & 175765 & 175 & 22.7 \\
1973 & 7849 & 239986 & 179 & 30.6 \\
1974 & 7784 & 294710 & 210 & 37.9 \\
1975 & 7411 & 221292 & 194 & 29.9 \\
1976 & 8196 & 256508 & 216 & 31.3 \\
1977 & 8013 & 307609 & 219 & 38.4 \\
1978 & 8132 & 260804 & 216 & 32.1 \\
1979 & 8592 & 302464 & 241 & 35.2 \\
\botrule
\end{tabular}
\caption{Total number of articles in the database with 10 authors or less per publication from 1946 to 1979. This data was consulted on the 16th of February 2024.}\label{table:global-papernumber1}%
\end{table}

\begin{table}[t!]
\begin{tabular}{@{}lllll@{}}
\toprule
Year & Citeable articles  & Citations & h-index & Citations/article (avg) \\
\midrule
1980 & 8522 & 320810 & 236 & 37.6 \\
1981 & 8979 & 309842 & 241 & 34.5 \\
1982 & 8699 & 322891 & 243 & 37.1 \\
1983 & 9618 & 385425 & 265 & 40.1 \\
1984 & 9492 & 334765 & 244 & 35.3 \\
1985 & 10054 & 358669 & 256 & 35.7 \\
1986 & 10190 & 362233 & 252 & 35.5 \\
1987 & 10810 & 373711 & 258 & 34.6 \\
1988 & 9767 & 319651 & 236 & 32.7 \\
1989 & 12198 & 359504 & 247 & 29.5 \\
1990 & 12299 & 407209 & 265 & 33.1 \\
1991 & 12428 & 383980 & 256 & 30.9 \\
1992 & 13935 & 432264 & 264 & 31 \\
1993 & 15188 & 489389 & 277 & 32.2 \\
1994 & 16860 & 539012 & 281 & 32 \\
1995 & 18352 & 610358 & 304 & 33.3 \\
1996 & 22345 & 730990 & 312 & 32.7 \\
1997 & 22345 & 730939 & 312 & 32.7 \\
1998 & 23689 & 895064 & 345 & 37.8 \\
1999 & 24546 & 858961 & 326 & 35 \\
2000 & 25685 & 897344 & 345 & 34.9 \\
2001 & 27763 & 889223 & 337 & 32 \\
2002 & 27116 & 962691 & 332 & 35.5 \\
2003 & 28607 & 1012780 & 357 & 35.4 \\
2004 & 29172 & 1004508 & 338 & 34.4 \\
2005 & 32044 & 980120 & 322 & 30.6 \\
2006 & 32389 & 1041917 & 333 & 32.2 \\
2007 & 34495 & 981784 & 317 & 28.5 \\
2008 & 34057 & 1070974 & 330 & 31.4 \\ 
2009 & 35388 & 1007662 & 321 & 28.5 \\
2010 & 38724 & 1044581 & 324 & 27 \\
2011 & 41214 & 1026718 & 312 & 24.9 \\
2012 & 41530 & 1075761 & 312 & 25.9 \\
2013 & 40290 & 1032334 & 316 & 25.6 \\
2014 & 36359 & 937763 & 287 & 25.8 \\
2015 & 38024 & 898160 & 274 & 23.6 \\
2016 & 36899 & 852117 & 277 & 23.1 \\
2017 & 37324 & 776436 & 241 & 20.8 \\
2018 & 33026 & 515254 & 179 & 15.6 \\
2019 & 39620 & 683985 & 227 & 17.3 \\
2020 & 36144 & 410667 & 139 & 11.4 \\
2021 & 41154 & 613821 & 191 & 14.9 \\
2022 & 52329 & 282253 & 98 & 5.4 \\
2023 & 54986 & 99411 & 56 & 1.8 \\
\botrule
\end{tabular}
\caption{Total number of articles in the database with 10 authors or less per publication from 1980 to 2023. This data was consulted on the 16th of February 2024.}
\label{table:global-papernumber2}%
\end{table}

In addition and for comparative reasons, we also present some statistics from a global perspective. We provide a comprehensive overview of the total number of articles in the database from all over the world, with 10 or fewer authors per publication from 1946 to 2023. This data encompasses various key metrics including the number of citeable articles, citations, h-index, and average citations per article. We have split this data into two tables to facilitate the formatting. Table \ref{table:global-papernumber1} provides coverage from 1946 until 1979 and Table \ref{table:global-papernumber2} from 1970 until 2023. This allowed us to make a comparative analysis of the scientific production in Latin America in contrast with the world production in Figs. \ref{fig:papers_world} and \ref{fig:citations_world}. In the early years (1946-1950), both the number of citeable articles and citations were relatively low. However, there was a gradual increase in both metrics over time, indicating a growing scientific activity. Then, the 1950s and 1960s witnessed significant growth in the number of citeable articles and citations. This period also saw a steady rise in the h-index, indicating an increase in the impact of published research. During the 1970s and 1980s, there was a marked growth in both the number of citeable articles and citations. In the 1990s and 2000s, although the growth in the number of citeable articles and citations persisted, it was at a slightly slower pace compared to previous decades. Nonetheless, the average number of citations per article remained relatively stable. Finally, in recent years (2010-2023), there has been a noticeable increase in the number of citeable articles and citations. 

In this study, we considered different normalisations for our results, for instance, the population of each country or the number of authors in each one. On the one hand, normalisation by population can be useful to obtain per-capita insights and allow for better comparability between countries of different sizes. Additionally, it provides us with an opportunity to shed light on policy insights to understand the overall engagement of society in scientific research, potentially informing education and research funding policies. On the other hand, normalising by number of authors enables us to directly quantify research productivity, which can give an intuition of the efficiency of the research workforce.

Furthermore, as a complementary socio-economic index, in Table \ref{table:HDI} we provide the average values of the HDI for each country, calculated to the greatest possible time extent from the United Nations Development Program: \url{https://hdr.undp.org/data-center/human-development-index#/indicies/HDI}.

\begin{table}[htp]
\begin{tabular}{@{}ll@{}}
\toprule
Country & Average HDI \\
\midrule
Argentina & 0.802 \\
Bolivia & 0.640 \\
Brazil & 0.698 \\ 
Chile & 0.790 \\
Colombia & 0.701 \\
Costa Rica & 0.742  \\
Cuba & 0.728 \\
Dominican Republic & 0.679 \\
Ecuador & 0.711 \\
El Salvador & 0.623 \\
Guatemala & 0.576 \\
Honduras & 0.572 \\ 
Mexico & 0.728 \\
Panama & 0.751 \\
Paraguay & 0.681 \\
Peru & 0.699 \\
Uruguay & 0.768 \\
Venezuela & 0.719 \\
\botrule
\end{tabular}
\caption{Average HDI of each country in our analysis.}
\label{table:HDI}
\end{table}
%

\section{Collective Latin American metrics}
\label{sec:coll_latam}

In this section, we present and analyse metrics that represent the scientific output of Latin America. By focusing on collective data, we aim to offer insight into the region's overall scientific contributions. This regional approach highlights the scientific trends, commonalities, and collaborative efforts across Latin American countries.

The analysis of collective metrics allows us to identify general patterns in scientific productivity, collaborations, and how Latin American research fits on the global stage. Furthermore, this perspective facilitates a broader understanding of how the efforts of the region's research align with or differ from global trends.

While this aggregated approach provides insights into the regional dynamics of scientific output, it does not focus on the nuances and specific characteristics of individual countries, which will be the focus of some of the following chapters. Instead, our aim here is to showcase Latin America’s contribution as a unified scientific community.

\begin{figure}[h!]%
\centering
\includegraphics[width=0.95\textwidth]{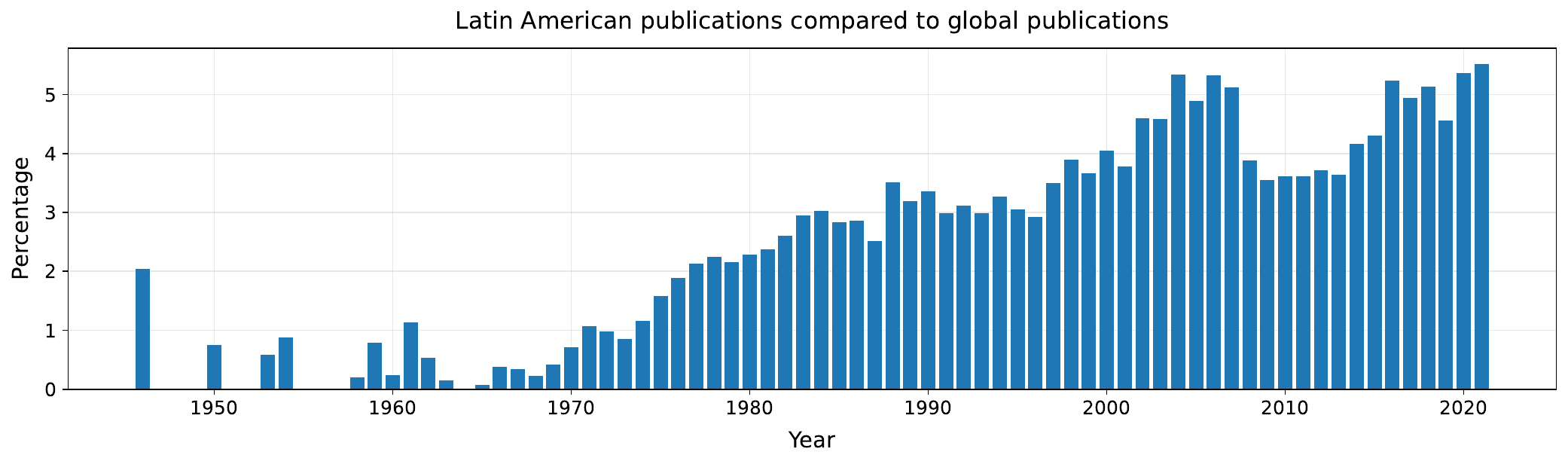}
\caption{\label{fig:papers_world} Percentage of Latin American publications compared to global publications per year.}
\end{figure}

Figure \ref{fig:papers_world} displays the percentage of Latin American publications relative to global publications (also restricted to those with 10 or fewer authors) from 1946, the year of the region's first publication, to 2021\footnote{The information on the global metrics was extracted from INSPIRE-HEP in November 2024.}.
%
In the earlier years, between the 1940s and 1960s, the percentage of Latin American publications is sporadic, with occasional increases, but remains relatively low overall. However, starting around the 1970s, there is a noticeable upward trend in scientific output. This trend continued through the 1980s and 1990s, with a steady increase in the percentage of global publications attributed to Latin America.

The figure shows that after the year 2000, Latin America's share of global publications reached a higher level, fluctuating between 4\% and 5\%, thus reflecting a more prominent role in global scientific contributions. In the late 2000s, we see a decrease in the number of publications to below 4\%. This number remains fairly constant until the early 2010s. The most recent data, around 2020, shows the highest recorded percentage, slightly exceeding 5\%. Although the number of publications has increased in recent years, the most recent number of publications is similar to the one of the mid-2000s. This overall trend suggests considerable growth in Latin American scientific output, particularly in the last few decades.
However, the stagnation observed in recent years indicates a need for continued efforts to sustain and further increase this progress.

\begin{figure}[h!]%
\centering
\includegraphics[width=0.95\textwidth]{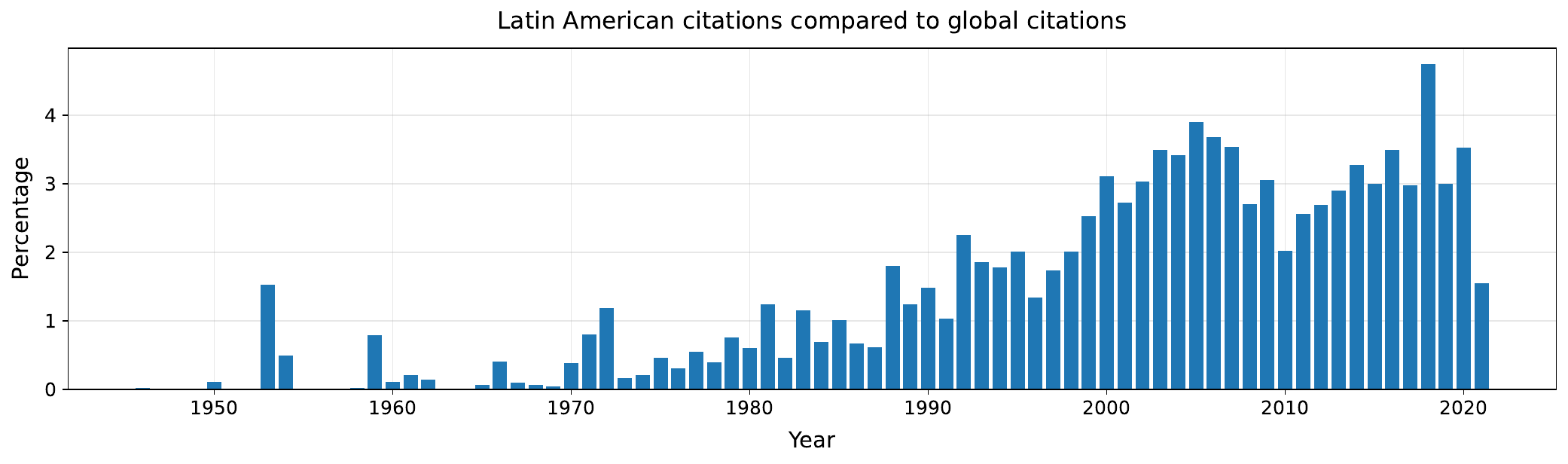}
\caption{\label{fig:citations_world} Percentage of Latin American citations compared to global publications per year.}
\end{figure}

In Fig. \ref{fig:citations_world} we show the percentage of the number of citations that articles published in a given year have, compared to the global citations. 
%
The results shown in this figure slightly reflect the trends observed in Fig. \ref{fig:papers_world}. The first non-zero result appears in 1950, indicating that publications from the 1940s have not been cited in the database. Until the 1970s, the percentage of citations fluctuates, with no consistent trend, but starting from that point, there is a general upward development. This increase continued until the late 2000s when we observed a drop in citations, followed by sustained growth afterwards.

The most recent years, however, show a notable decrease in citations, which can be attributed to the fact that these publications are relatively new, and there has not yet been sufficient time for them to accumulate citations. This time-lag effect is typical for recently published articles, as citations tend to build up over a longer period. Overall, the figure suggests that Latin American publications are receiving more citations over time, although recent articles have yet to achieve their full citation potential.

The collective results presented in this section highlight the growing scientific contribution of Latin America over the past several decades. While both the publication output and citation rates show a steady upward trajectory, there are periods of fluctuation that suggest room for further improvement. The recent trends, particularly the increase in both publications and citations, reflect the region’s increasing integration into the global scientific community. Though there is some stagnation in publication growth and a lag in citation accumulation for newer works, the overall trend is positive. With sustained effort, the region has the potential to strengthen its scientific presence on the global stage even further.

\section{Authors and ranks}
\label{sec:auths_ranks}

In this section, we address some of the relevant metrics that quantify the number of authors in each country as part of our study, and the ranks they have held at different stages of their research career in Latin America. The importance of authors in the fields of \ac{HECAP} in Latin America is paramount. Their role extends beyond just contributing to the body of research. They are integral to shaping the trajectory of these scientific disciplines. Through their work, authors not only advance knowledge but also foster the growth and development of the scientific community, influencing future research directions and collaborations. 

\begin{figure}[h!]%
\centering
\includegraphics[width=0.9\textwidth]{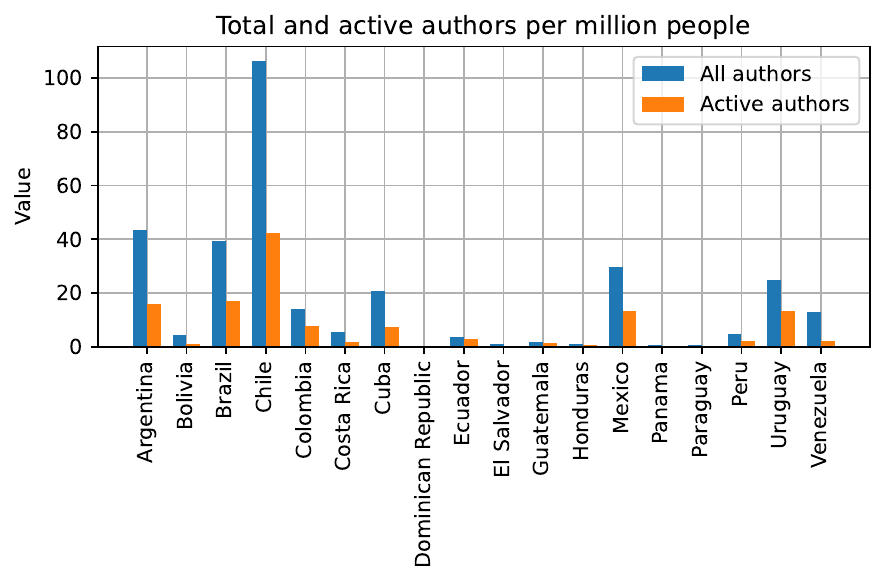}
\caption{Total and active authors.}
\label{fig:authors_plot} 
\end{figure}

In the context of our study, an \textit{author} is defined as an individual who has published at least one article while affiliated with an institution in Latin America. This definition recognises the diverse and interconnected nature of scientific research, where authors may have affiliations with multiple institutions across different countries. Consequently, if an author is associated with institutions in more than one country, their contributions are acknowledged in each of these countries, reflecting the cross-national collaboration and influence in our field. It should be noted that, in our analysis, the actual nationality or nationalities of the individuals are not considered. The emphasis is placed solely on their institutional affiliations, regardless of their national background. This approach allows us to focus on the impact and collaboration patterns within the \ac{HECAP} community based on institutional connections. In addition, we define an author as \textit{active} if they have at least one publication in the last 5 years. In table \ref{table:authors} we show the total number of authors and the subgroup of active authors that we used in the analysis for each country. In addition, we weighted these two categories \ac{PMI} to have a better metric considering the size of the population in each country. For a better visualisation of the difference in the size of both categories, Fig. \ref{fig:authors_plot} shows the total number of authors and compares it with the number of active authors for each country.   

\begin{table}[t]
\begin{tabular}{@{}lllll@{}}
\toprule
Country & \makecell{Total \\ authors} & \makecell{Active\\  authors} & \makecell{Total authors \\ \ac{PMI}} & \makecell{Active authors \\ \ac{PMI}}  \\
\midrule
Argentina & 1985 & 731 & 44 & 16 \\
Bolivia & 50 & 12& 4& 1 \\
Brazil & 8456 &  3659& 40& 17 \\ 
Chile & 2044 &  812& 106& 42\\
Colombia & 723 & 406& 14& 8\\
Costa Rica & 29 & 12& 6 & 2 \\
Cuba & 236 & 85& 21& 8\\
Dominican Republic & 1 & 0 & 0.1 & 0\\
Ecuador & 66&  54& 4& 3\\
El Salvador & 6 & 1 &1 & 0.2 \\
Guatemala & 32 & 28 & 2 & 2 \\
Honduras & 11 & 7& 1& 1 \\ 
Mexico & 3867 & 1749 &30 & 13 \\
Panama & 3 & 2& 1 & 0.5 \\
Paraguay & 5 & 3 & 1& 0.4\\
Peru & 156 & 68& 5& 2 \\
Uruguay & 87 & 47& 25 & 13\\
Venezuela & 370 & 58 &13 & 2\\
\botrule
\end{tabular}
\caption{Number of authors of each country and their subcategories.}
\label{table:authors}
\end{table}

%
Furthermore, we define \textit{rank} as an academic position occupied by a given author, and is an element of the set
\begin{description}
    \item[]  Ranks =  \{Senior, Junior, Staff, Postdoc, PhD, Master, Undergrad, Visitor, Other\}.
\end{description}
Notice that under this definition, one author can have more than a single rank and all of them will appear in the database if the author has declared it\footnote{It is worth mentioning that from the $29\,969$ unique authors in our database, only $7\,023$ have declared a profile with at least one rank, resulting in the total number in Table \ref{table:rank-country}. The quality of a profile is better if several ranks are declared. $5\,354$ authors have declared a profile with at least two ranks.}. Given the nature of this article, here we only discuss ranks that have taken place in Latin America and make its analysis threefold:

\begin{enumerate}
    \item Country: ranks declared as an affiliation in the academic trajectory of an author.

    \item Articles: ranks under which an article is published.
    
    \item Author-articles: relation between authors who have declared a rank obtained from a given country in Latin America, and the rank under which an article was published. 
\end{enumerate}
Each category is discussed individually in the following three subsections.

\paragraph{Ranks and countries} In this section, we discuss the rank declared by the authors if this has taken place or was obtained in Latin America, contributing to their academic trajectory. The list of declared ranks for each country is presented in Table \ref{table:rank-country}. As an example, in Mexico, 124 authors declared to have worked as postdocs, and 185 authors declared to have obtained their PhD in the country. This information is particularly relevant to understanding the number of students and academic jobs in the area for each country. In other words, this is a quantitative measurement of the production of scientific human capital. 
\begin{table}[htp]
\begin{tabular}{@{}llllllllll@{}}
\toprule
Country & Senior & Junior & Staff & Postdoc & PhD & Master & Undergrad & Visitor & Other \\
\midrule
 Argentina & 42 & 16 & 13 & 41 & 136 & 27 & 88 & 3 & 0 \\
 Bolivia &- & - &- & -& -& -& -&- & -\\
 Brazil & 232 & 65 & 46 & 345 & 445 & 239 & 254& 51 & 0\\ 
 Chile & 65 & 44& 4 & 110 & 108 & 46 & 77& 11 & 3\\
 Colombia & 32 & 19 & 5 & 15 & 15 & 21 & 48 & 1 & 0 \\
 Costa Rica & 4 & 0 & 1 & 0 & 0 & 0 & 1 & 0 & 0 \\
 Cuba & 3 & 1 & 2& 0& 2 &4 &11 &0 &0 \\
 Dom Rep & - & - & - &-  & - & - &- &- &- \\
 El Salvador & - & - & - &-  & - & - &- &- &- \\
 Ecuador & 1 & 0& 1 & 0& 1 & 0 & 2 &1 & 0\\
 Guatemala & 2 & 0 & 0& 1& 0 &0 & 4& 1& 0\\
 Honduras & 1 &0 & 0&0 &0 & 0& 0&0 &0 \\ 
 Mexico & 87 & 59 & 26 & 124 & 185 &71 & 85 & 10 & 1\\
 Panama & - & - & - &-  & - & - &- &- &- \\
 Paraguay & 1 & 0 &0 &0 &0 &0 &0 &0 &0\\
 Peru & 2 &0 & 0 &0 & 3 &7 & 18 &0 &0\\
 Uruguay & 1 & 1 & 2& 1& 2 & 3 & 5& 0& 0\\
 Venezuela & 15  & 5 &2 & 2 & 17 & 6& 34 &0 &0 \\ 
\botrule
\end{tabular}
\caption{Ranks obtained and/or occupied by authors for the different Latin American countries.}\label{table:rank-authors}%
\label{table:rank-country}
\end{table}

\paragraph{Rank and article production}

Here we discuss the relation between an article and the declared rank at the moment it was published. Table \ref{table:authors-2} shows the list of these ranks for each country. To illustrate this idea, 5339 articles have a senior author from Brazil at the moment of publication and 422 articles have a PhD author from Argentina. This data is important for identifying which academic position (rank) contributes the most to scientific production through publications.

\begin{table}[h!]
\begin{tabular}{@{}llllllllll@{}}
\toprule
Country & Senior & Junior & Staff & Postdoc & PhD & Master & Undergrad & Visitor & Other \\
\midrule
 Argentina & 726 & 135 & 120 & 101 & 422 & 14 & 34 & 0 & 0 \\
 Bolivia &- & - &- & -& -& -& -&- & -\\
 Brazil & 5339 & 593 & 675 & 1362 & 1721 & 405 & 66 & 309 & 0\\ 
 Chile & 923 & 441 & 1 & 772 & 323 & 87 & 19 & 34 & 4\\
 Colombia & 505 & 110 & 9 & 60 & 38 & 27 & 26 & 13 & 0 \\
 Costa Rica & 101 & 0 & 0 & 0 & 0 & 0 & 0 & 0 & 0 \\
 Cuba & 48 & 0 & 4 & 0 & 20 &16 &4 &0 &0 \\
 Dom Rep & - & - & - &-  & - & - &- &- &- \\
 El Salvador & - & - & - &-  & - & - &- &- &- \\
 Ecuador & 3 & 0& 0 & 0& 0 & 0 & 2 &0 & 0\\
 Guatemala & 3 & 0 & 0& 0& 0 &0 & 2& 1& 0\\
 Honduras & 3 &0 & 0&0 &0 & 0& 0&0 &0 \\ 
 Mexico & 1829 & 375 & 406 & 369 & 394 &66 & 29 & 37 & 0\\
 Panama & - & - & - &-  & - & - &- &- &- \\
 Paraguay & 1 & 0 &0 &0 &0 &0 &0 &0 &0\\
 Peru & 31 &0 & 0 &0 & 3 &14 & 2 &0 &0\\
 Uruguay & 0 & 0 & 1& 9& 7 & 1 & 1& 0& 0\\
 Venezuela & 117  & 6 &19 & 3 & 26 & 5 & 7 &0 &0 \\ 
\botrule
\end{tabular}
\caption{Number of articles that have a declared rank from a given country. An article can have a rank for each author in the best-case scenario.}
\label{table:authors-2}%
\end{table}

\paragraph{Rank and author-article analysis}

This analysis facilitates the monitoring of the author's mobility. For each author obtaining or occupying a rank in a given country within Latin America (refer to Table \ref{table:rank-authors}), it discerns whether a subsequent article was published within the same country, another Latin American country, or elsewhere around the world. To understand this category in more detail, we present the example of Colombia in Fig. \ref{fig:Colombia:rank-paper}, where we can visualise that 8 authors who have a rank from Colombia published as PhD students from Colombia, 21 as PhD students from another country in Latin America, and 9 authors as PhD students from the rest of the world. Therefore, someone who did all their academic degrees in a non-Latin American country, but occupied a postdoc position in Latin America, would contribute to the plots in the corresponding stage they publish an article, even though this was not associated with Latin America. The rest of the non-trivial plots for any Latin American country can be found in the appendix \ref{app:ranks}. 

\begin{figure}[t]%
\centering
\includegraphics[width=0.7\textwidth]{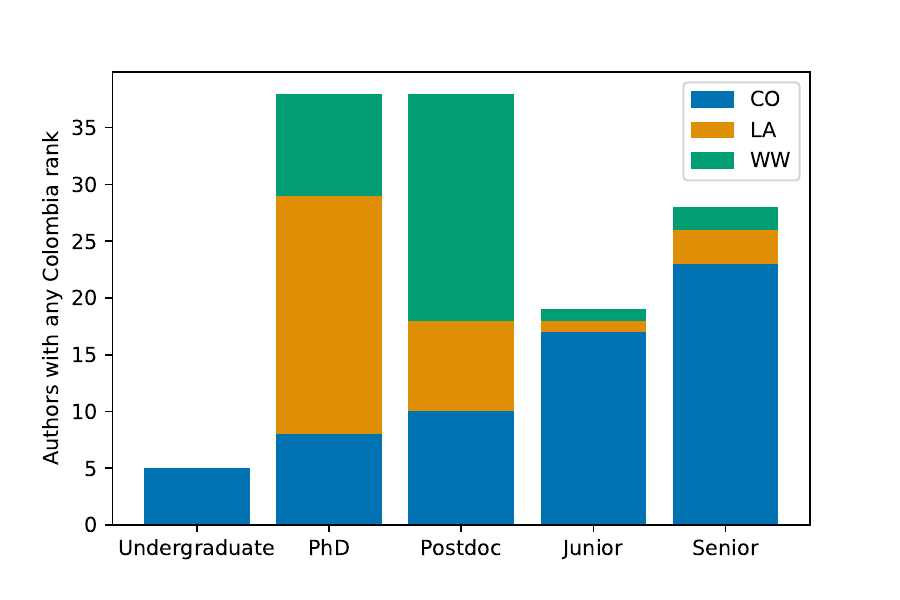}
\caption{Number of authors with Colombia ranks as in table \ref{table:rank-authors} who published an article under a given rank in Colombia, Latin America (LA) or elsewhere worldwide (WW).}
\label{fig:Colombia:rank-paper}
\end{figure}

An interesting conclusion that can also be inferred from the Colombian case, for example, where the number of authors that have published as PhD in that country is relatively low in comparison to the PhD publications from abroad (mostly in other Latin American countries), is the fact that most of the publications at the level of junior and senior positions take place in Colombia, which suggests that most of the researchers doing a postdoc abroad come back to the country to occupy these positions. A similar conclusion can be extrapolated to other countries such as Chile (Fig. \ref{fig:Chile:rank-paper}) and Mexico (Fig. \ref{fig:Mexico:rank-paper}), where the situation is different in terms of the origin of publications at the PhD and postdoc level. This can probably be explained in the context of the scholarships provided by these countries and the effective management of the work conditions to allow a positive return rate. The creation and support of postgraduate and postdoc programmes that can compete internationally will allow an increase in the local production of articles at that level, but only the implementation of correct scholarship policies and work conditions for permanent positions will incentivize the return of researchers after the postdoc level.

\section{Publications}
\label{sec:pubs}

\begin{figure}[h!]%
\centering
\includegraphics[width=0.9\textwidth]{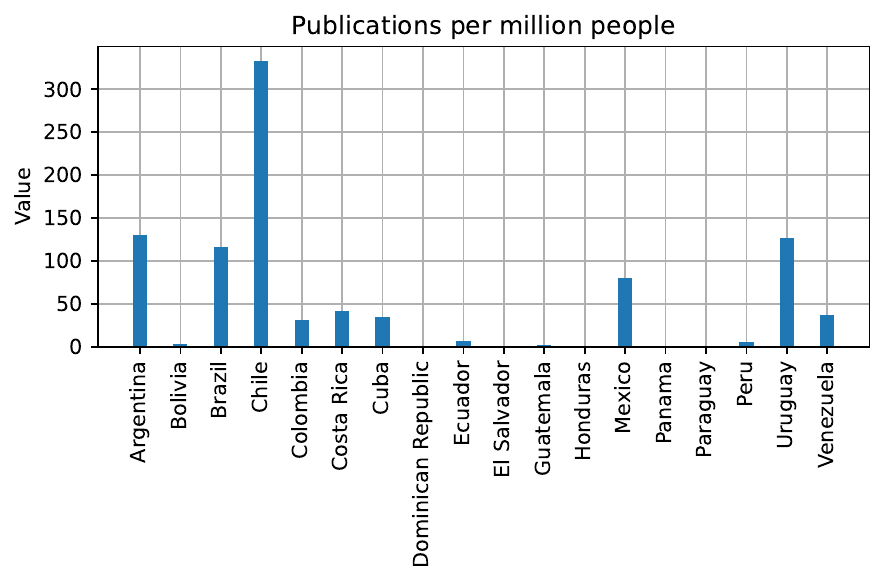}
\caption{Publications of each country \ac{PMI}.}
\label{fig:all_publications_pc} 
\end{figure}

A metric of scientific productivity is the number of articles that are published.
In this section, we provide a summary of the number of articles for each country in our study. An article has been published by a country if it has at least one author affiliated with an institution of that country. 
We present the summary of the number of articles of each country, normalised by population, in Fig.~\ref{fig:all_publications_pc}. 
The data shows significant variation across countries, with Chile showing the highest number of publications per capita, exceeding 300 per million inhabitants. Argentina, Brazil and Uruguay also exhibit relatively high publication rates, though lower than Chile. In contrast, several countries, such as Bolivia, Ecuador, Honduras, and others, display a low publication output per capita. The results indicate a varied landscape of scientific productivity across the region if we use the population of each country as a normalisation metric.

\begin{figure}[h!]%
\centering
\includegraphics[width=0.9\textwidth]{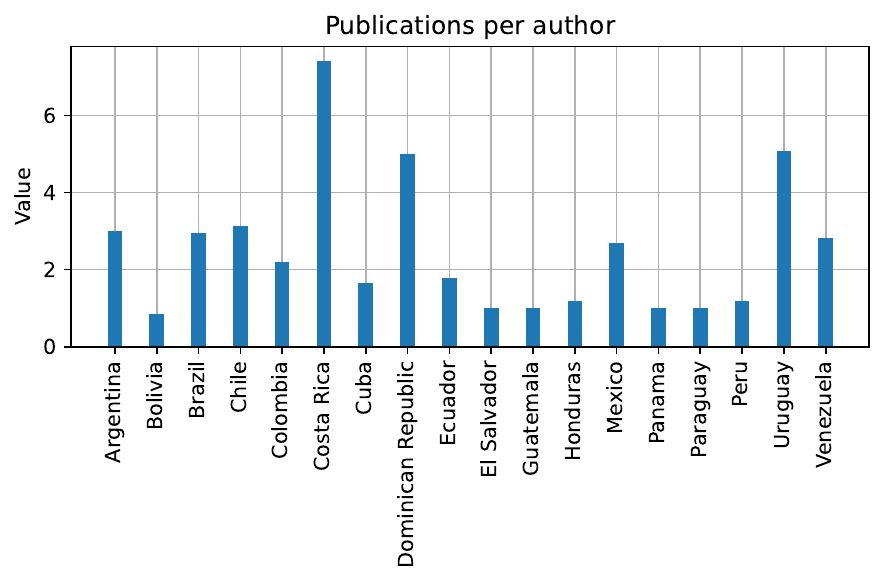}
\caption{Publications of each country in our study per authors.}
\label{fig:publications_per_author}
\end{figure}

Additionally, we present the number of publications of each country normalised by the number of total authors in Fig. \ref{fig:publications_per_author}.
In this case, we see that Costa Rica and Dominican Republic have the highest averages. However, in our analysis, the case of the Dominican Republic is \textit{unique}, as our database includes only a single author affiliated with the country, who published in the 1990s under that affiliation. This implies that the publications per author of the Dominican Republic are simply their absolute number of publications\footnote{We will encounter this feature with the Dominican Republic every time we address one of its metrics normalised by the number of authors.}. Uruguay also has high values (and has a high number of total authors). Other countries, such as Bolivia, Guatemala, and Paraguay, show lower averages. This variation suggests a diverse range of author productivity across the region.

In Fig. \ref{fig:all_publications_pc}, Chile leads in terms of publications \ac{PMI}, showcasing a high overall national scientific output. However, in Fig. \ref{fig:publications_per_author}, the leader shifts to Costa Rica when assessing productivity per author, indicating that, while Chile's scientific system produces a high volume of publications relative to its population, individual authors in Costa Rica tend to be more productive on average. Notice that, in the case of the Dominican Republic, only one author is indexed and, therefore, the number that is shown is equivalent to the absolute number of publications of the country. 

Uruguay ranks high in both metrics, reflecting a balance between national output and individual productivity. Several countries appear towards the bottom in both figures, indicating lower publication rates both at the national level and per author. This trend highlights the need for further development of scientific infrastructure and support for researchers in these nations.

These complementary normalisations provide a nuanced perspective, demonstrating how a high national output does not necessarily correlate with high productivity at the individual researcher level, and vice versa. The exact numbers of the total publications, total publications \ac{PMI}, and total publications per author can be found in Table \ref{table:publications} of App. \ref{app:pub}.

\begin{figure}[h!]%
\centering
\includegraphics[width=0.9\textwidth]{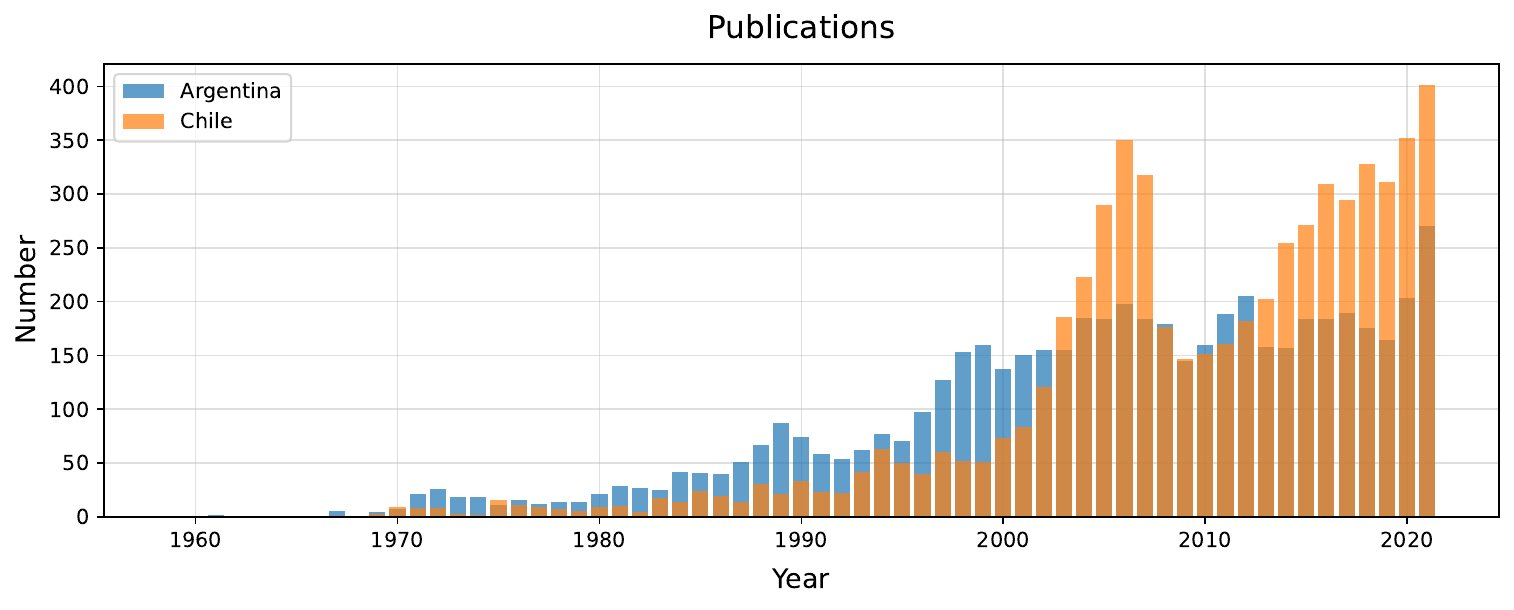}
\caption{Articles per year of Argentina and Chile.}
\label{fig:arg_cl} 
\end{figure}

In what follows, we compare the articles published by year for individual countries. To facilitate the display of the results, we split the comparison among pairs or triplets of countries whose number of publications per year is comparable. Fig. \ref{fig:arg_cl} shows the comparison between Argentina and Chile. We see that Argentina starts publishing articles earlier than Chile, and publishes consistently more articles until the early 2000s. In the mid-2000s the number of articles of Chile was higher, and they became similar around 2009. From then onwards, the number of publications of Chile has grown faster.

\begin{figure}[h!]%
\centering
\includegraphics[width=0.9\textwidth]{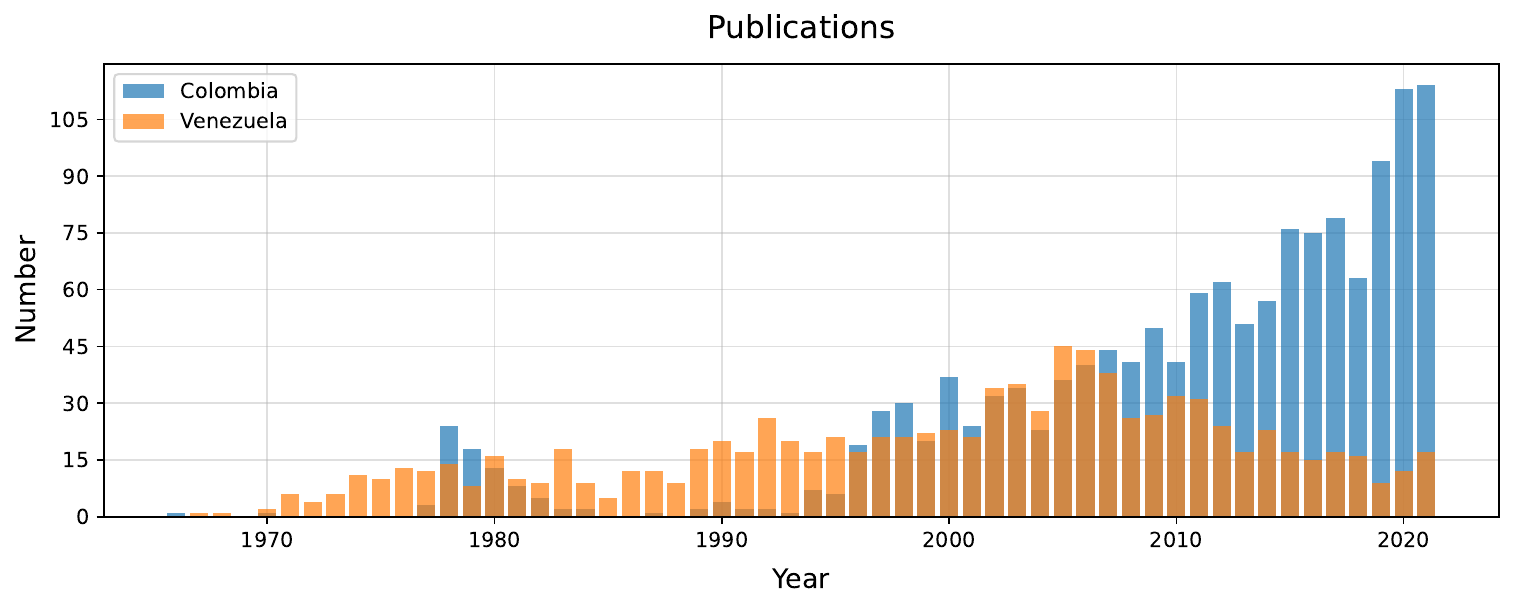}
\caption{\label{fig:ven_col} Articles per year of Venezuela and Colombia.}
\end{figure}

%
Fig. \ref{fig:ven_col} shows the trends in the number of articles published per year about Venezuela and Colombia from 1960 to 2021. Early publications in Venezuela began in the late 1960s and saw a gradual increase through the decades, with a rise starting in the late 1990s and continuing to grow steadily into the mid-2000s. Publications on Colombia were sparse until the mid-1970s. The number of articles on Colombia began to surpass those on Venezuela from around the late 1990s, continuing to rise until. The rate of publications of Venezuela starts to decrease from the mid-2000s onwards. The data indicates opposite tendencies in these countries, with Colombia receiving a higher volume of publications in recent years compared to Venezuela.
Comparisons between the remaining countries in the database are provided in App. \ref{app:pub}.

\section{Citations and h-index}
\label{sec:cits_hindex}

\begin{figure}[h!]
\centering
\includegraphics[width=0.9\textwidth]{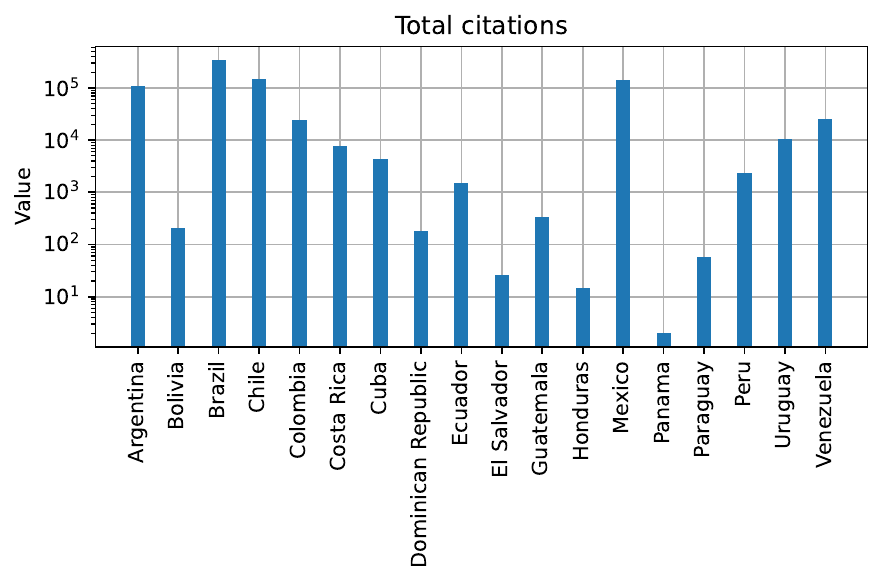}
\caption{\label{fig:cit_summary} Number of citations for each country in our study.}
\end{figure}

Citations are a crucial metric for evaluating the impact and reach of scientific publications.
In this section, we summarise the citation data for each country included in our study. A citation is attributed to a country if the cited article has at least one author affiliated with an institution of that country.
Fig.~\ref{fig:cit_summary} shows the total number of citations for each country included in our study.
The distribution of citations spans several orders of magnitude, reflecting significant variation across different countries. Notably, larger countries with more established research infrastructures, such as Brazil and Mexico, generally exhibit higher citation counts compared to smaller countries. This trend aligns with expectations, as larger research communities and funding capacities often correlate with higher research output and, consequently, more citations.

\begin{figure}[h!]%
\centering
\includegraphics[width=0.9\textwidth]{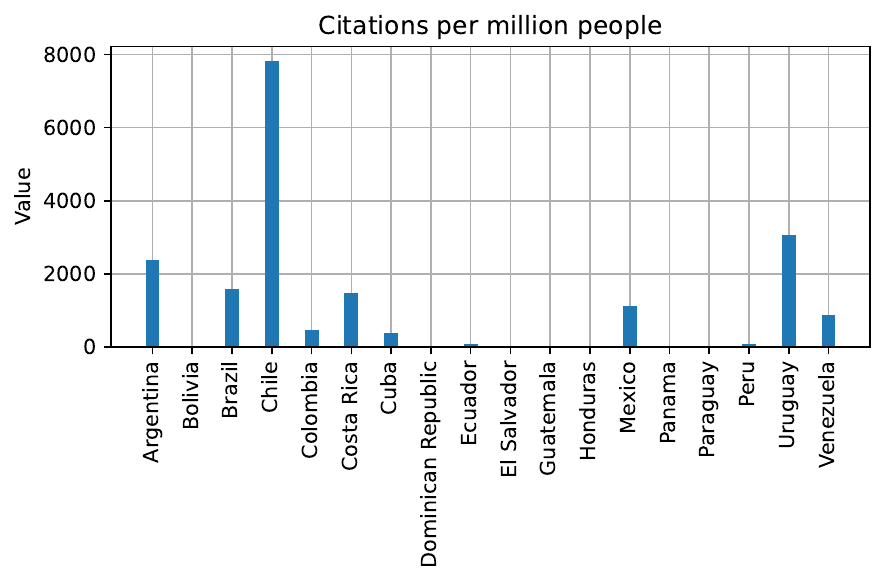}
\caption{\label{fig:cit_summary_pc} Number of citations per million people for each country.}
\end{figure}

Fig.~\ref{fig:cit_summary_pc} shows the total number of citations per million people for each country.
%
%
In this case, Chile and Uruguay stand out with significantly higher citations per million people, indicating that these countries, despite having smaller populations, maintain a notable research output relative to their size.

\begin{figure}[h!]%
\centering
\includegraphics[width=0.9\textwidth]{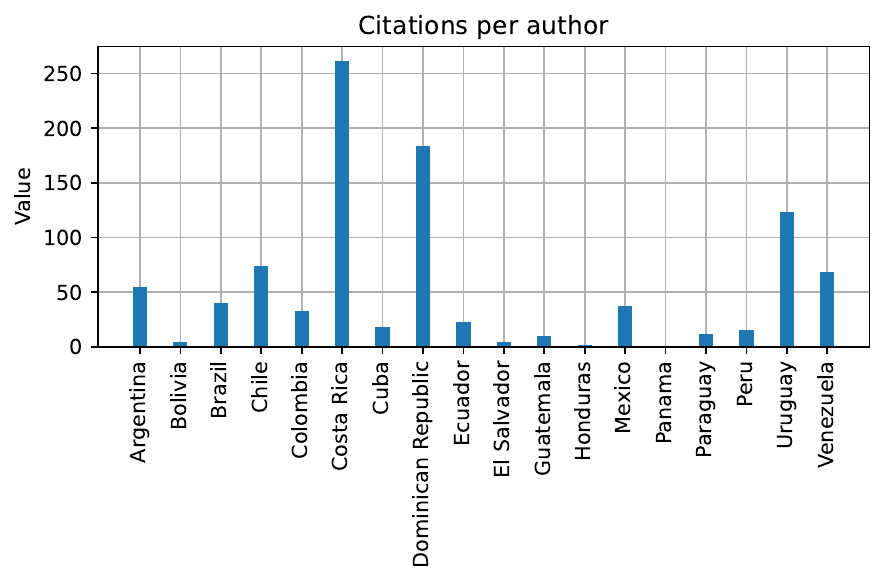}
\caption{\label{fig:cit_summary_per_author} Number of citations per author for each country.}
\end{figure}

Additionally, we show the number of citations per author for each country in Fig.~\ref{fig:cit_summary_per_author}. This metric provides insight into the average research impact of individual contributors within each country, complementing the broader measures of total and per capita citations.
From Fig.~\ref{fig:cit_summary_per_author}, we can see that some countries with moderate or smaller research communities, such as Costa Rica and the Dominican Republic, exhibit high citations per author, indicating a high average impact per researcher. However, as we indicated in the summary of publications per author, the case of the Dominican Republic is special as there is only a single author affiliated with the country, who published in the 1990s. The citations attributed to this author have accumulated over time, making this datapoint a particular case (rather than necessarily a reflection of broader national research performance). Uruguay also shows strong performance in this metric, consistent with its high citations per capita.

In contrast, larger countries which have high total citation counts and lower per capita rates, display moderate citations per author. This suggests that while their total output is substantial, the average impact per individual researcher may be diluted by the size of the research community.

The summary of the total citations, and their per capita and per author results, can be found in Table~\ref{table:citations} in App. \ref{app:cit}.

We now shift our focus to discussing the h-index of different countries. The h-index is a valuable metric that provides a combined measure of both the productivity and citation impact of a country's researchers. It is given by the number h of publications of a country that have been cited at least h times, offering a notion of sustained influence and quality of research output contributions.

\begin{figure}[h!]%
\centering
\includegraphics[width=0.9\textwidth]{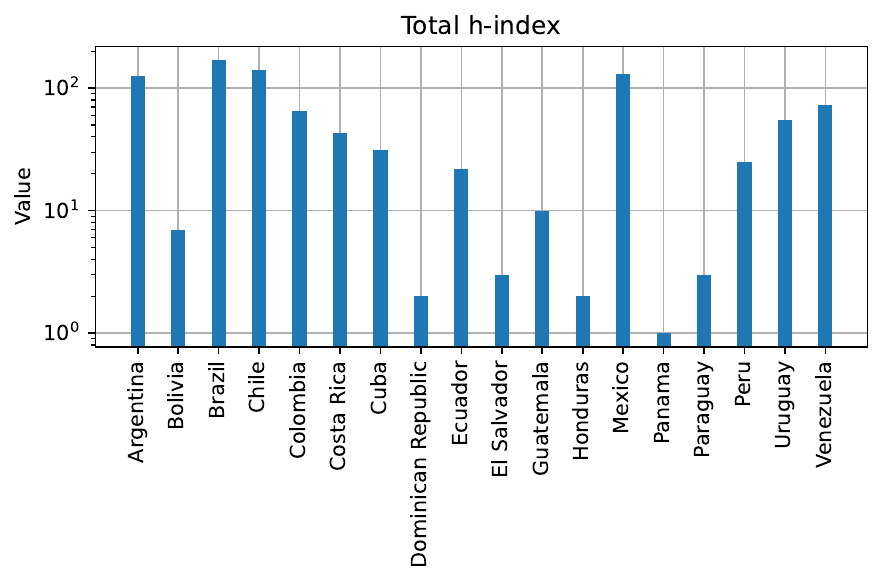}
\caption{\label{fig:hindex_total} Total h-index.}
\end{figure}

Fig.~\ref{fig:hindex_total} presents the total h-index for each country in our study. Brazil, Chile, Mexico, and Argentina exhibit high total h-indices. Countries with lower total h-indices, such as Panama, the Dominican Republic, and Honduras, reflect instances where publications may not yet reach the same level of citation consistency.

\begin{figure}[h!]%
\centering
\includegraphics[width=0.9\textwidth]{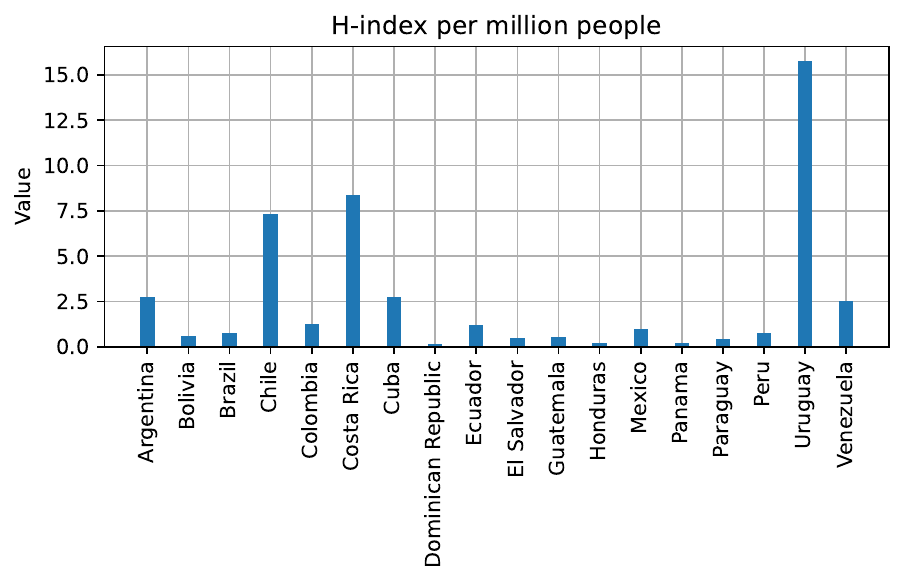}
\caption{\label{fig:hindex_pc} H-index \ac{PMI}.}
\end{figure}

As before, we also provide normalised h-index counts. Fig.~\ref{fig:hindex_pc} quantifies the h-index per capita for each country in our analysis.
We observe that Uruguay leads in terms of h-index \ac{PMI}, highlighting a strong research impact relative to their population size. Chile and Costa Rica also demonstrate relatively high h-index values per capita. In contrast, larger countries such as Brazil and Mexico show lower h-index values \ac{PMI}. This aligns with previous observations that while these countries have substantial total research output, the average impact per capita is more moderate. 

\begin{figure}[h!]%
\centering
\includegraphics[width=0.9\textwidth]{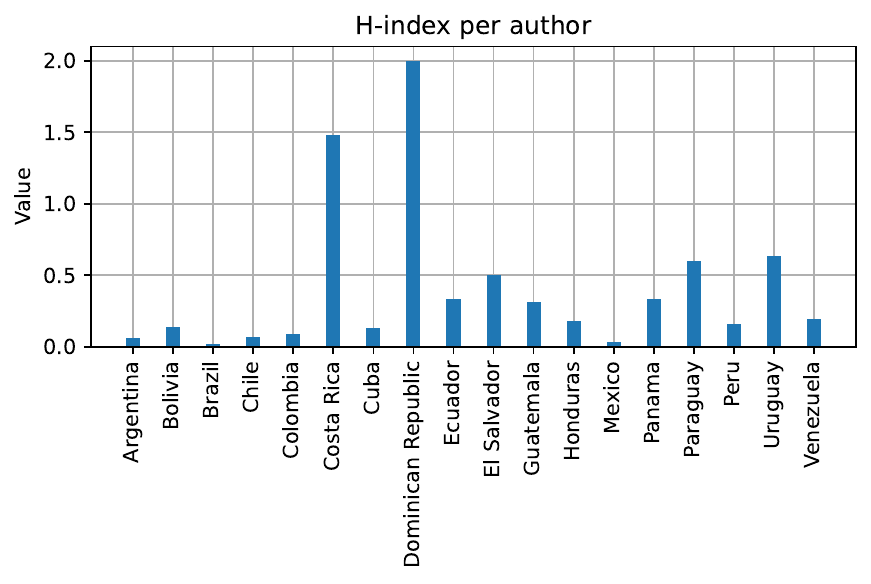}
\caption{\label{fig:hindex_per_author} H-index per author.}
\end{figure}

Additionally, Fig.~\ref{fig:hindex_per_author} shows the total h-index of each country normalised by their number of total authors.
As in the case of citations per author, here we also observe that the Dominican Republic has a high h-index per author. Again, this is because only one author with a Dominican Republic affiliation, whose articles have accumulated citations over time, is indexed in the database. 

Uruguay also shows high h-indices per author, indicating that individual researchers in these nations contribute significantly to impactful research. In contrast, larger countries like Brazil and Mexico exhibit lower h-index values per author, which aligns with earlier findings that while these countries produce a large volume of research, the average impact per researcher may be more distributed across a larger pool of authors.

The summary of the total h-index of each country, and their per capita and per author values, can be found in Table~\ref{table:h-index} in App. \ref{app:cit}.

The analysis of citation and h-index metrics provides a general view of the research impact and productivity across the different countries in our analysis. While total citation counts and h-indices reflect the overall contribution of each nation to the global research landscape, normalised metrics such as citations and h-index per capita and per author offer insights into the relative influence of research communities. These metrics underline both the collective achievements and the unique strengths of individual researchers within each country, offering a broader understanding of the diverse contributions to scientific progress across the region.

\section{Correlations with GDP in R$\&$D and \ac{HDI}}
\label{sec:corr_latam}

Understanding the relationship between the size of scientific communities, scientific productivity and impact, and national investment in \ac{RD} is crucial for evaluating a country's capacity to advance knowledge and foster innovation. A higher percentage of \ac{GDP} allocated to \ac{RD} typically reflects a nation’s prioritisation of science, technology, and innovation as key drivers of cultural and societal development and economic growth. By exploring the correlation between scientific metrics in \ac{HECAP} and the percentage of \ac{GDP} invested in \ac{RD}, we can gain insights into how resources dedicated to scientific research influence a country's scientific output. 

In this section, we will use the average \ac{GDP} values of Table \ref{table:PIB} as estimators of the level of resources that each country devotes to \ac{RD} activities in the aforementioned fields\footnote{Note that, as mentioned previously in the text, there is no information available on the World Bank for the Dominican Republic.}. It is important to acknowledge that the relationship between \ac{GDP} investment in \ac{RD} and scientific output is a multidimensional issue influenced by a variety of factors, including infrastructure, human capital, and institutional support. Additionally, the impact of changes in \ac{RD} investment on research productivity is not instantaneous as, for example, the effects of increased funding often take time to materialise in the form of new research and publications. Furthermore, the available data can be limited and sparse. Because of this, in this section we aim to provide a broad estimate of the correlation between scientific output in \ac{HECAP} and \ac{GDP} investment in \ac{RD}, recognising the inherent complexity of the issue. Our analysis offers a general perspective, which, while valuable, naturally \textit{cannot} capture all the intricate dynamics at play in the relationship between investment and output. Those aspects are beyond the scope of the present study.

To study the correlations we will provide a set of linear regression fits to the data along with their coefficient of determination, $R^2$, defined as  
\begin{equation*}
    R^2 = 1 - \frac{\sum_{i} (y_i - \hat{y}_i)^2}{\sum_{i} (y_i - \bar{y})^2},
\end{equation*}  
where $y_i$ are the observed values (authors, publications, citations), $\hat{y}_i$ are the predicted values from the regression, and $\bar{y}$ is the mean of the observed values. The $R^2$ coefficient provides a simple measure of the strength of the relationship between the dependent and independent variables. 

\begin{figure}[h!]%
\centering
\includegraphics[width=0.9\textwidth]{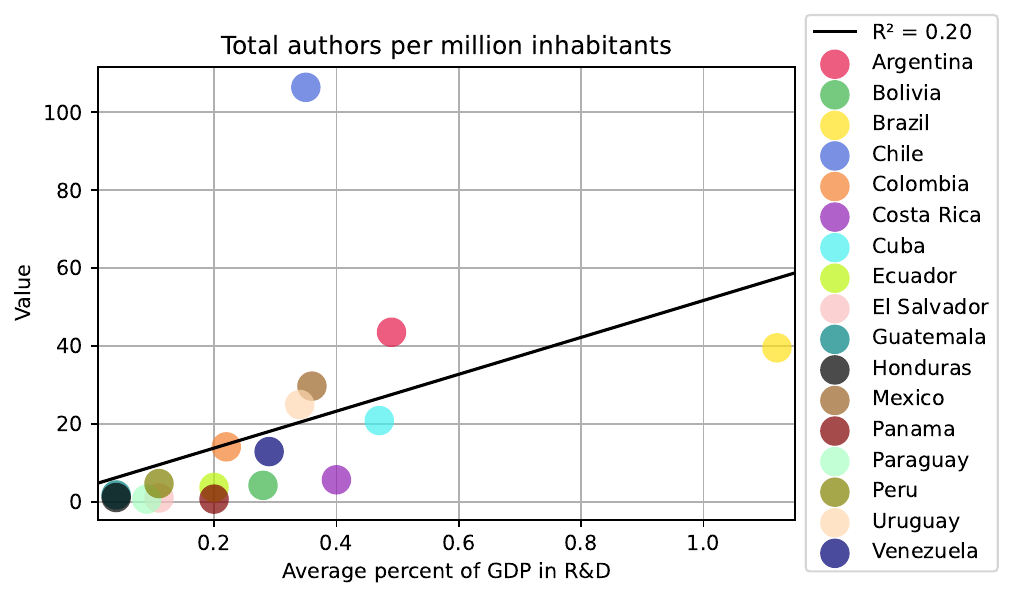}
\caption{\label{fig:authors_summary_sp} Number of authors \ac{PMI} versus the average percentage of \ac{GDP} invested in \ac{RD} for each country.}
\end{figure}

\begin{figure}[h!]%
\centering
\includegraphics[width=0.9\textwidth]{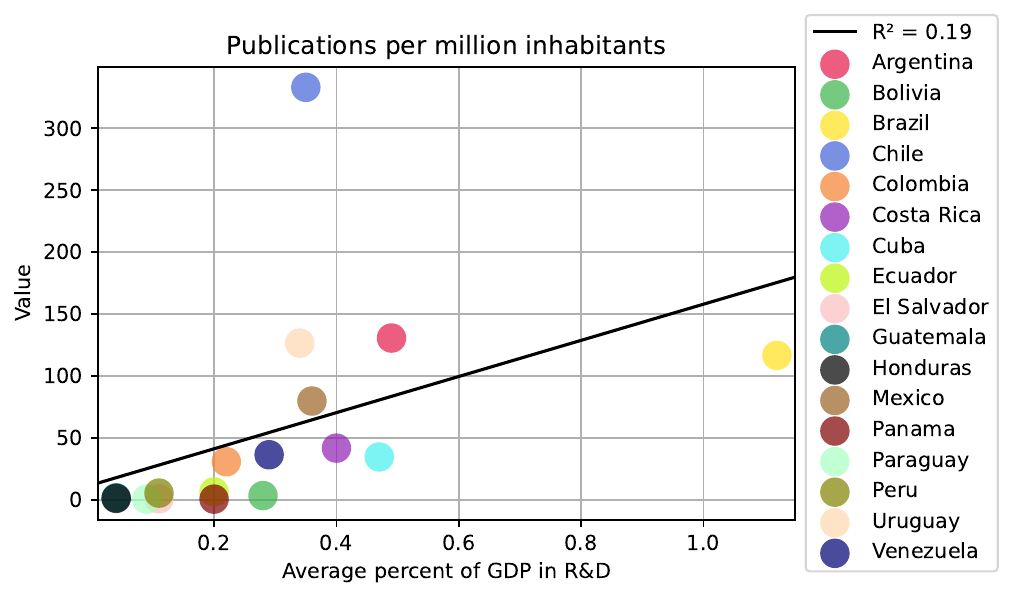}
\caption{\label{fig:publications_summary_sp} Number of publications \ac{PMI} versus the average percentage of \ac{GDP} invested in \ac{RD} for each country.}
\end{figure}

\begin{figure}[h!]%
\centering
\includegraphics[width=0.9\textwidth]{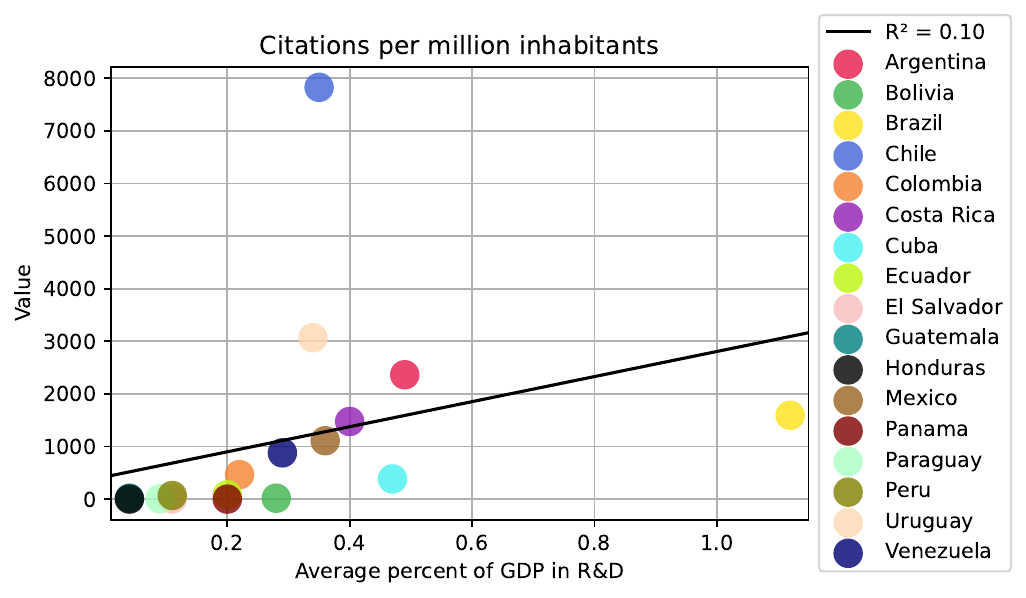}
\caption{\label{fig:cit_summary_sp} Number of citations \ac{PMI} versus the percentage of average \ac{GDP} invested in \ac{RD} for each country.}
\end{figure}

Figs.~\ref{fig:authors_summary_sp}, \ref{fig:publications_summary_sp}, and \ref{fig:cit_summary_sp} show respectively the correlation between the number of authors, publications, and citations \ac{PMI} and the average \ac{GDP} invested in \ac{RD} by each country. In all cases, a positive correlation is observed: higher national investment in \ac{RD} is associated with larger scientific communities, greater publication output, and increased research impact. We note that the 3 figures exhibit similar qualitative features. Chile and Brazil consistently emerge as outliers. It is not surprising that this trend suggests that national investments in research and development support larger scientific communities, higher volume of publications, and higher impact, which contributes to the overall development of scientific metrics in the region. This analysis quantifies the description of the trend.

We stress that the causal connection between scientific metrics and \ac{GDP} investment is highly complex. The linear correlations and $R^2$ values that we have provided aim to quantify their correlation in a simple way.

The analysis presented in this section shows a clear positive correlation between national investments in \ac{RD} and scientific output in terms of the size of the scientific community and number of researchers, the number of publications, and their citations. While increased \ac{RD} expenditure often leads to greater research productivity and higher scientific recognition, the relationship is complex and are influenced by additional factors that can escape the metrics we have chosen. Although this study provides a broad overview of the correlation between \ac{GDP} investment and scientific output, further work could explore the specific mechanisms through which these investments translate into research metrics.

Scientific productivity and impact do not exist in isolation from broader socio-economic conditions. A more comprehensive understanding of national research performance can be achieved by considering indicators that capture social well-being alongside economic investment. In this context, the \ac{HDI} serves as a valuable metric, as it integrates multiple dimensions of human development. It accounts for health, measured through life expectancy; education, quantified via the Education Index; and income, related to the  \ac{GDP} (total, not just in \ac{RD}) per capita. By incorporating these factors, the \ac{HDI} offers a more holistic perspective on the conditions that support scientific progress and national development, beyond purely economic indicators. We use the average \ac{HDI} values of Table \ref{table:HDI}.

\begin{figure}[h!]%
\centering
\includegraphics[width=0.9\textwidth]{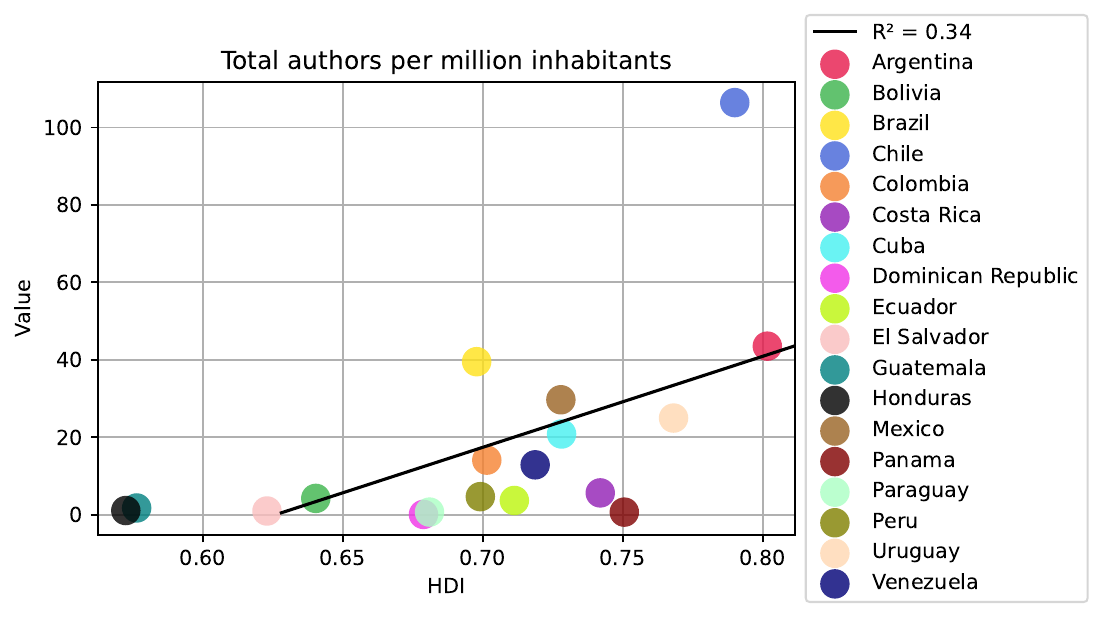}
\caption{\label{fig:auth_summary_sp_hdi} Linear regression with $R^2$ value and number of total authors \ac{PMI} versus the \ac{HDI} of each country.}
\end{figure}

\begin{figure}[h!]%
\centering
\includegraphics[width=0.9\textwidth]{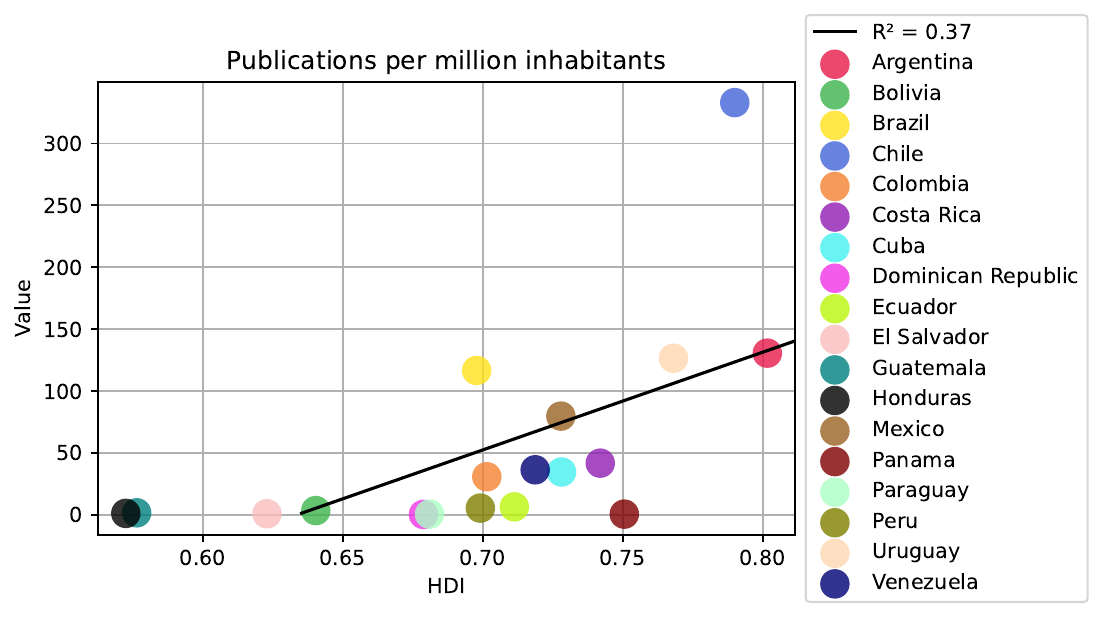}
\caption{\label{fig:pub_summary_sp_hdi} Linear regression with $R^2$ value and number of publications \ac{PMI} versus the \ac{HDI} of each country.}
\end{figure}

\begin{figure}[h!]%
\centering
\includegraphics[width=0.9\textwidth]{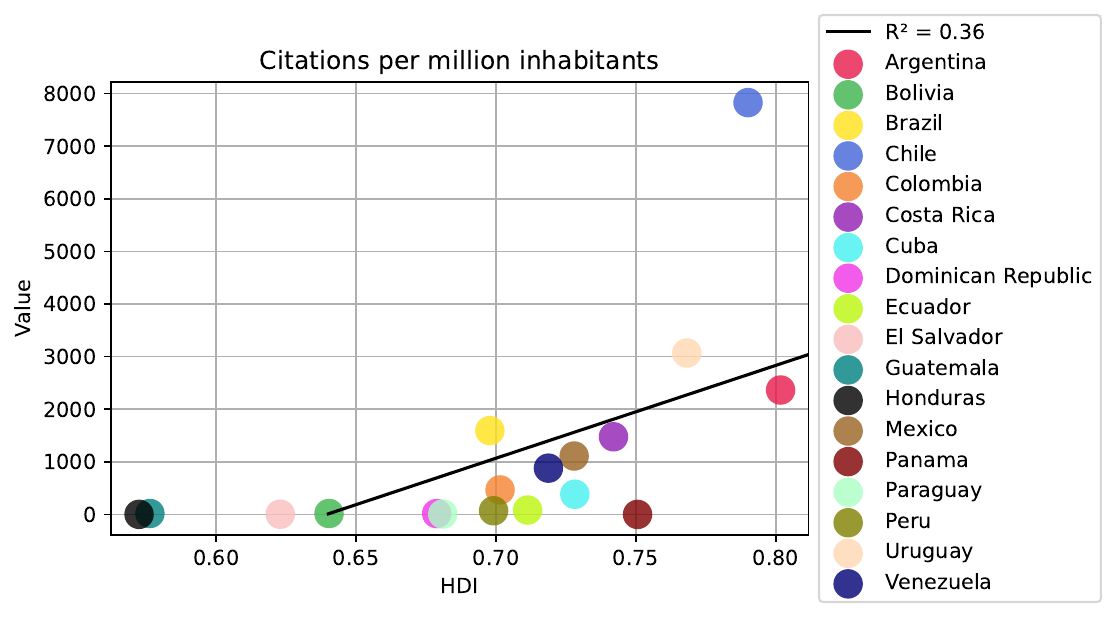}
\caption{\label{fig:cit_summary_sp_hdi} Linear regression with $R^2$ value and number of citations \ac{PMI} versus the \ac{HDI} of each country.}
\end{figure}

Figs. \ref{fig:auth_summary_sp_hdi}, \ref{fig:pub_summary_sp_hdi}, and \ref{fig:cit_summary_sp_hdi} show the correlation between the number of authors, publications, and citations per million inhabitants and the \ac{HDI} of each country, respectively. In each case, we describe this correlation using a linear regression model and report the coefficient of determination $R^2$ to assess the strength of the relationship. As expected, we observe a general positive correlation between the \ac{HDI} and scientific output indicators. Notably, the higher $R^2$ values compared to the GDP-based analysis suggest that the \ac{HDI} has a stronger correlation with national research performance. Naturally, these results show the relevance of broader socio-economic factors in shaping scientific productivity and impact. 

The results presented in this section show the dual role of economic investment in \ac{RD} and broader socio-economic conditions in shaping scientific output. While increased \ac{GDP} investment in research is a key driver of productivity and impact, the strong correlations observed with the \ac{HDI} suggest that, naturally, factors such as education, healthcare, and overall human development also play an important role. These findings indicate that a comprehensive approach—one that integrates both direct research funding and long-term investments in human capital—is essential to generate a robust and sustainable scientific ecosystem. While this study focuses on identifying correlations rather than causality, future research could further investigate the underlying mechanisms driving these relationships, offering a deeper understanding of how national policies can most effectively support scientific progress.

\section{Gender analysis}
\label{sec:gender}


To examine gender representation among researchers, we use the genderize.io~\cite{genderize.io} web service to determine the gender of first names. The service uses a dataset that contains around
a billion records which include associations with the vast majority of the countries.
For each first name and country, the service returns the probability of the assigned gender in the database. 
Accuracy analysis and comparison with other services are analysed\footnote{While this approach provides useful insights into broad trends in gender representation, it is important to acknowledge the inherent limitations of name-based classification methods and their potential biases.} for example in~\cite{arXiv:2411.07012,10.7717/peerj-cs.156,10.1073/pnas.1914221117}. We obtain the country associated with the first names, from the affiliation for each author. Moreover, we label the gender as unknown 
if the probability is less than 90\%~\cite{arXiv:2411.07012,10.1007/s10664-021-10080-8}.

To assess the reliability of this method, we validated the service against a dataset of 3629 first names from Colombia. After applying the previous criteria, 
3000 of them were classified as male, $m$, or female, $f$ with a probability greater than 90\%. The estimated coded error rate without 
unknowns (WU) is~\cite{10.32614/rj-2016-002}
\begin{align}
    \text{CER}_{\text{WU}} = \frac{m_f + f_m}{f_f + m_m + m_f + f_m } = 0.006,
\end{align}
where $m_f$ is the male first name with the associated country, predicted by the service as female, and so on. 

\begin{figure}[t]
    \centering
    \includegraphics[scale=0.8]{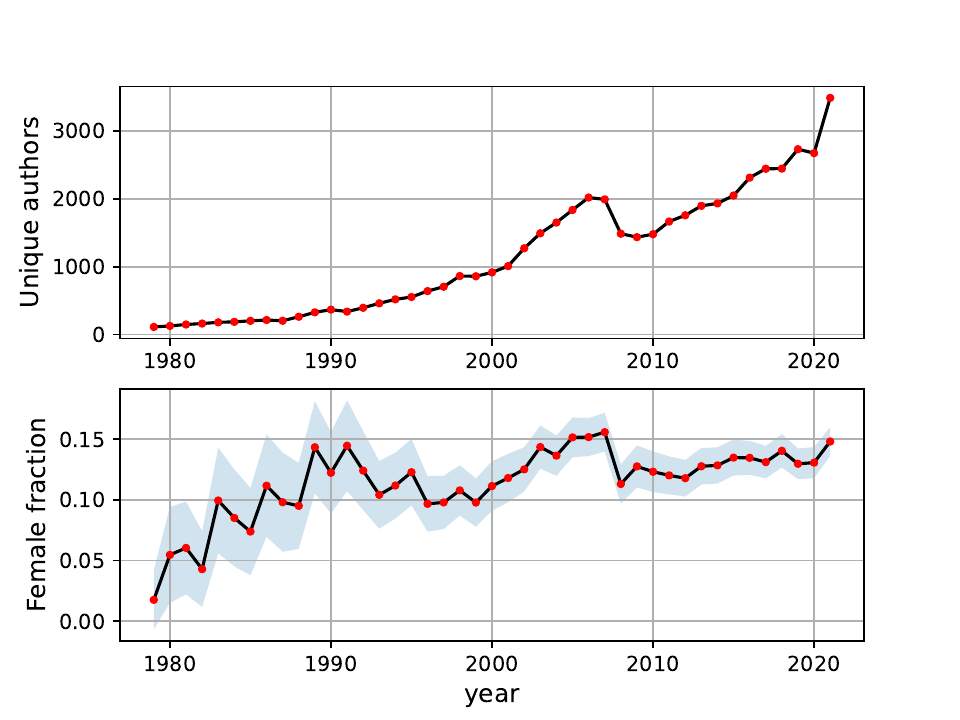}
    \caption{The top panel shows the number of unique authors per year for which the gender was identified. 
    The bottom panel shows the female fraction of the authors along with the margin of error at a 95\% confidence level.}
    \label{fig:fraction}
\end{figure}

From the almost $30\,000$ unique authors in our database, we were able to identify the gender, with a probability greater than $0.9$,
for $18\,000$ of them. For this reduced dataset,
we calculate the fraction of female authors for each year~\cite{10.1371/journal.pone.0066212,10.1371/journal.pbio.2004956} by selecting the \textit{unique authors}\footnote{This is different to active authors or total authors defined previously. It refers to the number of authors who published in a given year.} of all articles in each year.
The results are shown in Fig.~\ref{fig:fraction}. There in the top panel, we show the number of unique authors per year for which the gender was 
identified according to the previous criteria. In the bottom panel, we show the female fraction of the authors along with the band associated with the margin of error at $95\%$ confidence level of the fraction~\cite{diez2012openintro}. We can see that besides the increase of authors over the years, the female fraction 
is almost constant around $13\%$ in the last century. These results are consistent with previous results obtained for theoretical categories
of \ac{HEP}~\cite{10.1162/qss_a_00114}. 

In INSPIRE-HEP, one author can log in and update their profile. In the profile, the author can declare 
the ranks they have held in different stages of their research career. This information is particularly 
useful for understanding the academic trajectory and current status of researchers within the \ac{HECAP} community. 
The ranks that can be declared include positions such as Senior, Junior, Staff, Postdoc, PhD, Master, Undergraduate, Visitor, and Other in a range of years. 
This data can be systematically accessed through the INSPIRE-HEP REST API, allowing for a comprehensive analysis of the academic landscape in Latin America.
In particular, we can track if an author has declared a rank in a given year and if they have published an article in the same year. 
For the 18,000 authors with an identified gender, we have assigned this kind per year rank of publications to 3832 of them.
The number of authors with a declared rank per year of publication is shown in Table~\ref{tab:my_label}. We can see 
that the female fraction decreases through higher ranks~\cite{10.1371/journal.pone.0298736}.
This is the well-known ``scissor effect" which applies to many
areas of science~\cite{10.1371/journal.pbio.0040097}.

\begin{table}[h!]
\begin{tabular}{@{}lll@{}}
\toprule
Rank & All & Female Fraction  \\
\midrule
All ranks& 3832& $12\pm1$\\ \hline 
         Undergraduate, Master, PhD&  1268&   $14.4\pm 1.9$\\ \hline 
         Postdoc&  1498&   $13.5\pm 1.7$\\ \hline 
         Junior or Staff&  816&   $12.5\pm2.3$\\ \hline 
         Senior&  1592&   $7.9\pm1.3$\\
\botrule
\end{tabular}
\caption{Gender evolution through the ranks defined in the text. The ``All'' rank also includes  Visitor and Other with around 10 unique extra authors.}
\label{tab:my_label}
\end{table}

\section{Collaboration}
\label{sec:collab}

Collaborations are an important component of research activity. In this section, we examine the collaborative landscape among Latin American countries, focusing on both intra-regional collaboration and collaborations with countries in other parts of the world. We aim to identify some key partnerships and highlight networks that contribute to the region’s scientific integration and impact. 

\begin{figure}[h!]%
\centering
\includegraphics[width=1.0\textwidth]{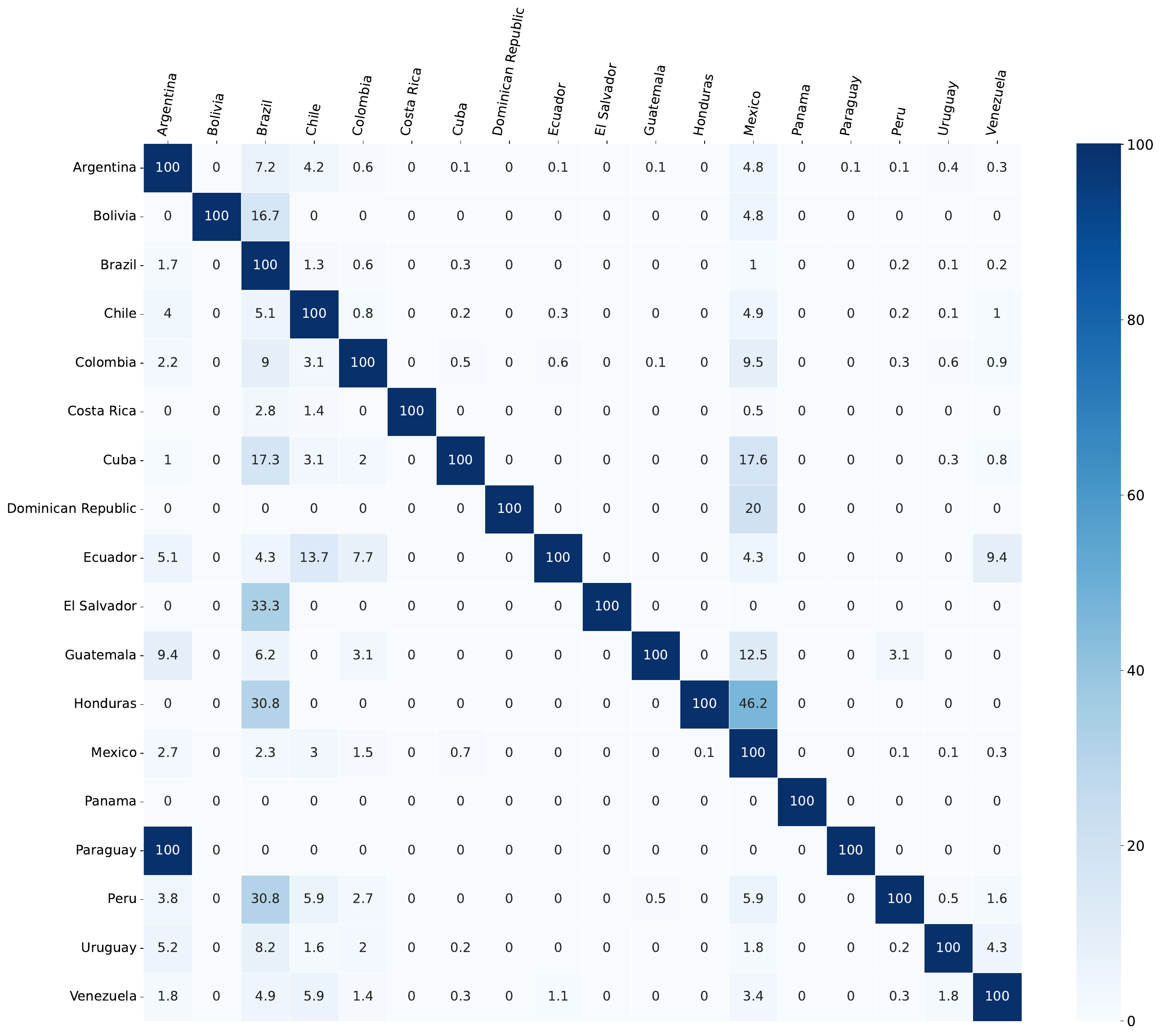}
\caption{\label{fig:normalised_collab_matrix} Percentage of collaborations between Latin American countries.}
\end{figure}

Fig.~\ref{fig:normalised_collab_matrix} illustrates the percentage of collaborations between Latin American countries, showing the relative strength of research ties within the region. Specifically, for a country in a given row, the figure displays the percentage of articles from that country that were co-authored with the country listed in the corresponding column.
In this way, for example, $16.7 \%$ of the articles of Bolivia were co-authored with Brazil. Notably, Brazil stands out as a strong collaborator across the region. Over $30 \%$ of the publications from three Latin American countries (El Salvador, Honduras, Peru) include Brazil as a co-author. This demonstrates Brazil's significant contribution to the regional scientific landscape, potentially influenced by its generous fellowship programmes and institutions like the ICTP-SAIFR.

Mexico and Argentina have also emerged as important collaborators with the rest of Latin America, showing the positive impact that larger countries can have on smaller countries' research efforts. For instance, all publications from Paraguay have been produced in collaboration with Argentina, reflecting a strong research relationship. 

Additionally, we study the collaborations between Latin American countries and Ibero-America\footnote{We consider important to highlight the collaborations including all Ibero-America not only because of the historical relation of the region with Spain and Portugal but also because it is the Ibero-American ministerial meetings that gave the mandate to the LASF4RI initiative and because of the existence of organisations like \ac{CYTED} \url{https://www.cyted.org/}}, as well as other countries of the world.
\begin{figure}[h!]%
\centering
\includegraphics[width=0.9\textwidth]{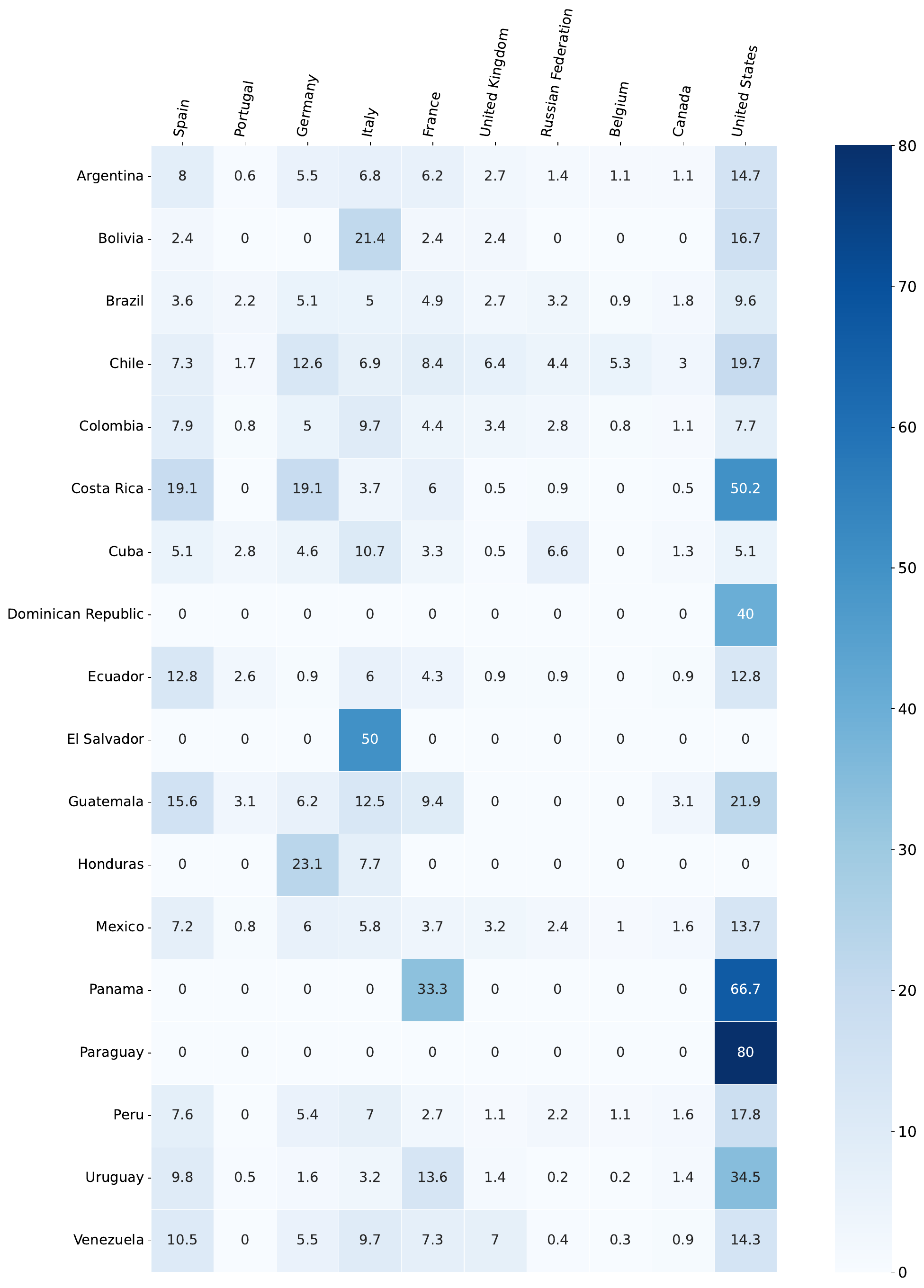}
\caption{\label{fig:normalised_collab_world_matrix} Percentage of collaborations between Latin American countries and other countries.}
\end{figure}
Fig.~\ref{fig:normalised_collab_world_matrix} shows the percentage of collaborations between Latin American countries and relevant scientific collaborators such as Spain, Portugal, Germany, Italy, France, the United Kingdom, the Russian Federation, Belgium, Canada, and the United States.

Collaborations with Spain have been significant for the region, with countries like Costa Rica and Guatemala benefiting from partnerships with this country. Germany has also established substantial collaborative ties, especially with Honduras, where over $20\%$ of its publications involve co-authors from German institutions. Italy shows a strong presence as well, with notable contributions to Bolivia and solid collaborative relationships with El Salvador and Guatemala. France has been an active collaborator in the region, particularly with Guatemala, Paraguay, and Panama, while the United Kingdom, Russia, Belgium, and Canada have collaborated with Latin America to a lesser extent compared to other countries. The United States stand out as the most significant collaborator for the region, with a substantial percentage of publications across nearly all Latin American countries. 

Collaboration in Latin American research demonstrates the vital role of partnerships, both within the region and beyond. Intra-regional collaborations, particularly with countries like Brazil and Mexico, show the importance of regional leaders in supporting scientific integration among neighbouring countries. Internationally, connections with Ibero-America, notably Spain, as well as strong ties with global research leaders such as the United States, Germany, and Italy, highlight the diverse networks that enhance scientific research in Latin America and the world. 

\section{Summary and conclusion}
\label{sec:summary}

We consider this to be the first systematic and quantitative approach towards collecting all information available regarding scientific production in the whole Latin American region, in the fields of high energy, cosmology and astroparticle physics. The existence of the INSPIRE-HEP database for many decades and the availability to freely extract information from this database has allowed us to address a set of questions relevant to better understanding the scientific development in the region. We are aware that there are other data sets of scientific publications such as Google Scholar. It would be interesting to start a full study covering all the areas of science and all regions of the world using these datasets.

We consider that the sample we have studied already gives us a good perspective on the scientific productivity of the region. It is reassuring that despite the many ups and downs in the financial and political situation of most of the countries in the region, there is an almost steady increase in productivity concerning the world average. It is worth noticing that even though on average there has been a continuous increase, the rate of increase has been reaching a plateau that indicates further stimuli are needed for the scientists in the region to improve their relative productivity. Furthermore, as can be seen from the figures in Sections \ref{sec:coll_latam} and \ref{sec:cits_hindex}, there is an interesting `gap' starting around 2010 that may need explanation. The world financial crisis and several internal crises may be behind the explanation, but a better understanding is needed. The good news is that whatever the reason, besides some notable exceptions, we are now back in the relatively slow-increasing trend.

Concretely, we addressed the 8 questions mentioned in the introduction and concluded that:
\begin{enumerate}
\item The scientific production shows uneven sequences of ups and downs concerning the rest of the world but it shows a steady increase especially during the decades of 1960-1990, with the slope flattening and sometimes reducing in later years. It would be interesting to understand in detail the reasons for these developments. It may be argued that the early work of individuals like Giambiagi, Leite Lopes and Moshinsky had a clear impact in establishing the culture of research in the region. This, complemented by the creation of institutions such as the \ac{CLAF}, the global impact of the Abdus Salam \ac{ICTP}, the establishment of the Latin American School of Physics and later on the \ac{CIF} in Colombia and the \ac{CURCAF} played a key role on this sustainable growth. Experimental physics initiatives such as the Lederman Fermilab programme and the HELEN programme from CERN (funded by the European Union) had a big impact on the experimental \ac{HEP} community, but this cannot be fully measured with our analysis since it is restricted to publications with no more than 10 authors. 

The relatively slow reaction to the global financial crisis in the first decade of this century and the usual fluctuations in political stability may have played a role in the slowing and, in some cases, periods of decreased relative productivity. However, a detailed analysis of this aspect is beyond the scope of this publication. 

\item The number of active scientists has clearly increased in the region which is a positive sign, despite the varying working conditions. The different political conditions of the dictatorship or military governments in Argentina, Brazil, Uruguay triggered a flow of scientists abroad, sometimes to countries from the region. More recently, the substantial decrease in the number of active scientists in Venezuela is worth noticing. The information we provided by country is important for several reasons; it shows a disparity in the number of active researchers from different countries, even when normalised by the population of the corresponding country. Chile stands out in this regard, and smaller countries such as the Central American and Caribbean (with the notable exception of Costa Rica)  as well as Bolivia and Paraguay do not yet reach a critical mass of active researchers. We hope this analysis will be informative when addressing funding issues related to the need to educate more scientists, but also to understand the current distribution of researchers when planning regional projects.
\item The evolution of publications per country was normalised in two ways. with respect to the population of the country and with respect to the number of active researchers. In the first case, Chile stands out and in the second case, Costa Rica does. It is worth noticing that even though the Dominican Republic appears high, this case is singular in the sense that there was at some point one single author who was very productive but later emigrated to another country.

\item 
The evolution over the years shows a steady increase, albeit with a decreased slope in most cases. It is worth remarking on the comparison between Chile and Argentina which shows very different evolution rates, especially towards recent times, but more remarkable is the difference between similar-sized countries like Colombia and Venezuela which shows a substantial difference in the evolution, particularly during this century.

\item In Section \ref{sec:cits_hindex} we considered the number of citations and the h-index (number $h$ of articles with more than $h$ citations) for each country throughout the period considered. We present the data with several measures. First, the total number of citations per country partially illustrates the impact that the scientific productivity of the country has. Then the number of citations by million inhabitants is an indirect measure of how well the country is performing relative to its population size. We also normalise the number of citations not by the total population of the country but by the number of authors. This is a measure of the impact that the scientific community of a country has. Similar considerations are made for the h-index.

We found that, as expected, the big countries dominate regarding the total number of citations and h-index which is clear since they have more total support and more total number of scientists. However, the relative numbers vary. Small countries, such as Uruguay and Costa Rica perform well but other small countries show they do not have enough scientists to make a difference. In some cases (like the Dominican Republic), one single scientist made a difference, illustrating the fact that in these cases the numbers are too small to extract concrete statistical information, rather than the obvious one that there is a large disproportion in the number of active scientists and the need for those countries to educate more scientists. The relative measures help to avoid the typical (false) excuse from policymakers of those countries arguing that they are too small and poor and therefore cannot compete. 

\item The productivity as a function of the \ac{GDP} also varies substantially. The investment in science varies substantially from the essentially $1\%$ in Brazil to less than $5$ per thousand in some Central American countries. The productivity as a function of the \ac{GDP} favours countries like Chile and Uruguay. We found, however, that comparing productivity with respect to the HDI (human development index) gives probably a better picture of the overall conditions for research. We presented both cases for the completeness of the information. 

Using a linear regression model to describe the correlation between scientific productivity and the HDI, we find an overall positive correlation between the HDI and the indicators of scientific output, as expected. The fact that the HDI has a stronger impact on scientific productivity than the GDP indicates that to be more efficient, an increase in research funding should be combined with an overall initiative to improve all aspects of human development such as education, health and security.

\item Based on our method to approximately estimate the gender of different authors we managed to extract useful information regarding the important gender gap in this scientific community. Our analysis not only shows the overall estimate of order $15\%$ of female authors but also the fact that despite the community being fully aware of this disparity over the years, the percentage has not improved. Probably more worrisome, we found that this percentage decreases throughout the career of a given scientist with fewer and fewer female scientists remaining active as they age. We hope this quantitative information can be used to promote programmes to reduce this gender gap.


\item Finally, we studied the different collaborations of Latin American countries, both within the region and outside the region. The results show some clear patterns illustrating the fact that the scientists from smaller countries find their main collaborations in the bigger Latin American countries or outside the region.

Within the region, Brazil stands out as the country with the most collaborations in the region. This is because Brazil has the largest scientific community, but also partly because of its generous support to smaller countries by offering post-graduate fellowships to students from the smaller countries of the region, despite the large geographical distance from some of the countries, as is the case for Central America. A similar trend was also followed by Mexico and more recently by Argentina.

Outside the region, the USA stands out as the main collaborator, which is natural due to the proximity of some countries and the large amount of scientific possibilities in that country. But also European countries keep a scientific presence in the region. In part, due to the various cooperation programs in the region, such as DAAD from Germany and ICTP in Italy. The results show that healthy international collaborations are the best way to succeed in research. Especially for small countries that may lack a critical mass of researchers. Strengthening the collaborations among neighbouring countries would be desirable since in some cases it is absent, especially among the small countries. Furthermore, the relatively weak collaboration with countries like the UK, China and Russia may be explained by the weak historical connection with those countries (and their priorities regarding international cooperation programmes); by this reasoning, closer collaboration with countries like Spain and Portugal that do have a close historical and cultural connection to the region would be more than welcome.

\end{enumerate}

The different questions addressed regarding the comparison among the different countries reflect some general trends. The bigger countries dominate even the relative productivity partly because there is already an almost critical mass of scientists, whereas the progress in the smaller countries depends much more on the effort of individuals and the amount of support from the government. The cases of Chile, Uruguay and in some cases, Costa Rica are worth emphasising. Chile has benefited in part by being the location of many astronomical observatories (which can be seen to some extent in the primary arXiv categories of App. \ref{app:arxiv}), among other positive measures that have been taken recently. While these steps have contributed to its scientific output, further investment could allow Chile to fully capitalise on its research potential. It is important to point out the negative situation in Venezuela in which scientific productivity has been in decline for the past decade. In particular in comparison with similar size countries such as Colombia.

All in all, our study can be used to illustrate scientific development, to inform policymakers of the importance of investing in science in the long term and for the scientific community to make an effort to work together and support each other's initiatives. Although just investing more of the percentage of \ac{GDP} is not enough to guarantee success, it is an important fact that has been known for a while and we quantify it here in different ways.

Our analysis can be expanded to all areas of science, including large experimental collaborations that are typical in \ac{HECAP} and other disciplines, but also to all regions of the world. It will be important for policymakers and even the private sector to have a global view of the existence of active scientists generating high-level research results in all parts of the world, and science is not only an activity concentrated in the richest countries.

We hope that our work can trigger further quantitative research to fully cover the science datasets and that in the end it can be used to guide scientists and policymakers to have a broader view on how science can be supported and play a key role in the development of nations.

\vfill\eject

\subsection*{Acknowledgments}

We thank the members of the \ac{HECAP} Latin American community for many discussions that motivated this work. We are particularly thankful to Rafael Anta, José L. López-Gómez, Marta Losada, Sandro Scandolo, Rogerio Rosenfeld and Galileo Violini for their valuable input. MR-H thanks the New York University Physics Department in Abu Dhabi for their hospitality.











\appendix

\section{Publications}
\label{app:pub}
In this first appendix, we provide the procedure and exact numbers of the citation count for all the countries in our study.

In Table \ref{table:publications}, we present the total number of publications, publications \ac{PMI}, and publications per author for each country in our analysis.

\begin{table}[h!]
\begin{tabular}{@{}llll@{}}
\toprule
Country & Total & Total P. M. I. & Total per author \\
\midrule
Argentina & 5958 & 131 & 3\\
Bolivia & 42 & 4 & 1\\
Brazil & 24962 &  117 & 3 \\ 
Chile & 6396 &  333 & 3\\
Colombia & 1582 & 31 & 2\\
Costa Rica & 215 & 42 & 7\\
Cuba & 393 & 35 & 2\\
Dominican Republic & 5 & 0.4 & 5 \\
Ecuador & 117 &  7 & 2\\
El Salvador & 6 & 1 & 1 \\
Guatemala & 32 & 2 & 1 \\
Honduras & 13 & 1 & 1\\ 
Mexico & 10398 & 80 & 3  \\
Panama & 3 & 1 & 1  \\
Paraguay & 5 & 1 & 1 \\
Peru & 185 & 6 & 1 \\
Uruguay & 441 & 127 & 5\\
Venezuela & 1047 & 36 & 3\\
\botrule
\end{tabular}
\caption{Number of publications of each country, and \ac{PMI} and per author normalisations.}
\label{table:publications}
\end{table}

\begin{figure*}
        \centering
        \begin{subfigure}[b]{0.9\textwidth}
            \centering
            \includegraphics[width=\textwidth]{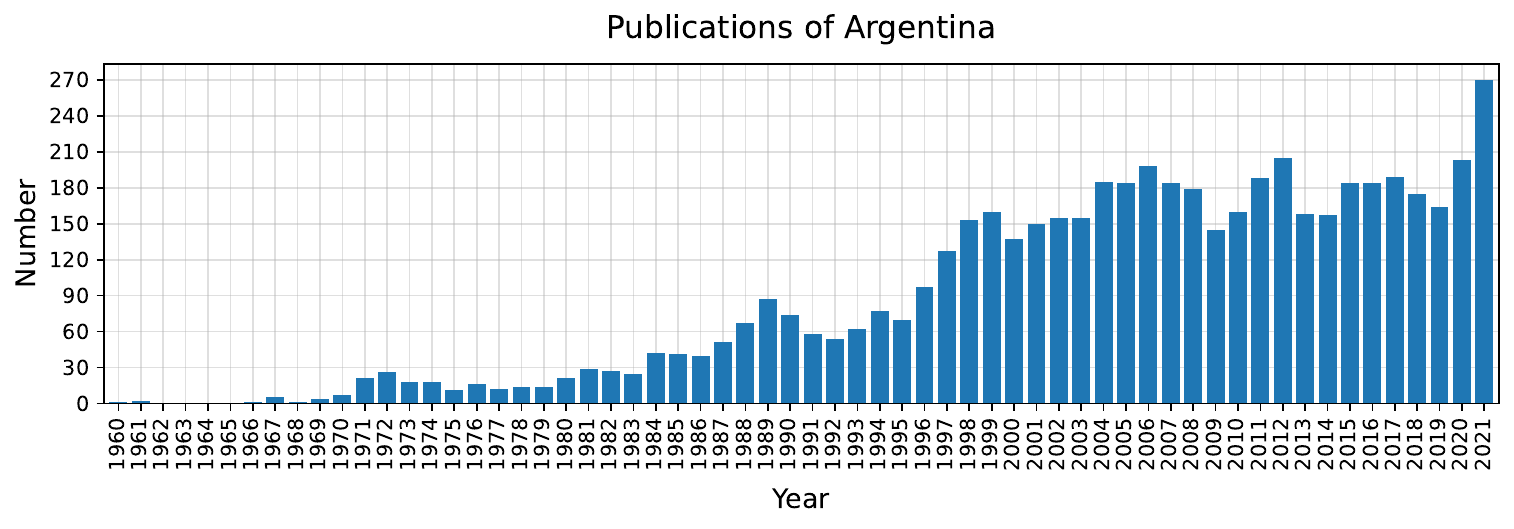}
           \label{fig:papers_year_argentina}
        \end{subfigure}
        \hfill
        \begin{subfigure}[b]{0.9\textwidth}  
            \centering 
            \includegraphics[width=\textwidth]{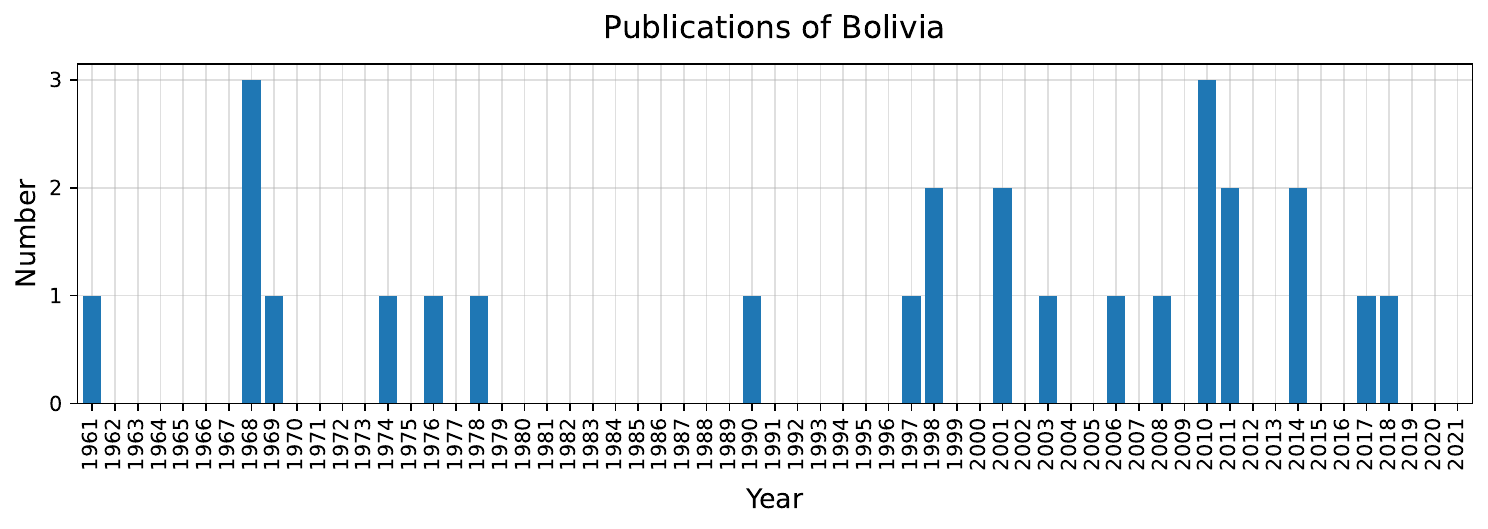}
            \label{fig:papers_year_bolivia}
        \end{subfigure}
        \hfill
        \begin{subfigure}[b]{0.9\textwidth}   
            \centering 
            \includegraphics[width=\textwidth]{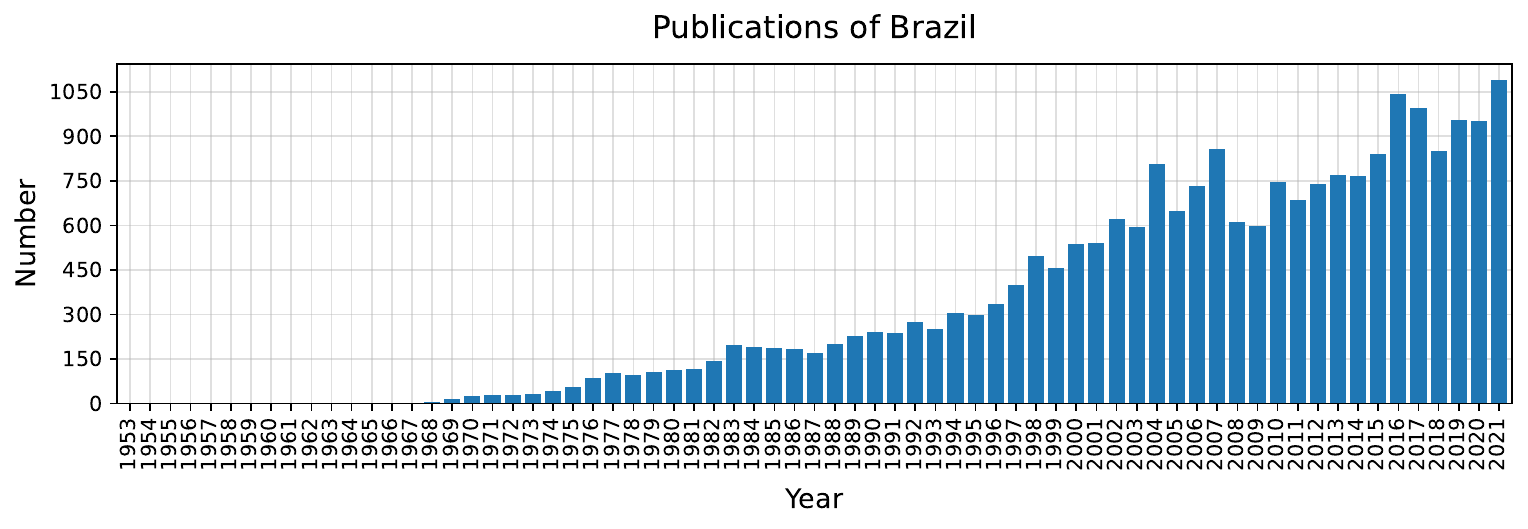}
            \label{fig:papers_year_brazil}
        \end{subfigure}
        \hfill
        \begin{subfigure}[b]{0.9\textwidth}   
            \centering 
            \includegraphics[width=\textwidth]{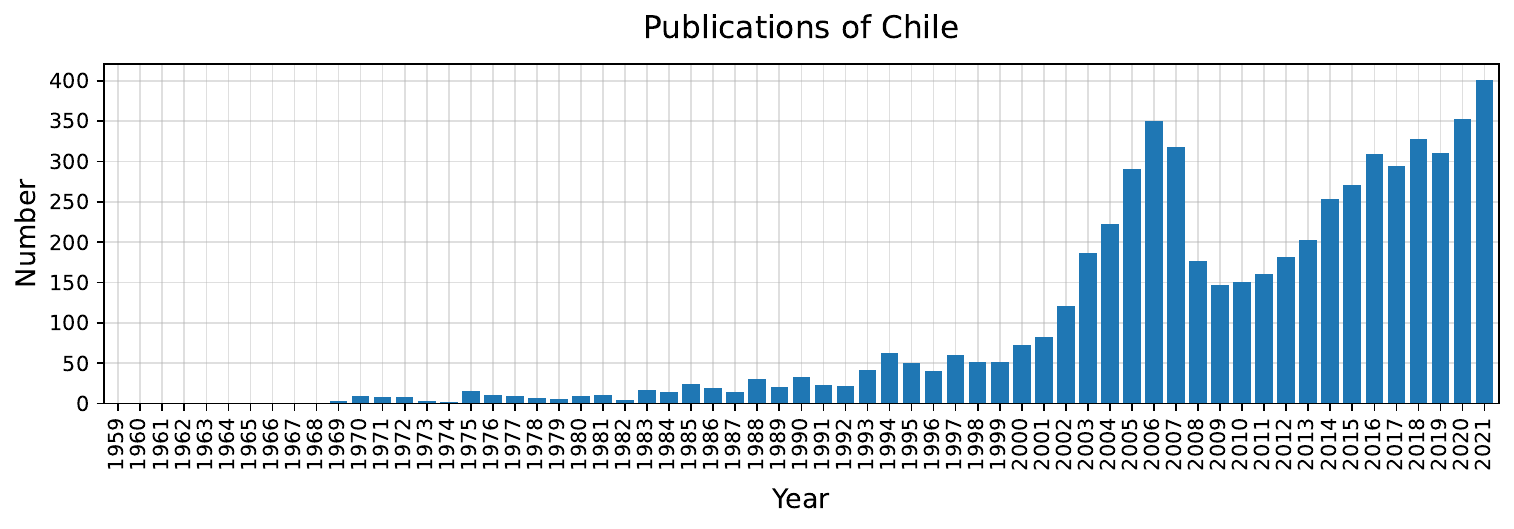}
            \label{fig:papers_year_chile}
        \end{subfigure}
        \caption[]
        {\small Histogram of the number of publications of Argentina, Bolivia, Brazil and Chile from their respective first year of publication in the database until 2021.} 
        \label{fig:Pub_Argentina-Chile}
    \end{figure*}

\begin{figure*}
        \centering
        \begin{subfigure}[b]{0.9\textwidth}
            \centering
            \includegraphics[width=\textwidth]{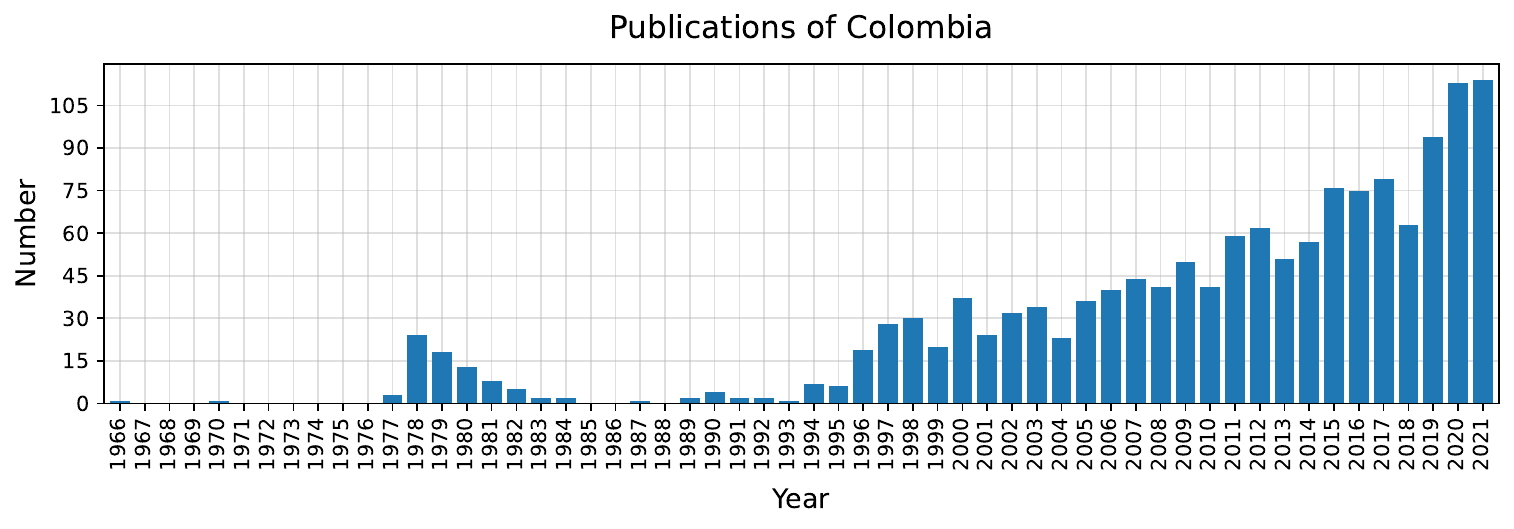}
           \label{fig:papers_year_colombia}
        \end{subfigure}
        \hfill
        \begin{subfigure}[b]{0.9\textwidth}  
            \centering 
            \includegraphics[width=\textwidth]{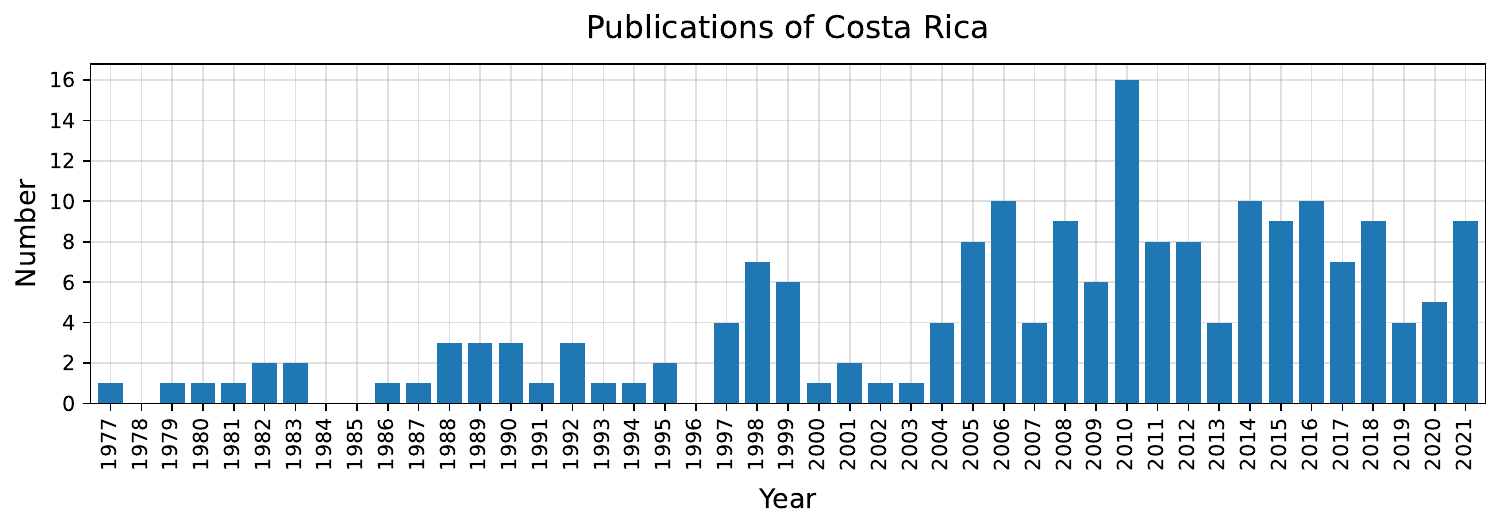}
            \label{fig:papers_year_costarica}
        \end{subfigure}
        \hfill
        \begin{subfigure}[b]{0.9\textwidth}   
            \centering 
            \includegraphics[width=\textwidth]{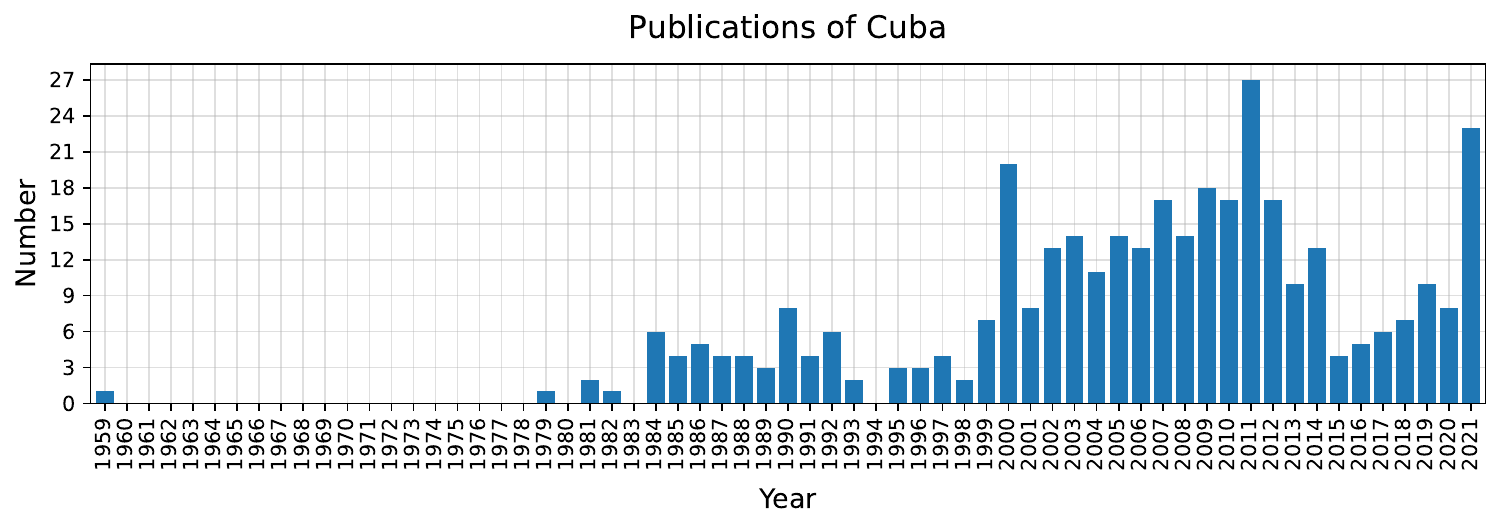}
            \label{fig:papers_year_cuba}
        \end{subfigure}
        \hfill
        \begin{subfigure}[b]{0.9\textwidth}   
            \centering 
            \includegraphics[width=\textwidth]{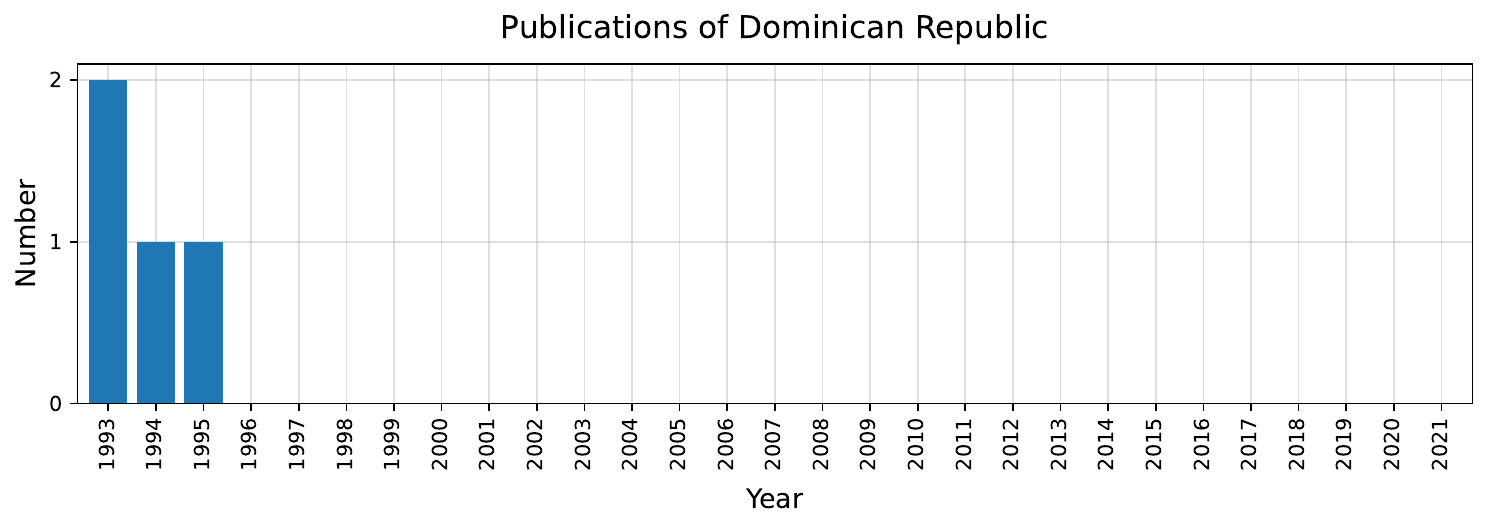}
            \label{fig:papers_year_dominicanrepublic}
        \end{subfigure}
        \caption[]
        {\small Histogram of the number of publications of Colombia, Costa Rica, Cuba and Dominican Republic from their respective first year of publication in the database until 2021.} 
        \label{fig:Pub_Colombia-DomRep}
    \end{figure*}

\begin{figure*}
        \centering
        \begin{subfigure}[b]{0.9\textwidth}
            \centering
            \includegraphics[width=\textwidth]{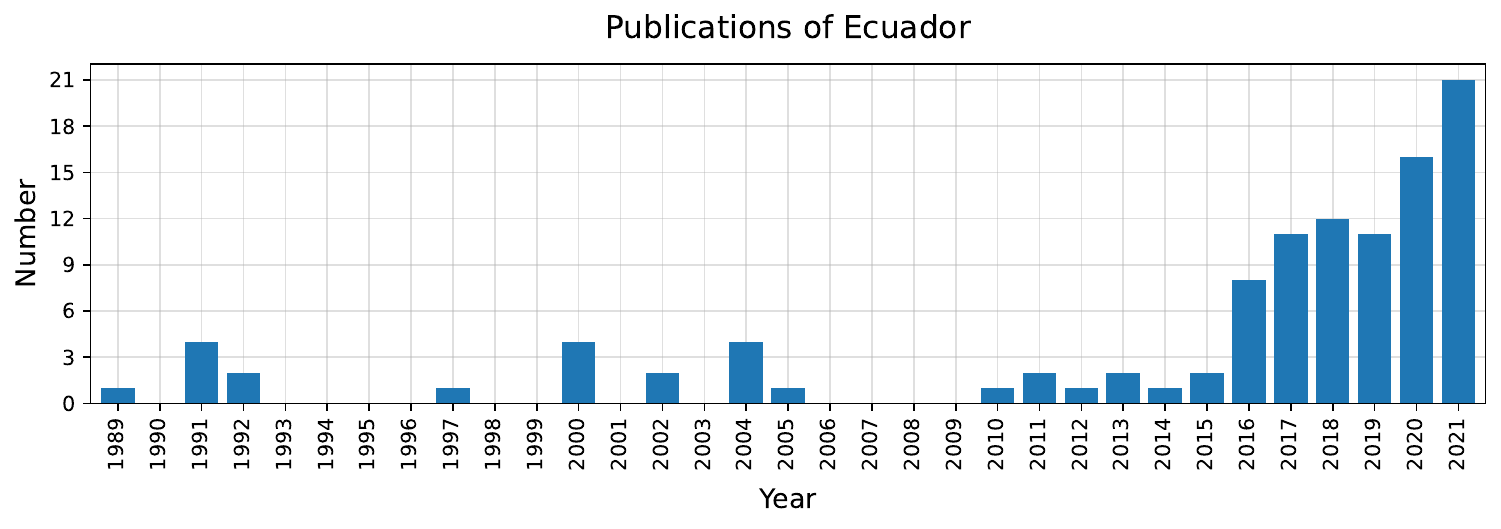}
           \label{fig:papers_year_ecuador}
        \end{subfigure}
        \hfill
        \begin{subfigure}[b]{0.9\textwidth}  
            \centering 
            \includegraphics[width=\textwidth]{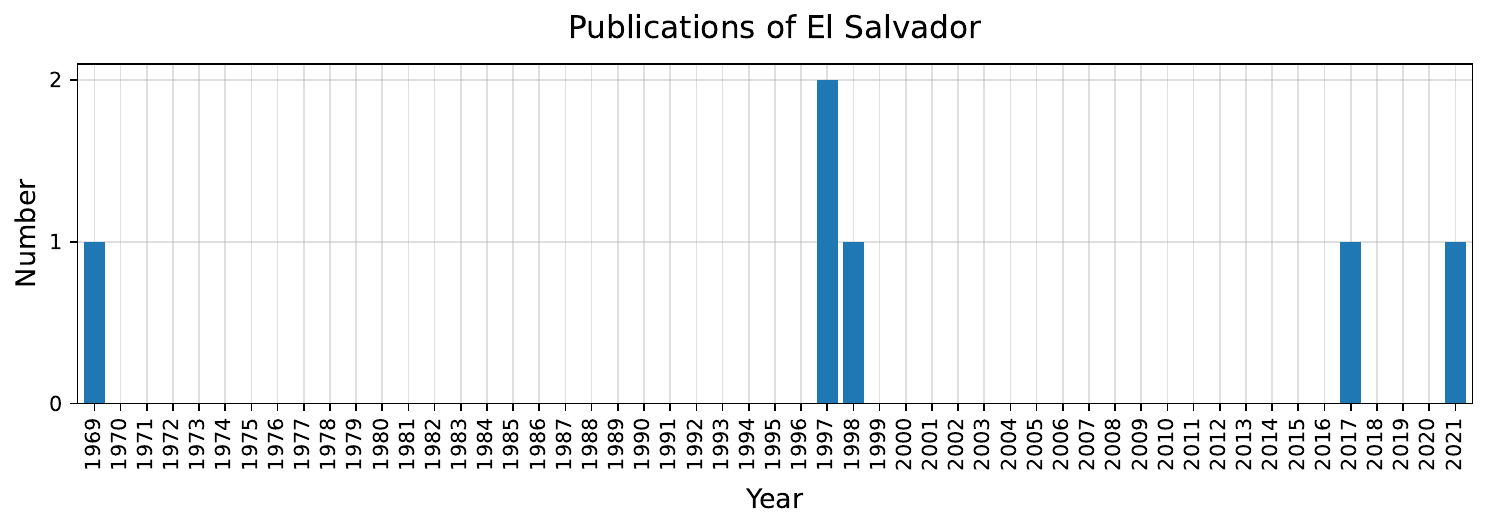}
            \label{fig:papers_year_elsalvador}
        \end{subfigure}
        \hfill
        \begin{subfigure}[b]{0.9\textwidth}   
            \centering 
            \includegraphics[width=\textwidth]{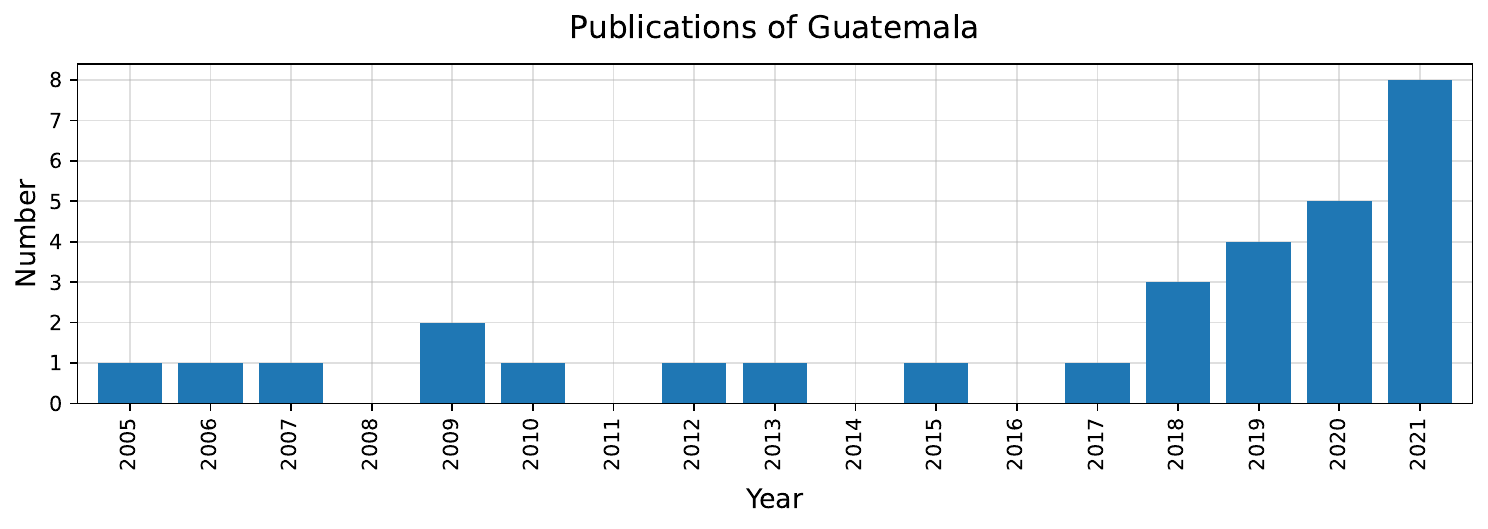}
            \label{fig:papers_year_guatemala}
        \end{subfigure}
        \hfill
        \begin{subfigure}[b]{0.9\textwidth}   
            \centering 
            \includegraphics[width=\textwidth]{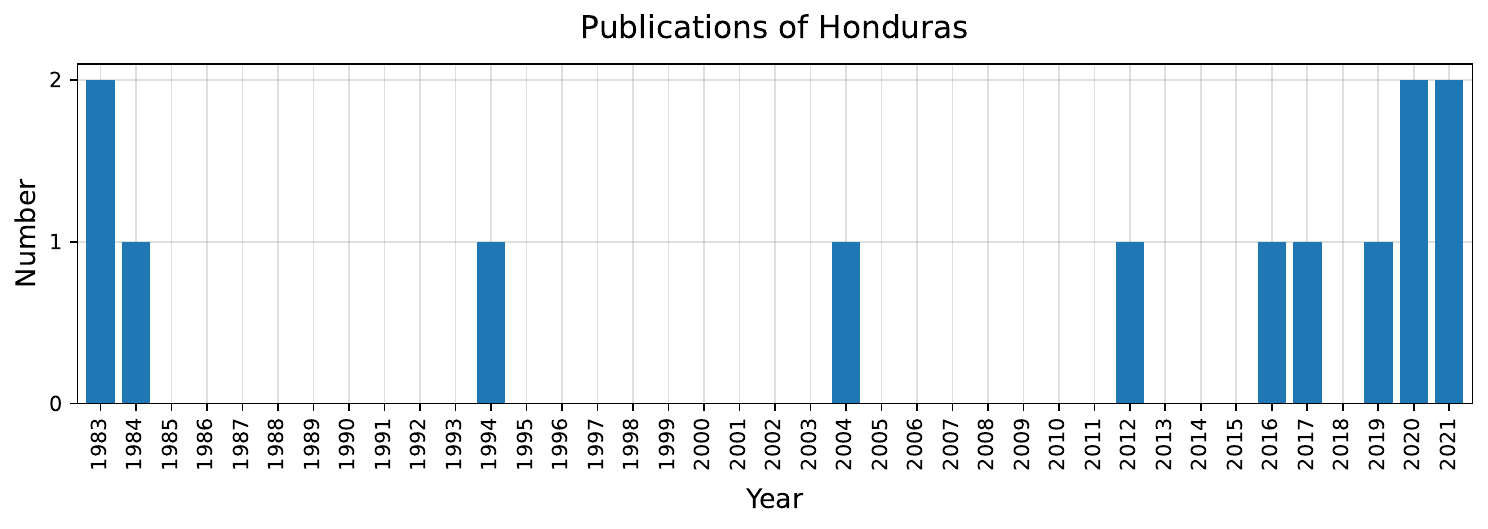}
            \label{fig:papers_year_honduras}
        \end{subfigure}
        \caption[]
        {\small Histogram of the number of publications of Ecuador, El Salvador, Guatemala and Honduras from their respective first year of publication in the database until 2021.} 
        \label{fig:Pub_Ecuador-Honduras}
    \end{figure*}

\begin{figure*}
        \centering
        \begin{subfigure}[b]{0.9\textwidth}
            \centering
            \includegraphics[width=\textwidth]{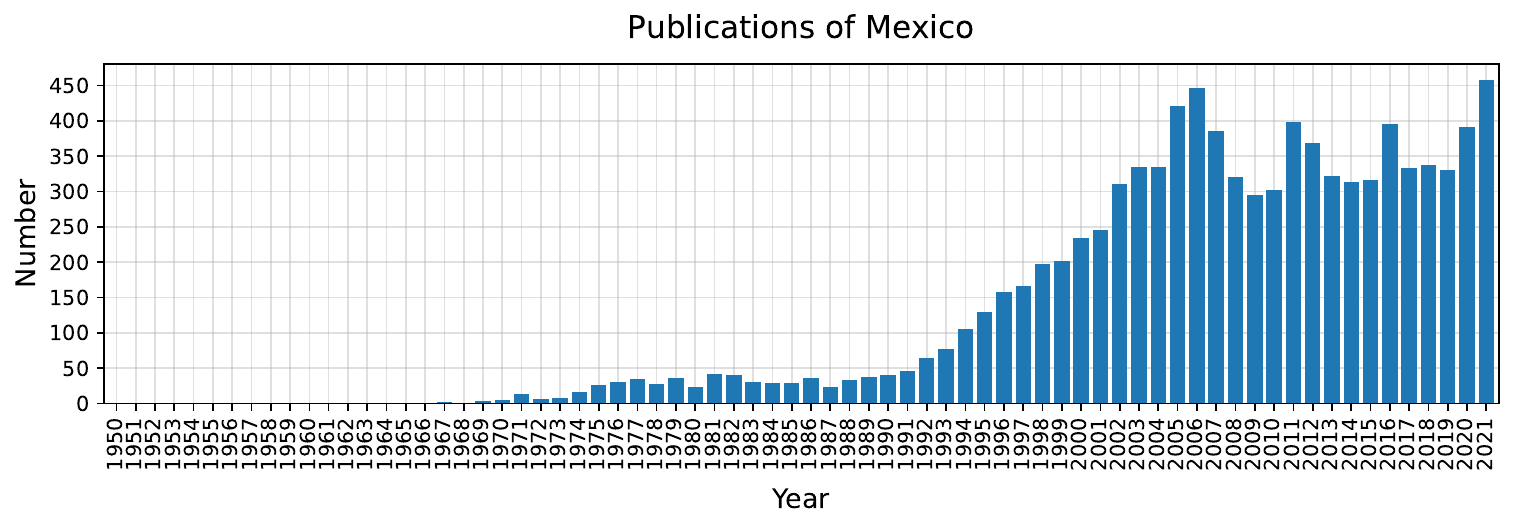}
           \label{fig:papers_year_mexico}
        \end{subfigure}
        \hfill
        \begin{subfigure}[b]{0.9\textwidth}  
            \centering 
            \includegraphics[width=\textwidth]{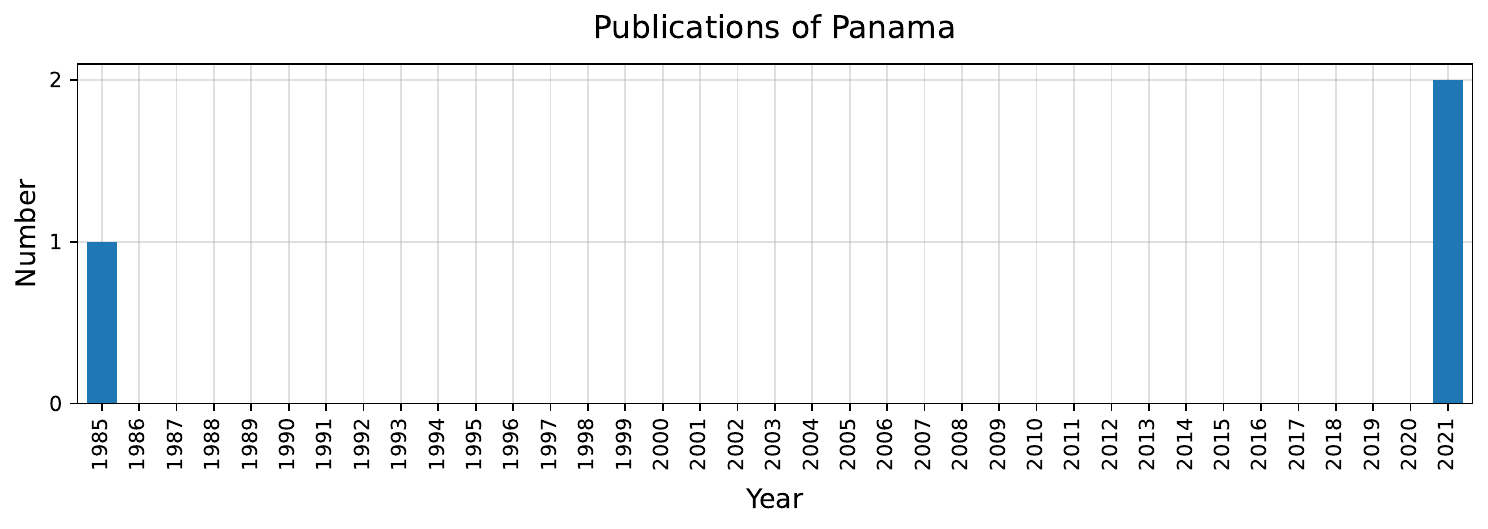}
            \label{fig:papers_year_panama}
        \end{subfigure}
        \hfill
        \begin{subfigure}[b]{0.9\textwidth}   
            \centering 
            \includegraphics[width=\textwidth]{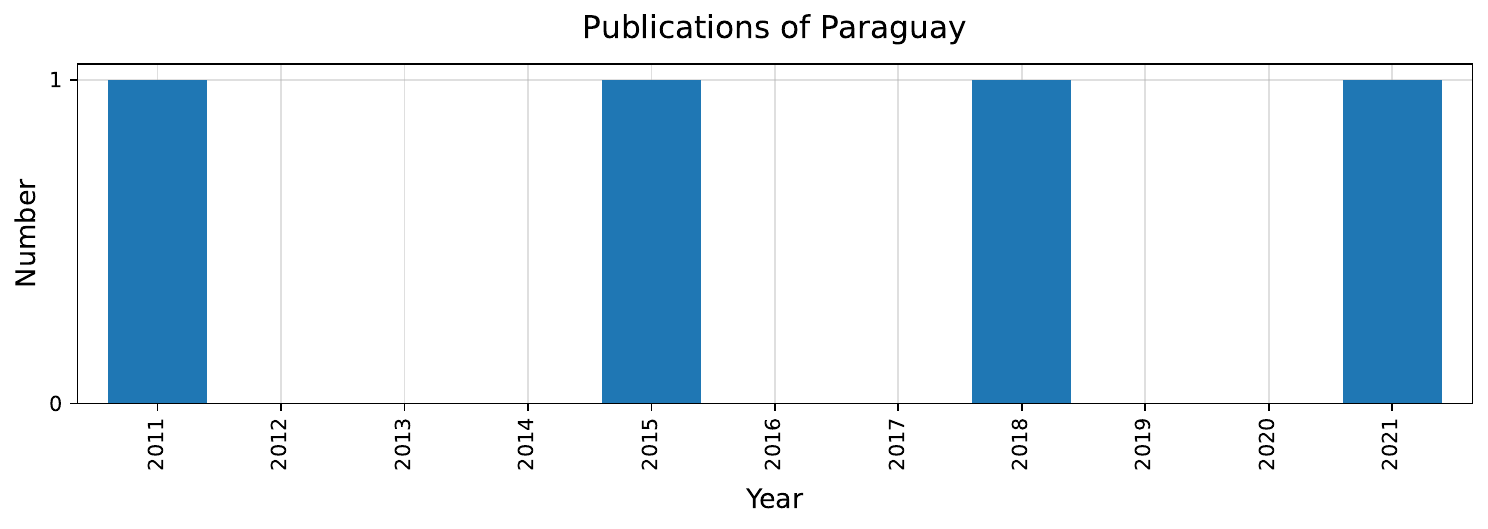}
            \label{fig:papers_year_paraguay}
        \end{subfigure}
        \hfill
        \begin{subfigure}[b]{0.9\textwidth}   
            \centering 
            \includegraphics[width=\textwidth]{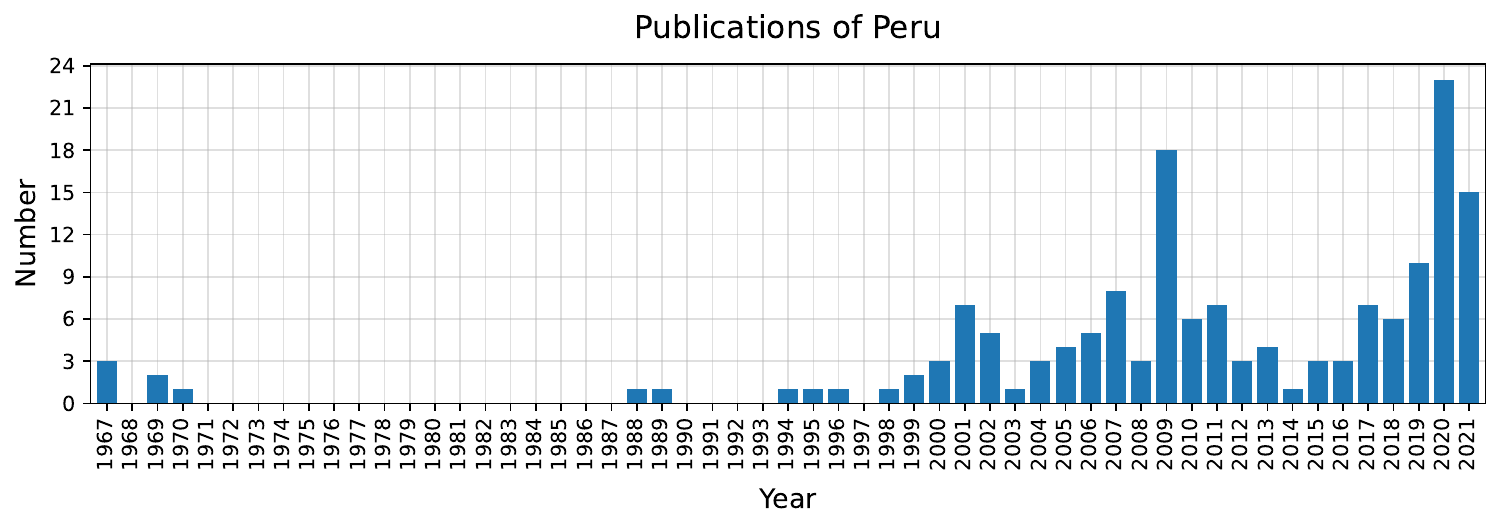}
            \label{fig:papers_year_peru}
        \end{subfigure}
        \caption[]
        {\small Histogram of the number of publications of Mexico, Panama, Paraguay and Peru from their respective first year of publication in the database until 2021.} 
        \label{fig:Pub_Mexico-Peru}
    \end{figure*}

\begin{figure*}
        \centering
        \begin{subfigure}[b]{0.9\textwidth}
            \centering
            \includegraphics[width=\textwidth]{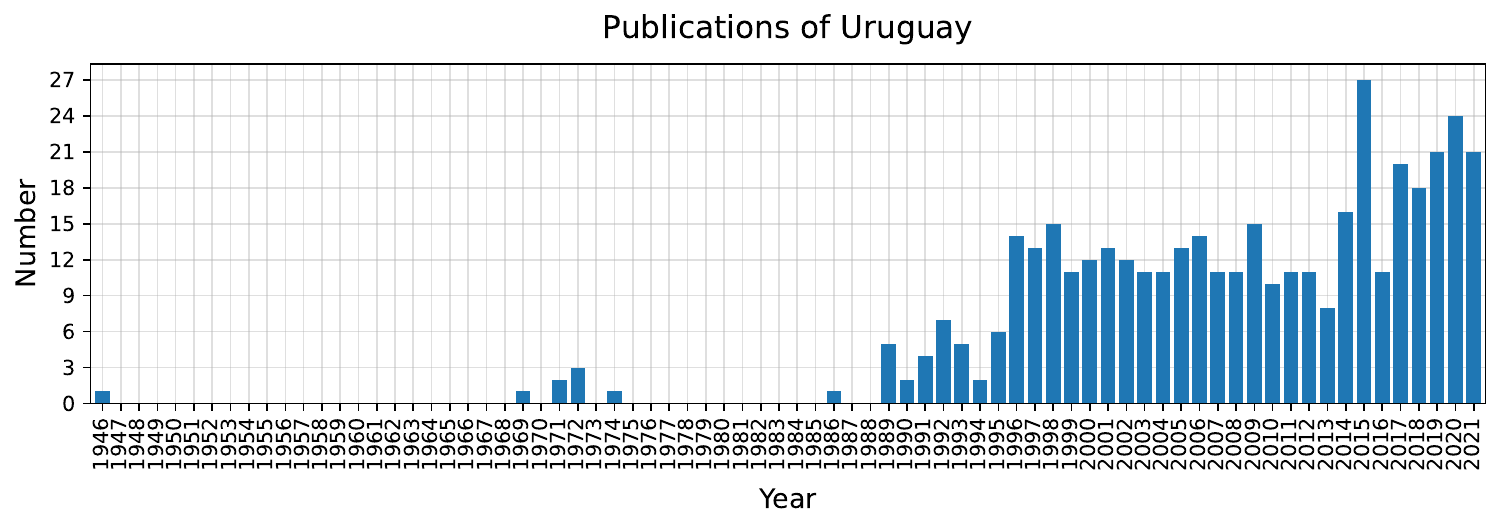}
           \label{fig:papers_year_uruguay}
        \end{subfigure}
        \hfill
        \begin{subfigure}[b]{0.9\textwidth}   
            \centering 
            \includegraphics[width=\textwidth]{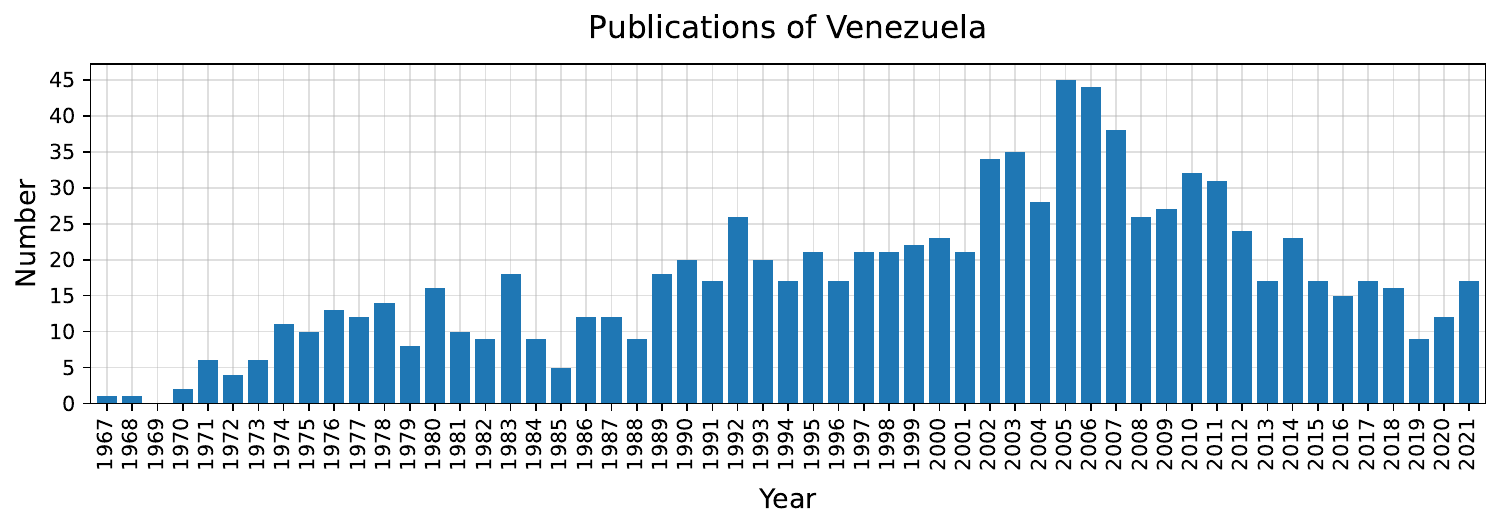}
            \label{fig:papers_year_venezuela}
        \end{subfigure}
        \caption[]
        {\small Histogram of the number of publications of Uruguay and Venezuela from their respective first year of publication in the database until 2021.} 
        \label{fig:Pub_Uruguay-Venezuela}
    \end{figure*}

Now, regarding bibliographical production, we present histograms of the annual number of publications from Latin American countries with available data in the INSPIRE-HEP database. Countries are grouped alphabetically, and each histogram starts from the publication year of the earliest recorded article for that country:

\begin{itemize}
    \item Figure \ref{fig:Pub_Argentina-Chile}: Argentina, Bolivia, Brazil, and Chile.
    \item Figure \ref{fig:Pub_Colombia-DomRep}: Colombia, Costa Rica, Cuba, and the Dominican Republic.
    \item Figure \ref{fig:Pub_Ecuador-Honduras}: Ecuador, El Salvador, Guatemala, and Honduras.
    \item Figure \ref{fig:Pub_Mexico-Peru}: Mexico, Panama, Paraguay, and Peru.
    \item Figure \ref{fig:Pub_Uruguay-Venezuela}: Uruguay and Venezuela.
\end{itemize}

%
%

%

\begin{figure}[h!]%
\centering
\includegraphics[width=0.9\textwidth]{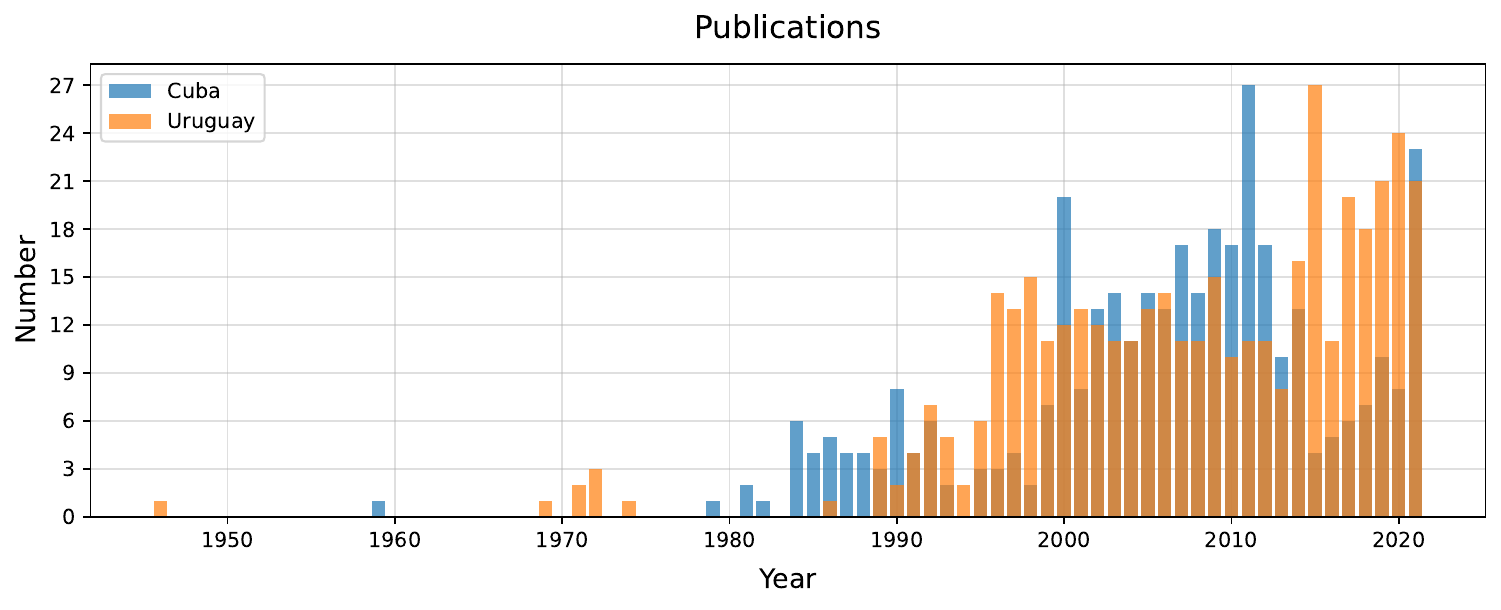}
\caption{Articles per year of Cuba and Uruguay.}
\label{fig:cub_ur} 
\end{figure}
\begin{figure}[h!]%
\centering
\includegraphics[width=0.9\textwidth]{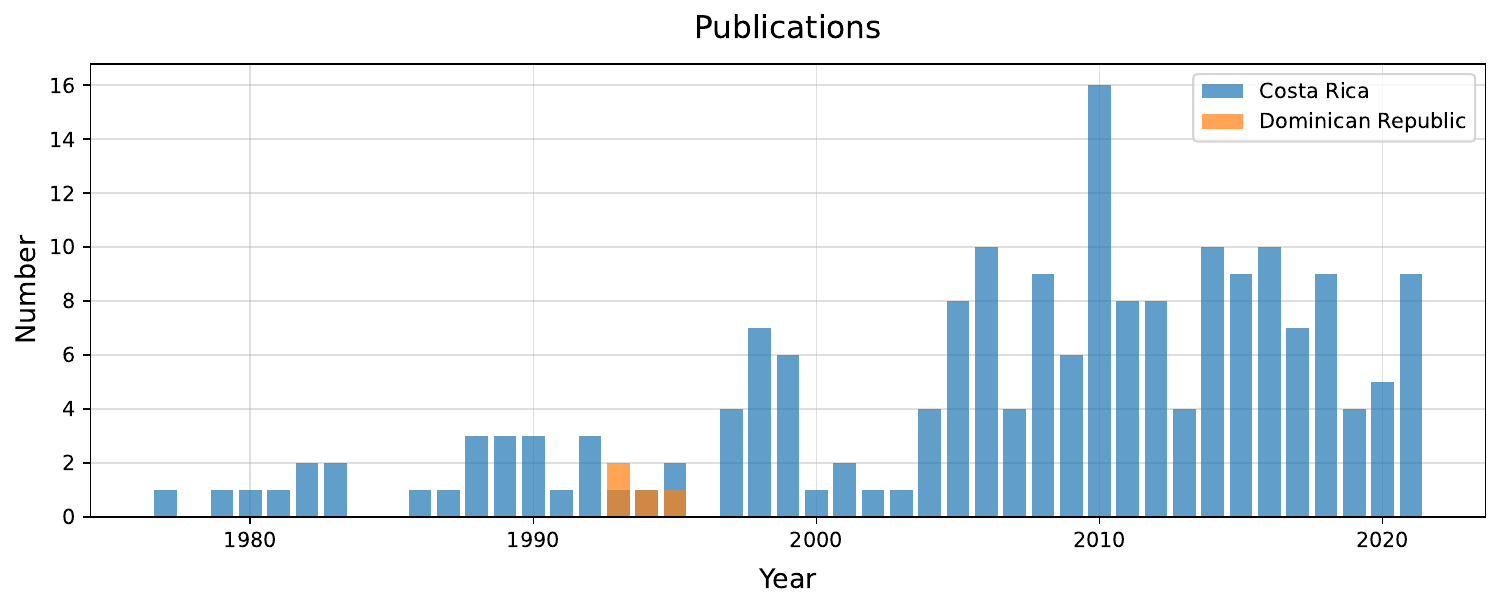}
\caption{\label{fig:dom_cr} Articles per year of Dominican Republic and Costa Rica.}
\end{figure}
\begin{figure}[h!]%
\centering
\includegraphics[width=0.9\textwidth]{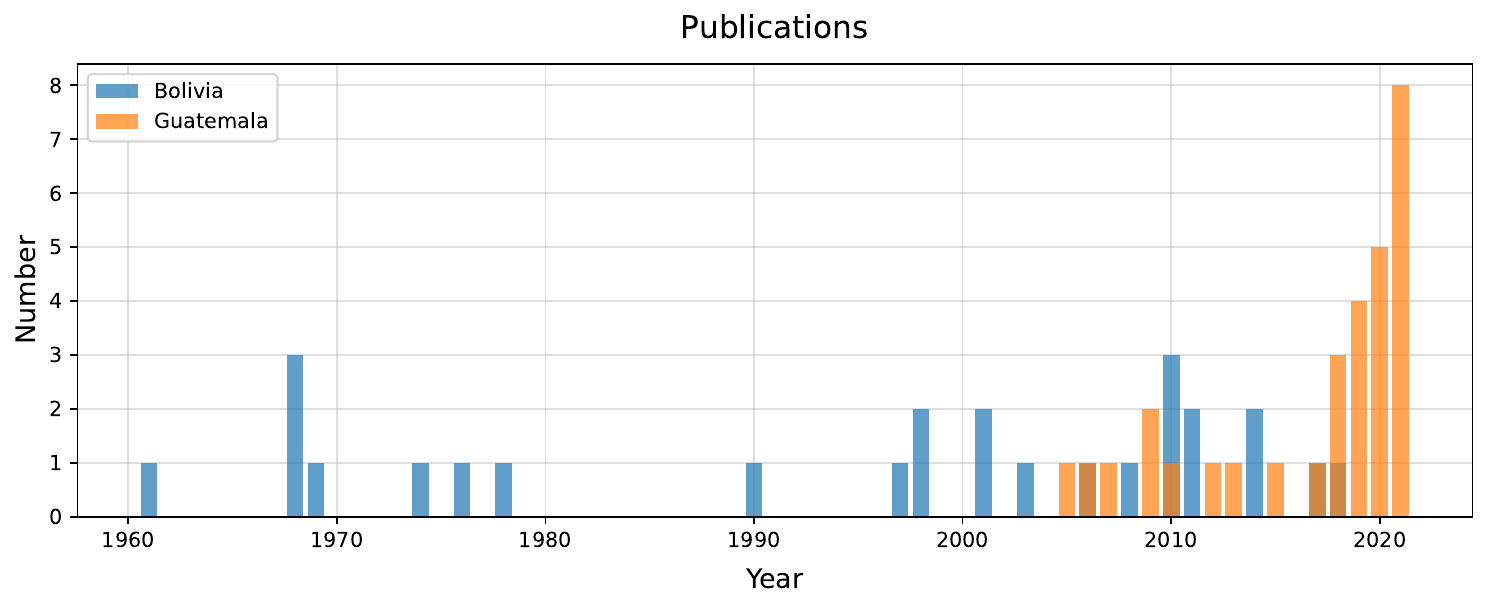}
\caption{\label{fig:gua_bol} Articles per year of Guatemala and Bolivia.}
\end{figure}
\begin{figure}[h!]%
\centering
\includegraphics[width=0.9\textwidth]{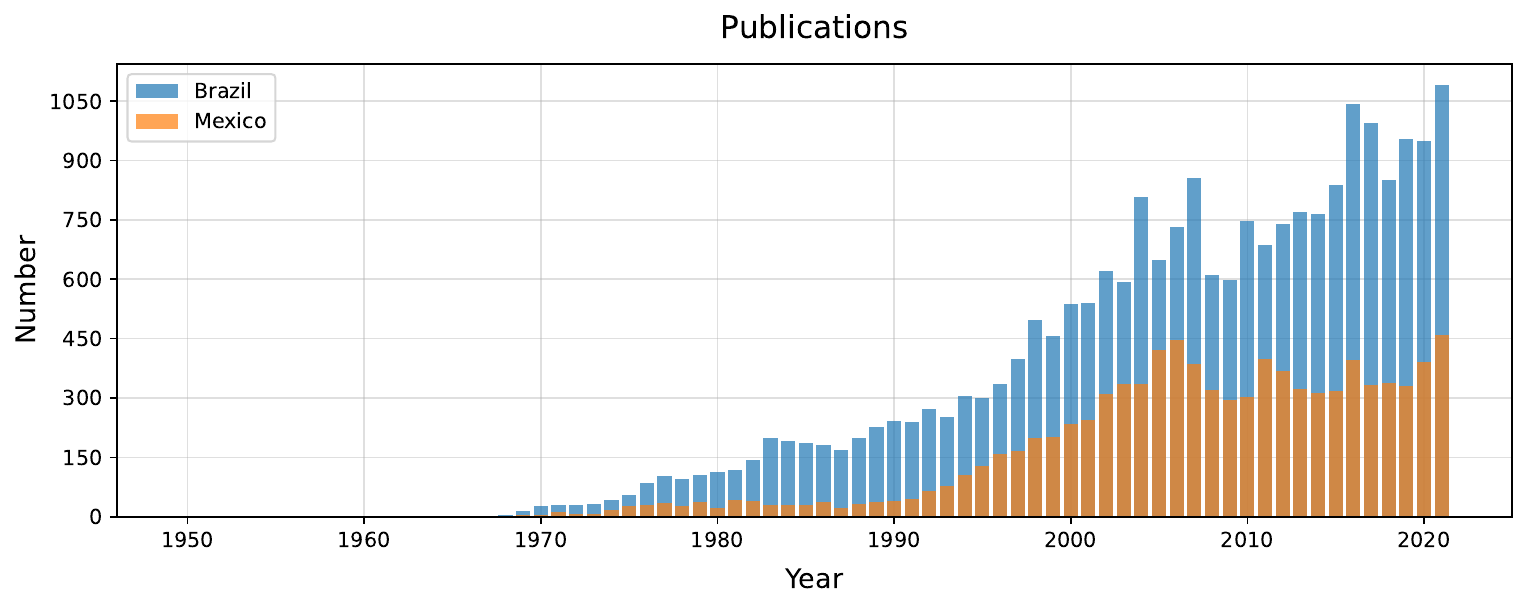}
\caption{\label{fig:mex_br} Articles per year of Mexico and Brazil.}
\end{figure}
\begin{figure}[h!]%
\centering
\includegraphics[width=0.9\textwidth]{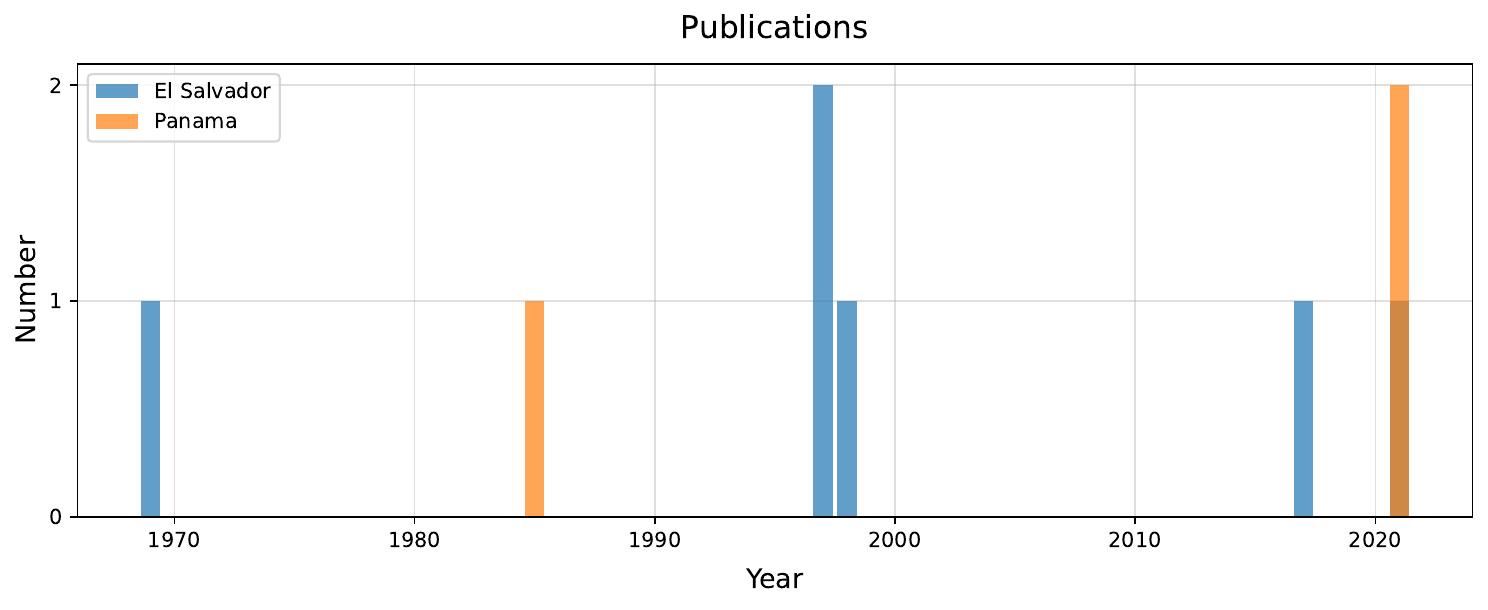}
\caption{\label{fig:pan_elsal} Articles per year of Panama and El Salvador.}
\end{figure}
\begin{figure}[h!]%
\centering
\includegraphics[width=0.9\textwidth]{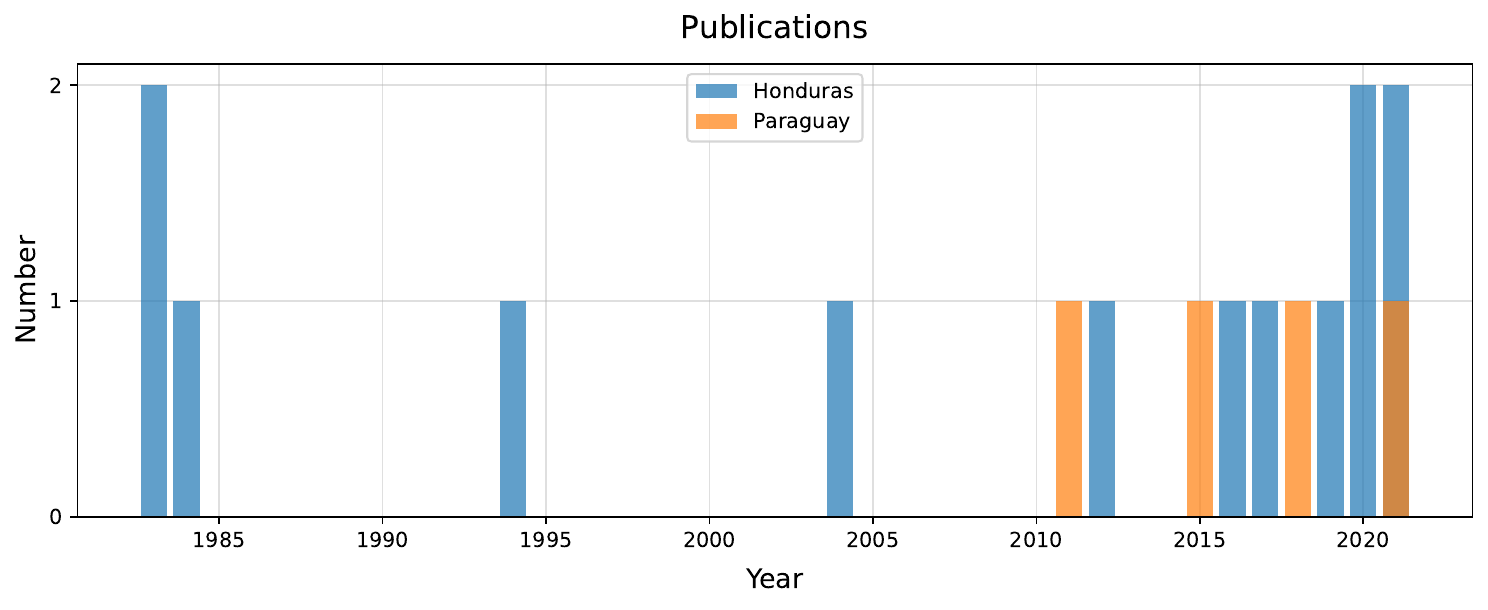}
\caption{\label{fig:par_hon} Articles per year of Paraguay and Honduras.}
\end{figure}
\begin{figure}[h!]%
\centering
\includegraphics[width=0.9\textwidth]{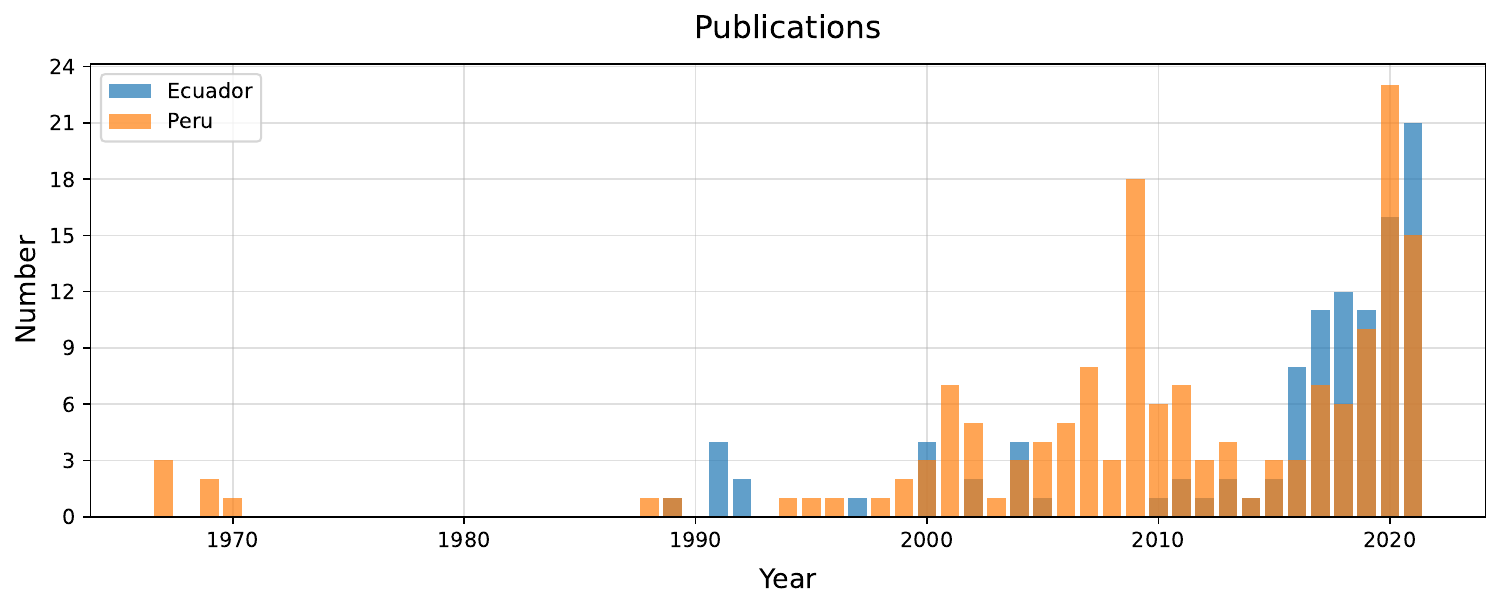}
\caption{\label{fig:per_ec} Articles per year of Peru and Ecuador.}
\end{figure}

As an additional resource for comparison, we provide combined plots showing the annual number of publications for different country groupings. Countries have been grouped based on similar publication scales, as closely as possible, to enhance visual clarity and facilitate comparison:

\begin{itemize}
    \item Figure \ref{fig:cub_ur}: Cuba and Uruguay.
    \item Figure \ref{fig:dom_cr}: Dominican Republic and Costa Rica.
    \item Figure \ref{fig:gua_bol}: Guatemala and Bolivia.
    \item Figure \ref{fig:mex_br}: Mexico and Brazil.
    \item Figure \ref{fig:pan_elsal}: Panama and El Salvador.
    \item Figure \ref{fig:par_hon}: Paraguay and Honduras.
    \item Figure \ref{fig:per_ec}: Peru and Ecuador.
\end{itemize}

%


\section{Citations and h-index}
\label{app:cit}
In this appendix we provide the exact numbers of the total citations and h-index, and the normalised results for each country in our study. Table~\ref{table:citations} and Table~\ref{table:h-index} respectively show the information related to the citations and to the h-index in our analysis.

\begin{table}[h!]
\begin{tabular}{@{}llll@{}}
\toprule
Country & Total Citations & \makecell{Total Citations \\ \ac{PMI}.} & \makecell{Total Citations \\ Per Author} \\
\midrule
Argentina & 107849 & 2365 & 54\\
Bolivia & 211 & 18 & 4\\
Brazil & 341064 &  1594 & 40 \\ 
Chile & 150398 &  7828 & 74\\
Colombia & 23877 & 466 & 33\\
Costa Rica & 7587 & 1476 & 262\\
Cuba & 4351 & 384 & 18\\
Dominican Republic & 184 & 17 & 184 \\
Ecuador & 1492&  83 & 23\\
El Salvador & 26 & 4 & 4 \\
Guatemala & 330 & 18 & 10 \\
Honduras & 15 & 1 & 1\\ 
Mexico & 144645 & 1110 & 37  \\
Panama & 2 & 0.4 & 0.6  \\
Paraguay & 57 & 8 & 11 \\
Peru & 2335 & 70 & 15 \\
Uruguay & 10693 & 3068 & 123\\
Venezuela & 25259 & 880 & 68\\
\botrule
\end{tabular}
\caption{Citations of each country.}
\label{table:citations}
\end{table}

\begin{table}[h!]
\begin{tabular}{@{}llll@{}}
\toprule
Country & h-index & h-index \ac{PMI}. & h-index Per Author \\
\midrule
Argentina & 126 & 3 & 0.06\\
Bolivia & 7 & 1 & 0.14\\
Brazil & 169 &  1 & 0.02 \\ 
Chile & 141 &  7 & 0.07\\
Colombia & 65 & 1 & 0.09\\
Costa Rica & 43 & 8 & 1.5\\
Cuba & 31 & 3 & 0.13\\
Dominican Republic & 2 & 0.1 & 2 \\
Ecuador & 22 &  1 & 0.34\\
El Salvador & 3 & 0.4 & 1 \\
Guatemala & 10 & 1 & 0.31 \\
Honduras & 2 & 0.1 & 0.18\\ 
Mexico & 130 & 1 & 0.03  \\
Panama & 1 & 0.2 & 0.34  \\
Paraguay & 3 & 0.4 & 0.6 \\
Peru & 25 & 1 & 0.16 \\
Uruguay & 55 & 16 & 0.63\\
Venezuela & 73 & 3 & 0.19\\
\botrule
\end{tabular}
\caption{h-index of each country.}
\label{table:h-index}
\end{table}

\section{Ranks and mobility}
\label{app:ranks}

\begin{figure*}
        \centering
        \begin{subfigure}[b]{0.496\textwidth}
            \centering
            \includegraphics[width=\textwidth]{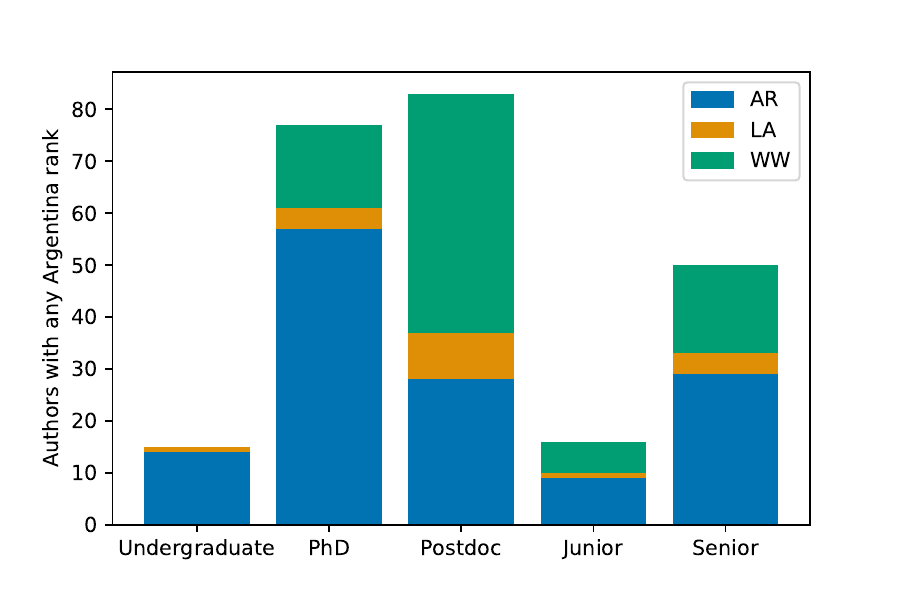}
            \caption[]%
            {{\small  Argentina.}}  
           \label{fig:Argentina:rank-papero}
        \end{subfigure}
        \hfill
        \begin{subfigure}[b]{0.496\textwidth}  
            \centering 
            \includegraphics[width=\textwidth]{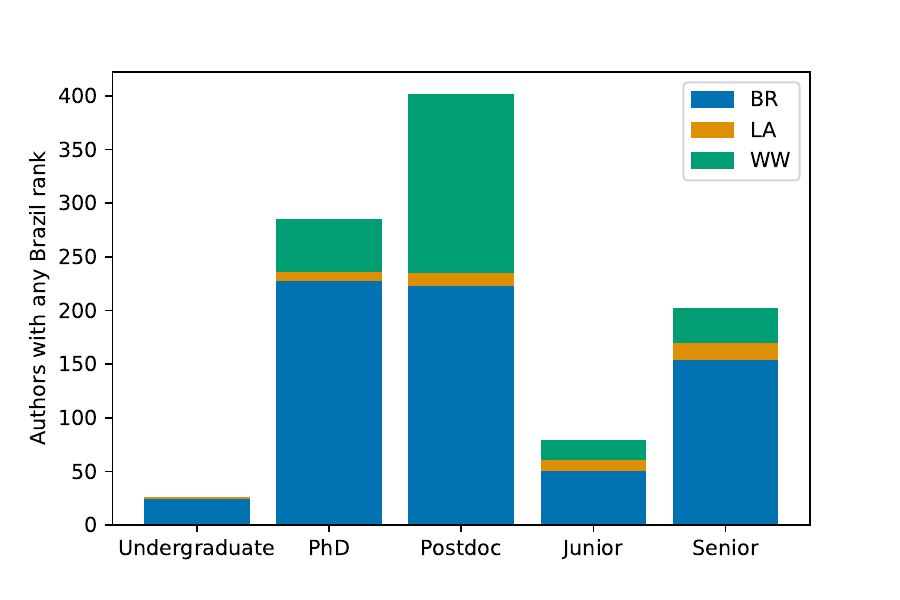}
            \caption[]%
            {{\small Brazil.}}    
            \label{fig:Brazil:rank-paper}
        \end{subfigure}
        \hfill
        \begin{subfigure}[b]{0.496\textwidth}   
            \centering 
            \includegraphics[width=\textwidth]{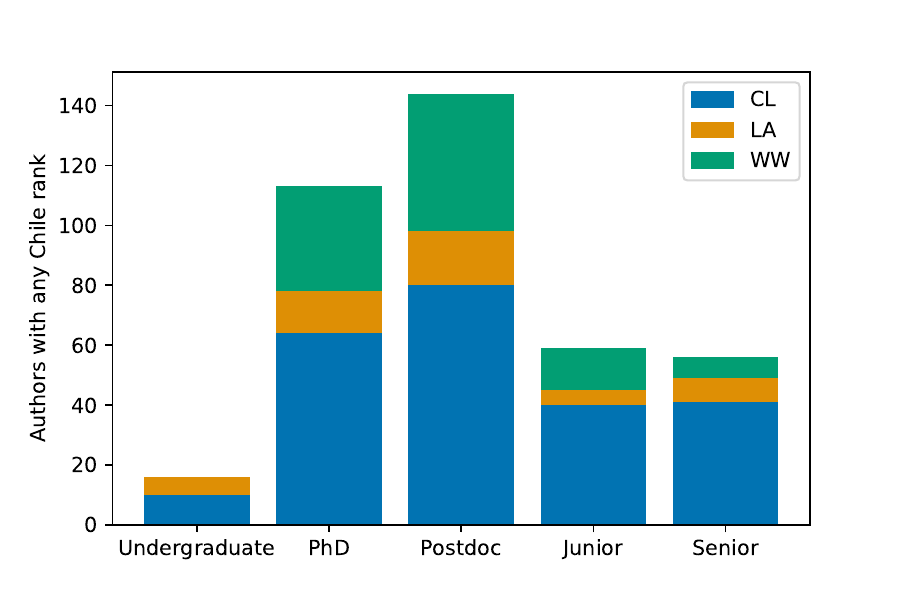}
            \caption[]%
            {{\small Chile.}}    
            \label{fig:Chile:rank-paper}
        \end{subfigure}
        \hfill
        \begin{subfigure}[b]{0.496\textwidth}   
            \centering 
            \includegraphics[width=\textwidth]{Figures/ranks/Colombia_rank-paper.pdf}
            \caption[]%
            {{\small Colombia.}}    
            \label{fig:Colombia:rank-paper}
        \end{subfigure}
        \caption[]
        {\small Number of authors with a given rank in given country as in table \ref{table:rank-authors} who published an article under a given rank in the same country, Latin America or elsewhere worldwide.} 
        \label{fig:Rank_Argentina-Colombia}
    \end{figure*}

\begin{figure*}
        \centering
        \begin{subfigure}[b]{0.496\textwidth}
            \centering
            \includegraphics[width=\textwidth]{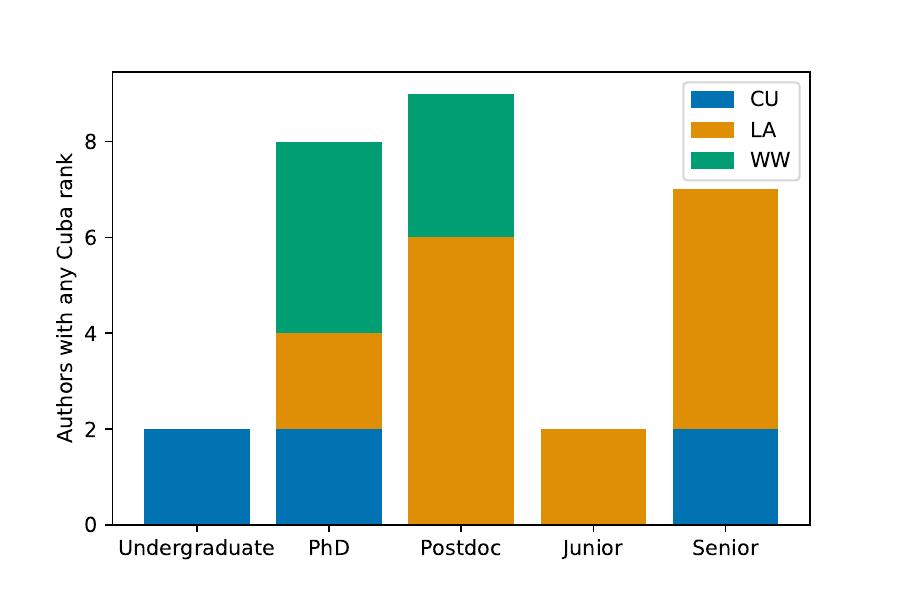}
            \caption[]%
            {{\small  Cuba.}}  
           \label{fig:Cuba:rank-paper}
        \end{subfigure}
        \hfill
        \begin{subfigure}[b]{0.496\textwidth}  
            \centering 
            \includegraphics[width=\textwidth]{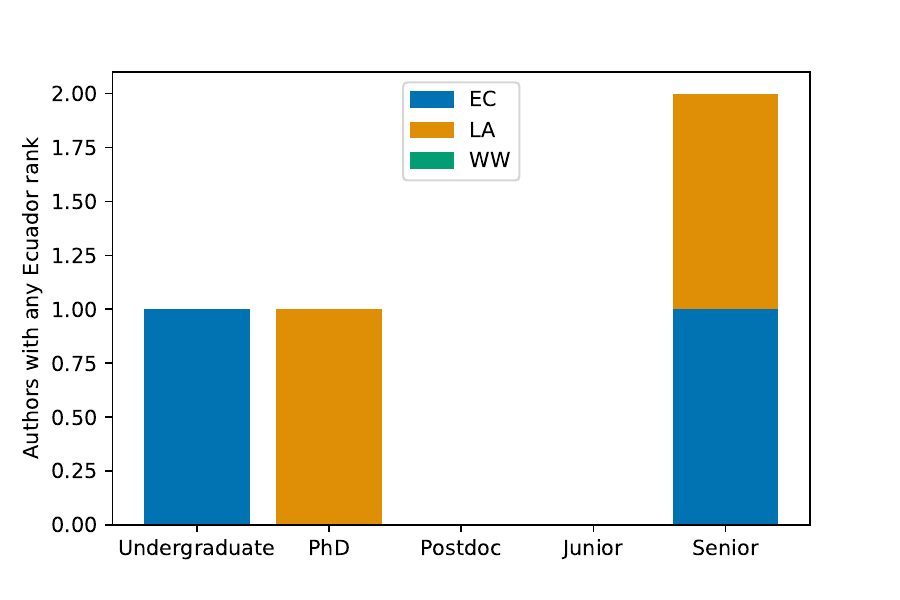}
            \caption[]%
            {{\small Ecuador.}}    
            \label{fig:Ecuador:rank-paper}
        \end{subfigure}
        \hfill
        \begin{subfigure}[b]{0.496\textwidth}   
            \centering 
            \includegraphics[width=\textwidth]{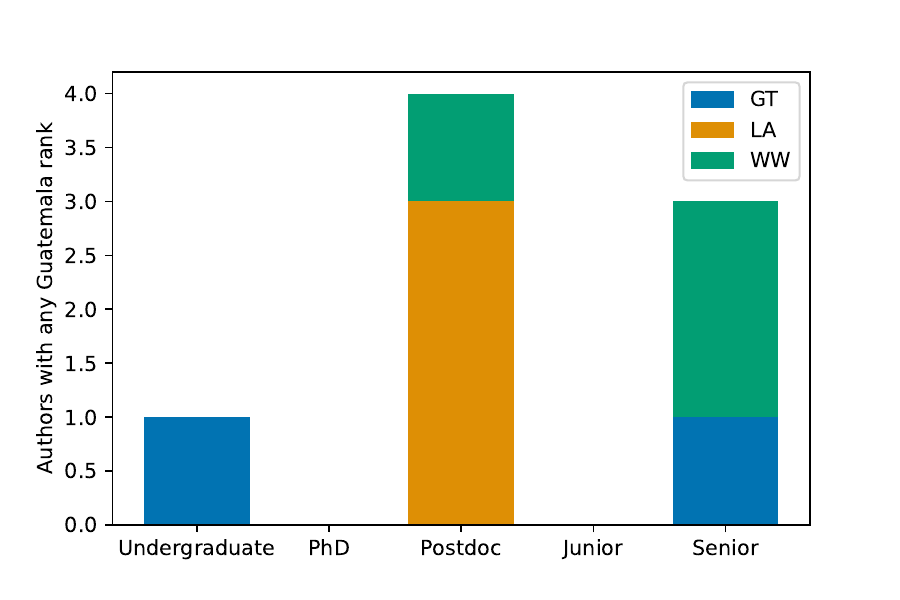}
            \caption[]%
            {{\small Guatemala.}}    
            \label{fig:Guatemala:rank-paper}
        \end{subfigure}
        \hfill
        \begin{subfigure}[b]{0.496\textwidth}   
            \centering 
            \includegraphics[width=\textwidth]{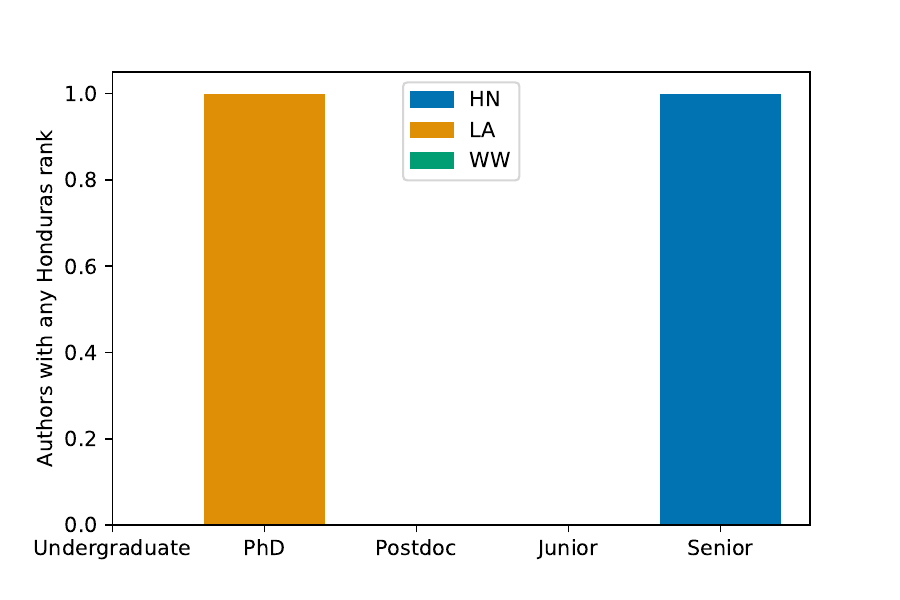}
            \caption[]%
            {{\small Honduras.}}    
            \label{fig:Honduras:rank-paper}
        \end{subfigure}
        \caption[]
        {\small Number of authors with a certain rank in a given country as in table \ref{table:rank-authors} who published an article under a given rank in the same country, Latin America or elsewhere worldwide.} 
        \label{fig:Rank_Cuba-Honduras}
    \end{figure*}

\begin{figure*}
        \centering
        \begin{subfigure}[b]{0.496\textwidth}
            \centering
            \includegraphics[width=\textwidth]{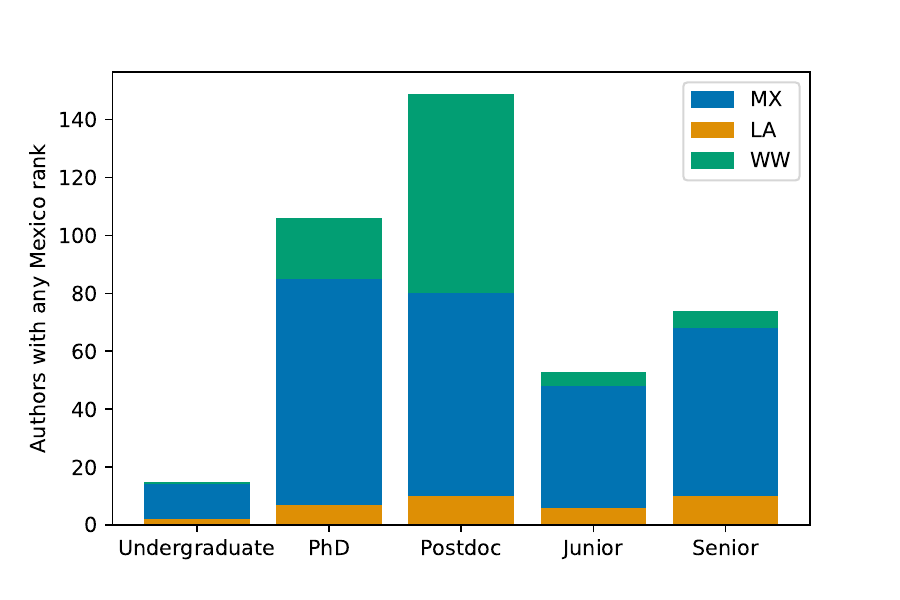}
            \caption[]%
            {{\small  Mexico.}}  
         \label{fig:Mexico:rank-paper}
        \end{subfigure}
        \hfill
        \begin{subfigure}[b]{0.496\textwidth}  
            \centering 
            \includegraphics[width=\textwidth]{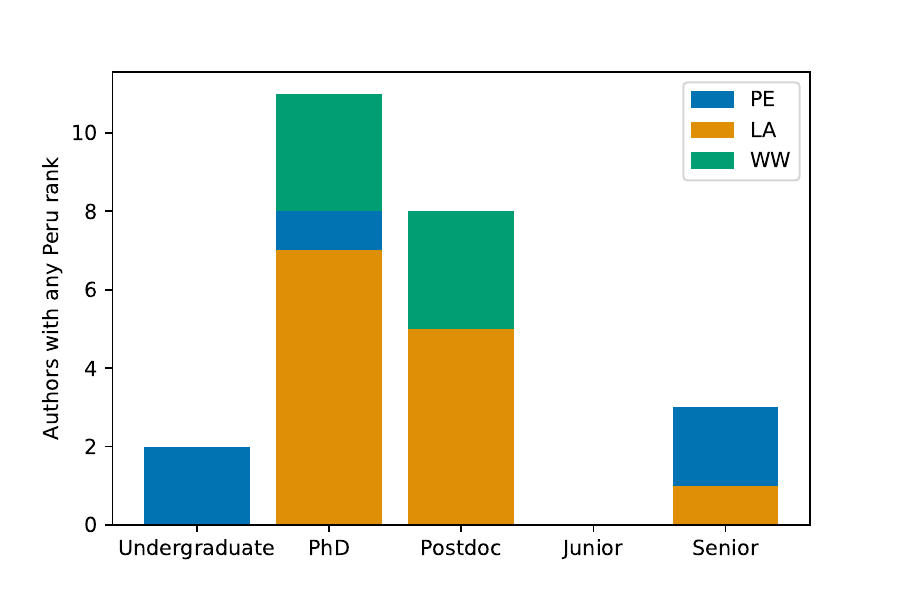}
            \caption[]%
            {{\small Peru.}}    
            \label{fig:Peru:rank-paper}
        \end{subfigure}
        \hfill
        \begin{subfigure}[b]{0.496\textwidth}   
            \centering 
            \includegraphics[width=\textwidth]{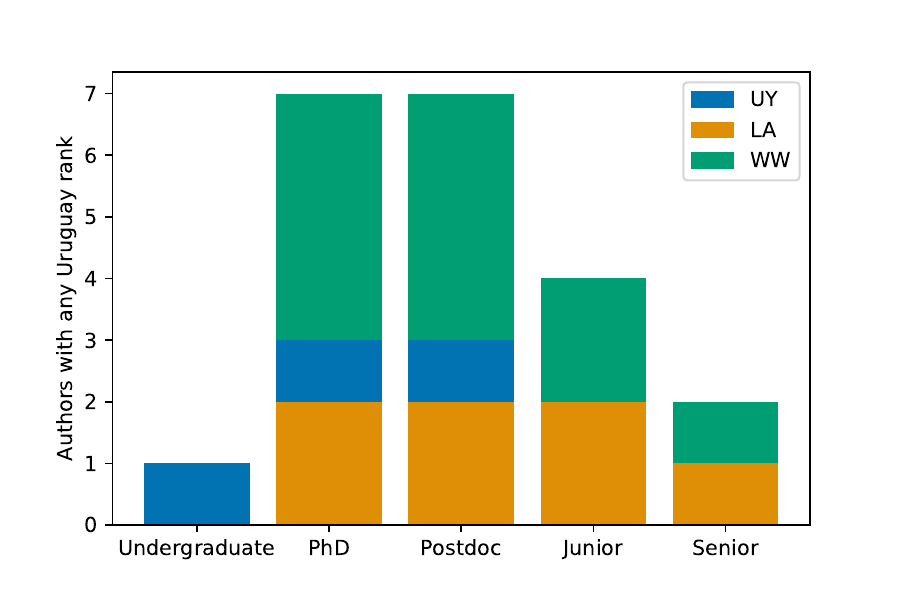}
            \caption[]%
            {{\small Uruguay.}}    
            \label{fig:Uruguay:rank-paper}
        \end{subfigure}
        \hfill
        \begin{subfigure}[b]{0.496\textwidth}   
            \centering 
            \includegraphics[width=\textwidth]{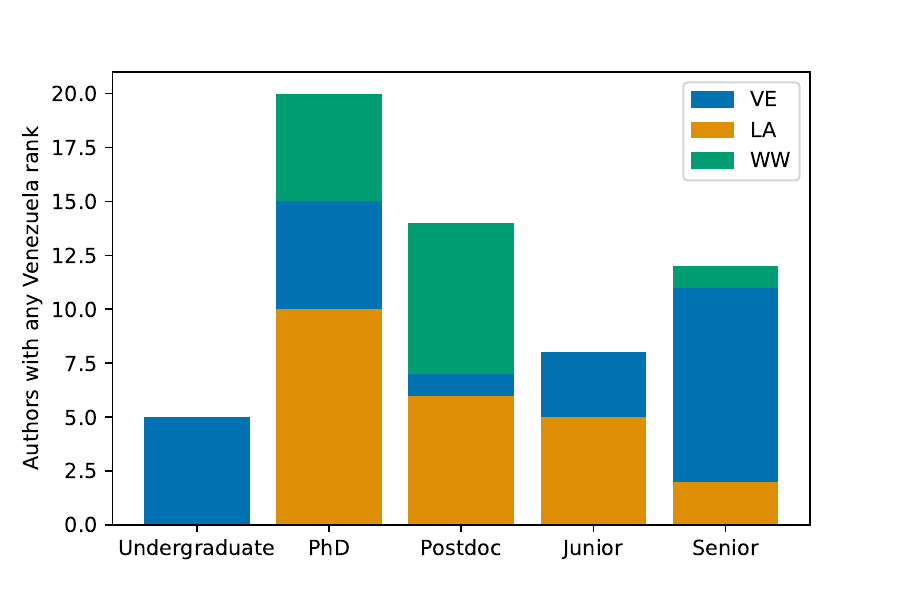}
            \caption[]%
            {{\small Venezuela.}}    
            \label{fig:Venezuela:rank-paper}
        \end{subfigure}
        \caption[]
        {\small Number of authors with a certain rank in a given country as in table \ref{table:rank-authors} who published an article under a given rank in the same country, Latin America or elsewhere worldwide.} 
        \label{fig:Rank_Mexico-Venezuela}
    \end{figure*}

In this appendix, we present the non-trivial plots for each country in the database, based on the rank author-article analysis discussed in Section \ref{sec:auths_ranks}:  

\begin{itemize}
    \item Figure \ref{fig:Rank_Argentina-Colombia}: Argentina, Brazil, Chile, and Colombia.
    \item Figure \ref{fig:Rank_Cuba-Honduras}: Cuba, Ecuador, Guatemala, and Honduras.
    \item Figure \ref{fig:Rank_Mexico-Venezuela}: Mexico, Peru, Uruguay, and Venezuela.
\end{itemize}

%

%
%

\section{Primary arXiv category}
\label{app:arxiv}

See Tables \ref{tab:arXiv:LA} and \ref{tab:arXiv:Chile}.

\begin{table}[]
    \centering
\begin{tabular}{lr}
\toprule
 & count \\
Primary arXiv category &  \\
\midrule
None & 16946 \\
hep-th & 8616 \\
hep-ph & 6941 \\
gr-qc & 6058 \\
astro-ph & 3564 \\
nucl-th & 1365 \\
quant-ph & 1169 \\
astro-ph.CO & 907 \\
math-ph & 620 \\
astro-ph.HE & 545 \\
cond-mat.stat-mech & 214 \\
hep-lat & 197 \\
cond-mat & 191 \\
physics.gen-ph & 181 \\
astro-ph.GA & 158 \\
hep-ex & 135 \\
cond-mat.str-el & 133 \\
physics & 124 \\
cond-mat.mes-hall & 118 \\
\bottomrule
\end{tabular}
\caption{Primary arXiv category for all the Latin-American countries with at least 50 publications.}
    \label{tab:arXiv:LA}
\end{table}

\begin{table}[]
    \centering
\begin{tabular}{lr}
\toprule
 & count \\
Primary arXiv category &  \\
\midrule
hep-th & 1454 \\
astro-ph & 1189 \\
None & 1135 \\
hep-ph & 1107 \\
gr-qc & 881 \\
astro-ph.CO & 154 \\
quant-ph & 70 \\
astro-ph.GA & 66 \\
math-ph & 61 \\
nucl-th & 58 \\
astro-ph.HE & 44 \\
astro-ph.SR & 24 \\
cond-mat.mes-hall & 22 \\
physics.gen-ph & 14 \\
cond-mat & 10 \\
astro-ph.IM & 10 \\
\end{tabular}
    \caption{Primary arXiv category for Chile  with at least 10 publications.}
    \label{tab:arXiv:Chile}
\end{table}


\section{Acronyms}
\label{app:acro}
\begin{acronym}[]

\acro{CIF}{International Centre of Physics}
\acro{CLAF}{Latin-American Centre for Physics}
\acro{CURCAF}{Central American Course of Physics}
\acro{CYTED}{(Ibero-American Programme for Science and Technology for Development}
\acro{GDP}{Gross Domestic Product}
\acro{HDI}{Human Development Index}
\acro{HECAP}{ High Energy Physics, Cosmology and Astroparticle Physics}
\acro{HLSG}{High Level Strategic Group}
\acro{HEP}{High Energy Physics}
\acro{ICTP}{International Centre for Theoretical Physics}
\acro{PMI}{per million inhabitants}
\acro{RD}[R\&D]{Research and Development}
\acro{UN}{United Nations}

\end{acronym}







\bibliography{sn-bibliography}

\end{document}